\documentclass[longauth]{aa} 
\usepackage{graphicx}
\usepackage{epsfig}
\usepackage[]{placeins}

\usepackage{txfonts}
\usepackage{natbib}
\usepackage[usenames]{color}
\usepackage{subfig}
\usepackage{multirow}
\newcommand{\funits}{10$^{-16}$ erg~s$^{-1}$~cm$^{-2}$~\AA$^{-1}$}

\DeclareRobustCommand{\ion}[2]{%
\relax\ifmmode
\ifx\testbx\f@series
{\mathbf{#1\,\mathsc{#2}}}\else
{\mathrm{#1\,\mathsc{#2}}}\fi
\else\textup{#1\,{\mdseries\textsc{#2}}}%
\fi}

\newcommand{\hh}{\ion{H}{ii}~}

\newcommand{\nii}{[\ion{N}{ii}]}

\newcommand{\oiii}{[\ion{O}{iii}]}

\newcommand{\kms}{km\,s$^{-1}$} 
\newcommand{\lam}{$\lambda$}

\newcommand{\ha}{H$\alpha$} 
\newcommand{\hb}{H$\beta$}

\usepackage[ps2pdf,pdfpagemode=UseThumbs,colorlinks=false,bookmarks=true]{hyperref}

\begin{document}
  \title{The effects of spatial resolution on Integral Field Spectrograph surveys at different redshifts $-$ The CALIFA perspective. }

   \author{D. Mast\inst{1,2}
    \and
    F.~F.~Rosales-Ortega\inst{3,4}
     \and
     S.~F.~S\'anchez\inst{2,1}
     \and
     J.~M.~V\'{i}lchez\inst{2}
     \and
     J.~Iglesias-Paramo\inst{2,1}
     \and
     C.~J.~Walcher\inst{5}
     \and
     B.~Husemann\inst{5}
     \and
     I.~M\'arquez\inst{2}
     \and
     R.~A.~Marino\inst{6}
     \and
     R.~C.~Kennicutt\inst{7}
     \and
     A.~Monreal-Ibero\inst{2,5}
     \and
     L.~Galbany\inst{8}
     \and
     A.~de~Lorenzo-C\'aceres\inst{9}
     \and
     J.~Mendez-Abreu\inst{10,11}
     \and
     C.~Kehrig\inst{2}
     \and
     A.~del Olmo\inst{2}
     \and
     M.~Rela\~no\inst{12}
     \and
     L.~Wisotzki\inst{5}
     \and
    E.~M\'armol-Queralt\'o\inst{10,11}
     \and
    S.~Bekerait\`e\inst{5}
    \and
     P.~Papaderos\inst{13}
     \and
      V.~Wild \inst{9}
    \and
     J.~A.~L.~Aguerri\inst{10,11}
     \and
     J.~Falc\'on-Barroso\inst{10,11}
    \and
    D.~J.~Bomans\inst{14,15}
    \and
    B.~Ziegler\inst{16}
    \and
    B.~Garc\'{i}a-Lorenzo\inst{10,11}
    \and
    J.~Bland-Hawthorn\inst{17}
    \and
    \'A.~R.~L\'opez-S\'anchez\inst{18,19}
        \and
   G. van~de~Ven\inst{20}
  }

   \institute{Centro Astron\'omico Hispano Alem\'an, Calar Alto, (CSIC-MPG),
  C/Jes\'us Durb\'an Rem\'on 2-2, E-04004 Almer\'{i}a, Spain.
   \email{dmast@caha.es}
   \and
   Instituto de Astrof\'\i sica de Andaluc\'\i a (CSIC), Glorieta de la Astronom\'{i}a s/n, E18008, Granada, Spain
   \and
   Instituto Nacional de Astrof{\'i}sica, {\'O}ptica y Electr{\'o}nica, Luis E. Erro 1, 72840 Tonantzintla, Puebla, Mexico
   \and
   Departamento de F{\'i}sica Te{\'o}rica, Universidad Aut{\'o}noma de Madrid, 28049 Madrid, Spain.
   \and
   Leibniz Institute for Astrophysics Potsdam, An der Sternwarte 16, D-14482 Potsdam, Germany
   \and
   CEI Campus Moncloa, UCM-UPM, Departamento de Astrof\'{i}sica y CC$.$ de la Atm\'{o}sfera, Facultad de CC$.$ F\'{i}sicas, Universidad Complutense de Madrid, Avda.\,Complutense s/n, 28040 Madrid, Spain
   \and
   Institute of Astronomy, University of Cambridge, Madingley Road, Cambridge CB3 0HA, UK6
   \and
   CENTRA Ð Centro Multidisciplinar de Astrof\'{i}sica, Instituto Superior T\'ecnico, Av. Rovisco Pais 1, 1049-001 Lisbon, Portugal
   \and
   School of Physics and Astronomy, University of St Andrews, North Haugh, St Andrews, KY16 9SS, U.K (SUPA)
   \and
   Instituto de Astrof\'{i}sica de Canarias (IAC), Glorieta de la Astronom\'{i}a S/N, La Laguna, S/C de Tenerife, Spain
   \and
   Depto. Astrof\'{i}sica, Universidad de La Laguna (ULL), 38206 La Laguna, Tenerife, Spain 
   \and
   Dep. F\'{i}sica Te\'orica y del Cosmos, Campus de Fuentenueva, Universidad de Granada, 18071 Granada, Spain
   \and
   Centro de Astrof\'{i}sica and Faculdade de Ci\^encias, Universidade do Porto, Rua das Estrelas, 4150-762 Porto, Portugal
   \and
   Astronomisches Institut, Ruhr-Universit\"at Bochum, Universit\"atsstr. 150, 44801 Bochum, Germany 
   \and
   RUB Research Department Plasmas with Complex Interactions, Germany
   \and
   University of Vienna, T\" urkenschanzstrasse 17, 1180 Vienna, Austria
   \and
   Sydney Institute for Astronomy, School of Physics A28, University of Sydney, 2006 NSW, Australia
   \and
   Australian Astronomical Observatory, PO Box 915, North Ryde, NSW 1670, Australia 
   \and
   30 Department of Physics and Astronomy, Macquarie University, NSW 2109, Australia 
   \and
   Max Planck Institute for Astronomy, K\"onigstuhl 17, 69117 Heidelberg, Germany
 }
 
 \date{Received ----- ; accepted ---- }

 
  \abstract
  {Over the past decade, 3D optical spectroscopy has become the preferred tool
      for understanding the properties of galaxies and is now increasingly used to 
      carry out galaxy surveys.   Low redshift surveys include SAURON, DiskMass, ATLAS3D, 
      PINGS and VENGA. At redshifts above 0.7, surveys such as MASSIV, SINS, GLACE, and 
      IMAGES have targeted the most luminous galaxies to study mainly their kinematic
      properties. The on-going CALIFA survey ($z\sim0.02$) is the first
      of a series of upcoming Integral Field Spectroscopy (IFS) surveys with
      large samples representative of the entire population of
      galaxies. Others include SAMI and MaNGA at lower redshift and the
      upcoming KMOS surveys at higher redshift. Given the importance of
      spatial scales in IFS surveys, the study of the effects of spatial
      resolution on the recovered parameters becomes important.}
   { We explore the capability of the CALIFA survey and a hypothetical higher redshift survey to reproduce the properties of a sample of objects observed with better spatial resolution at lower redshift.}
   {Using a sample of PINGS galaxies, we simulate observations at different redshifts. We then study the behaviour of different parameters as the spatial resolution degrades with increasing redshift. }
   {We show that at the CALIFA resolution, we are able to measure and map common observables in a galaxy study: the number and distribution of \hh regions (\ha\, flux structure), the gas metallicity (using the O3N2 method), the gas ionization properties (through the \nii/\ha\, and \oiii/\hb\, line ratios) and the age of the underlying stellar population (using the D4000 index). This supports the aim of the survey to characterise the observable properties of galaxies in the Local Universe. Our analysis of simulated IFS data cubes at higher redshifts highlights the
importance of the projected spatial scale per spaxel as the most
important figure of merit in the design of an integral field survey.}
   {}

   \keywords{ techniques: spectroscopic -- galaxies:
     abundances -- stars: 
     formation -- galaxies: 
     ISM -- galaxies: stellar content
               }
               
 \titlerunning{The effects of spatial resolution on Integral Field Spectrograph surveys at different redshifts.}

 \maketitle
%

\section{Introduction}

The \textit{Calar Alto Legacy Integral Field Area Survey} (CALIFA, $0.005 < z < 0.03$; \citealt{Sanchez:2012a}) aims to characterise spectroscopically the galaxy population in the Local Universe, and on completion will be the largest and most comprehensive wide-field integral field spectrograph (IFS) survey carried out to date. With its statistically significant sample of $\sim 600$ galaxies, CALIFA will form the bridge between large single aperture surveys and detailed studies of individual galaxies.

Other surveys in the Local Universe using the power of integral-field
spectrophotometers for a detailed study of nearby galaxies include the SAURON
project \citep{2002MNRAS.329..513D}  and its extension ATLAS3D ($z < 0.01$ ;
\citealp{2011MNRAS.413..813C}), VIRUS-P (VENGA,
\citealp{2010ASPC..432..180B}), the DiskMass Survey
\citep{2010ApJ...716..198B} and the PINGS survey
(\citealp{2010MNRAS.405..735R}, $z \sim 0.002$, hereafter RO10).  Upcoming
surveys are SAMI \citep{2012MNRAS.421..872C} and
MaNGA\footnote{\url{http://www.sdss3.org/future/manga.php}} ($z \sim
0.05$). At high redshift there is the \textit{SINS} (Spectroscopic Imaging
survey in the Near-infrared with SINFONI) survey ($z\sim2$;
\citealp{2009umei.confE..44F}), the \textit{MASSIV} (Mass Assembly Survey with
SINFONI in VVDS) survey ($z\sim1$; \citealp{2012A&A...546A.118V}), the
\textit{GLACE} (GaLAxy Cluster Evolution) survey ($z\sim0.6$;
\citealp{2011hsa6.conf..353S})  and the MAss Galaxy Evolution Sequence
  (IMAGES) ($z\sim1$; \citealp{2007A&A...465.1099R}).  

In order to map the properties of local galaxies, a compromise must be made between the observing time invested on each object and the number of objects of the sample. The mosaicking approach designed for the PINGS survey, which mapped the \hh regions over the whole extension of the galaxies and explored the 2-D metallicity structure of disks, involved more than 30 pointings and three years of observation. In contrast, the SAURON project opted for single pointings of the central part of early-type galaxies and bulges of spirals. 

The CALIFA survey was designed with the aim of obtaining a balance between a detailed study of each galaxy and a statistically significant sample to fully characterise the optical properties of galaxies in the Local Universe. With this in mind, the sample was selected to have optical sizes that fit within the field-of-view (FoV) of the PPak instrument \citep{2005PASP..117..620R,2004AN....325..151V}, allowing us to map their physical properties over their whole optical extent in just one pointing. In this context, one of the aims of the present paper is to determine the capability of CALIFA to measure the main properties of the observed galaxies despite resolution limitations.

\begin{figure*}[ht]
  \centering	
  \includegraphics[width=18cm]{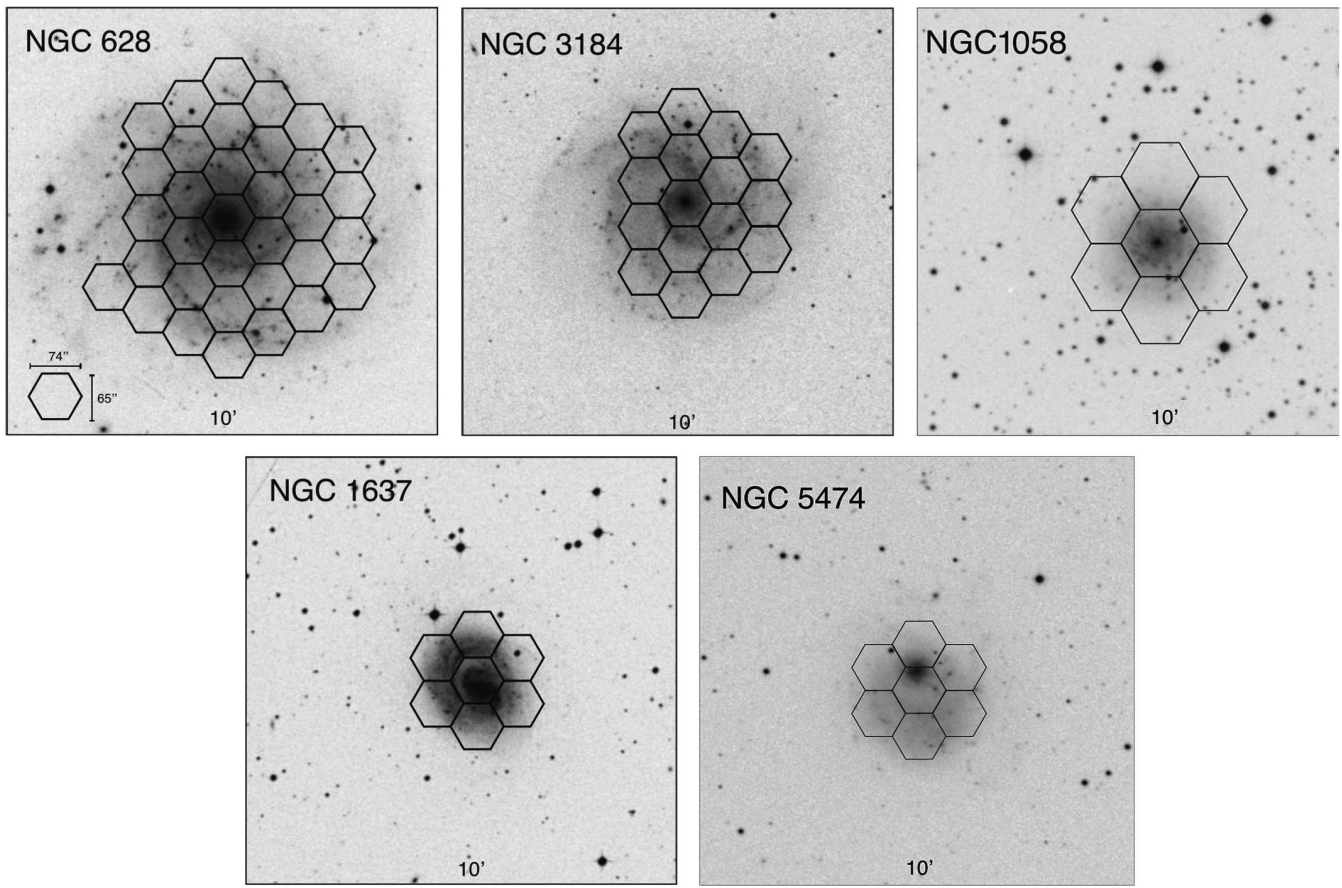}
  \caption{Digital Sky Survey $R$-band images of the five galaxies selected
    from PINGS for this study (adapted from RO10). The IFS mosaicking is
    indicated by the superposition of the hexagonal PPak FoVs. Images are
    $10\times10$ arcmin$^2$. North is up, East is to the left.}
  \label{fig:sample}
\end{figure*}

Looking ahead to the next decade when new proposed surveys will explore the intermediate and high redshift regime, in this paper we also study how the information loss due to spatial resolution degradation will affect an hypothetical Integral Field Spectroscopy (IFS) survey at higher redshift. Taking advantage of the superb spatial sampling of a subsample of PINGS galaxies ($z\sim 0.0009-0.0023, \approx 0.04$ kpc/\arcsec), we create two simulations: simulation $Z1$ represents CALIFA at $z\sim0.02$,  and simulation $Z2$ represents an hypothetical survey at $z\sim0.05$, or other redshift ranges depending on the instrument. In what follows, we refer to the PINGS redshift regime at an average $z\sim0.002$ as $Z0$. We will analyse the behaviour of some common observables used in galaxy studies as the spatial resolution degrades with increasing redshift. Due to the low spectral resolution of PINGS data, we are unable to study the effect of spatial resolution on kinematics because the simulated  cubes would not be representative of the higher spectral resolution CALIFA data.

Previous studies of this issue include \cite{2012A&A...538A.144V} with a detailed study of UGC 9837 ($z\sim0.008863$) repeating the analysis with simulated versions of the galaxy at higher redshifts ($z\approx1-2$). \cite{2007A&A...473..761K} model disc galaxies using N-body/hydrodynamic simulations to investigate distortions in the velocity fields at different redshifts from z=0 to z=1. \cite{1996AJ....112..369G} simulate Hubble Space Telescope ($HST$) images at cosmological distances using images from the Ultraviolet Imaging Telescope ($UIT$), or the more recent work of \cite{2008ApJS..175..105B}  where they simulate galaxies at redshifts $0.1<z<1.1$ using images from the SDSS in the u, g, r, i, and z filter bands as input. Most of these studies have focussed on the degradation of images with redshift. Examples of studies focussing on spectral effects are \cite{2006AJ....131...70L} who mapped the velocity fields of high-redshift galaxies, and \cite{2013ApJ...767..106Y} who utilized the \nii/\ha\, ratio to investigate how an inferred metallicity gradient is altered by the loss of angular resolution.

The structure of this paper is as follows. In Sect. \ref{sec:sample} we present the sample. Sect. \ref{sec:obs} summarises the observations and data reduction while Sect. \ref{sec:method} explains how we simulate the different redshift regimes. The analysis of both gaseous and stellar components of the sample is presented in Sect. \ref{analysis}. In Sect. \ref{conclusions} we review the main results of this work.

Throughout this paper, we assume a standard $\Lambda$CDM cosmology consistent with WMAP results \cite{2003ApJS..148....1B} with $\Omega_{m} = 0.3, \Omega_{A} = 0.7$ and H$_0 = 70$ h$_{70}$ km$^{-1}$ s$^{-1}$.  

\section{Sample}\label{sec:sample}

The PINGS survey is a project designed to construct 2D spectroscopic mosaics of a representative sample of 17 nearby spiral galaxies, using the Postdam Multi Aperture Spectrograph,
PMAS \citep{2005PASP..117..620R}  in the PPak mode (Verheijen et al. 2004; Kelz et al. 2006; Kelz \& Roth 2006)
at the Centro Astron\'omico Hispano Alem\'an (CAHA) at Calar Alto, Spain. PINGS is one
of the most detailed and largest IFS surveys of individual galaxies at
low redshift ($z\sim0.001$). The data cover most of the optical
extent of the galaxies, down to R$_{25}$ (the radius of the galaxy at the isophotal level of 25 mag/arcsec$^2$ in $B-band$), with one of the best spatial resolutions
and spectroscopic coverage achieved so far in any IFS survey. From the PINGS sample, we selected the five galaxies (Fig. \ref{fig:sample}) with the largest angular size
 (see Table \ref{tab:sample} for details of each galaxy). The NGC 628 mosaic subtends 34 arcmin$^2$ and was constructed from 34 different pointings, the NGC 1058 mosaic subtends $3.0\times2.8$ arcmin$^2$ and was constructed from 7 different fields, the NGC 1637 mosaic was built from one central pointing and a concentric ring of 6 pointings covering a total area of $4.0\times3.2$ arcmin$^2$, the NGC 3184 mosaic covers an area of $7.4\times6.9$ arcmin$^2$ and was constructed from 16 IFS pointings, and finally the NGC 5474 mosaic covers an area of $4.8\times4.3$ arcmin$^2$ and was built from of one central pointing and a concentric ring of 6 pointings.

\begin{table}[t]
    \caption{PINGS subsample}             
    \label{tab:sample}      
    \centering 
    \renewcommand{\footnoterule}{}         
    \begin{tabular}{l c c c c }     
      \hline
      \hline       
      Object &	Type & Dist. [Mpc] &	size [arcmin$^2$]\tablefootmark{a} & z\\
      \hline
      NGC 628	  &        SA(s)c	& 9.3	  & $10.5 \times 9.5$	& 0.00219 \\
      NGC 1058 &	SA(rs)c	& 10.6 &	$3.0\times2.8$ & 0.001728 \\
      NGC 1637 &	SAB(rs)c	& 12.0 &	 $4.0\times3.2$	& 0.00239 \\ 
      NGC 3184 &	SAB(rs)cd	 & 11.1 &	$7.4\times6.9$	 & 0.001975 \\
      NGC 5474 &	SA(s)cd	&  6.8  &	 $4.8\times4.3$	&  0.000911 \\
      \hline 
      \hline 
    \end{tabular}
  \tablefoottext{a}{Size of the galaxy down to R$_{25}$ \citep{1992yCat.7137....0D}}                 
\end{table}

\section{Observations and data reduction}\label{sec:obs}

The observations were carried out at
the 3.5m telescope of the Calar Alto observatory with the PMAS in the PPak mode.  The PPak unit features a central hexagonal bundle with 331 densely packed optical fibres to sample an astronomical object with a resolution of 2.7\arcsec/fibre, over an area of $74 \times 65$ arcsec$^2$, with a filling factor of 65 \% ($<$100\% due to
gaps in between the fibres). The sky background is sampled by 36 additional fibres distributed in 6 mini bundles of 6 fibres each, which encircle the central hexagon at a distance of $\sim$90\arcsec. All galaxies were observed using the same telescope and instrument set-up. The V300 grating was used,
covering a wavelength range of  $3700 - 7100$ \AA\, with a resolution of $\sim10$ \AA\,  FWHM, corresponding to $\sim$600 \kms. The final product of the data reduction is a set of Row-Staked-Spectra (RSS), a 2D FITS image where the X and Y axes contain the spectral and spatial information respectively, regardless of their position in the sky. An additional file links the different spatial elements to position on the sky. RSS spectra present a discontinuous sampling of the sky, so we performed an interpolation to obtain a regularly sampled 3D cube (datacube). Not all PINGS datacubes have the same final sampling. NGC 1637, NGC 1058, NGC 3184 and  NGC 5474 have 1\arcsec/spaxel sampling, while NGC628 has a sampling of 2\arcsec/spaxel. For more details on the observing strategy, data reduction (similar to that used for the CALIFA data) and sample selection of the PINGS survey, see RO10.

\section{Methodology}\label{sec:method}

The aim of this work is to study how the loss of spatial resolution due to redshift affects the derivation of physical parameters obtained from IFS observations at different redshifts. We study three different spatial scales: (i) $Z0$ is at the original resolution of the PINGS data set, i.e. sampling of $\sim0.045$ kpc/\arcsec, or 80-100 spatial elements across the optical extent; (ii) $Z1$ is at the typical spatial resolution and sampling of a CALIFA galaxy of diameter $\sim60$\arcsec, with 25-30 spatial elements out to $\sim2-3$ effective radii (Re); (iii) $Z2$ corresponds to a galaxy sampled at 1~kpc/\arcsec\ or $\sim7-10$ spatial elements out to $\sim2-3$ Re i.e. a $z\sim0.05$ galaxy and instrument with $\sim1$\arcsec spatial elements.
We should note that the key parameter is the number of spatial elements across the extent of the galaxy, the fiducial redshift of the galaxy will depend on the precise survey design. We then performed the same analysis on the three sets of galaxies (i.e. the original PINGS galaxies and both simulated sets) focusing on a set of commonly measured observables. In this way, we can quantify the extent to which we can recover observables of interest, and thus achieve the aim of characterising the properties of galaxies in the Local Universe, in both CALIFA and a hypothetical higher redshift IFS survey.

\begin{table}
\caption{Spatial sampling at Z0, Z1 and Z2}             
\label{tab:simu}      
\centering 
\renewcommand{\footnoterule}{}         
\begin{tabular}{cllll|}     
\hline
\hline       
\multicolumn{4}{ c  }{NGC 628  [2\arcsec/pixel]} \\
\hline
z &	0.00219 & 0.0135 & 0.056   \\ 
pixel [kpc/pix.] & 0.09 & 0.27 & 1.08 \\
\hline
\hline 
\multicolumn{4}{ c  }{NGC 1637  [1\arcsec/pixel]} \\
\hline
z &	0.002392 & 0.007 & 0.028   \\ 
pixel [kpc/pix.] & 0.05 & 0.144 & 0.562 \\
\hline
\hline
\multicolumn{4}{ c  }{NGC 1058 [1\arcsec/pixel]} \\
\hline
z &	0.001728 & 0.005 & 0.02   \\ 
pixel [kpc/pix.] & 0.036 & 0.103 & 0.405 \\
\hline
\hline
\multicolumn{4}{ c  }{NGC 3184 [1\arcsec/pixel]} \\
\hline
z &	0.001975 & 0.004 & 0.015   \\ 
pixel [kpc/pix.] & 0.041 & 0.083 & 0.306 \\
\hline
\hline
\multicolumn{4}{ c  }{NGC 5474 [1\arcsec/pixel]} \\
\hline
z &	0.000911 & 0.003 & 0.01   \\ 
pixel [kpc/pix.] & 0.019 & 0.062 & 0.205 \\
\hline 
\hline 
\end{tabular}
\end{table}

We decided to focus this study only on resolution effects, and in particular on spatial resolution degradation due to redshift. We have not considered either surface brightness dimming or increase of noise, i.e. we take for granted that the exposure time is sufficient to reach a similar depth in all cases. Also, we have not taken into account the effects of PSF, since, so far, the ability to spatially resolve a feature in an IFS dataset is more strongly affected by the sampling than by PSF size.

\begin{figure*}
  \centering
  \includegraphics[width=0.32\linewidth,clip=true,angle=-90]{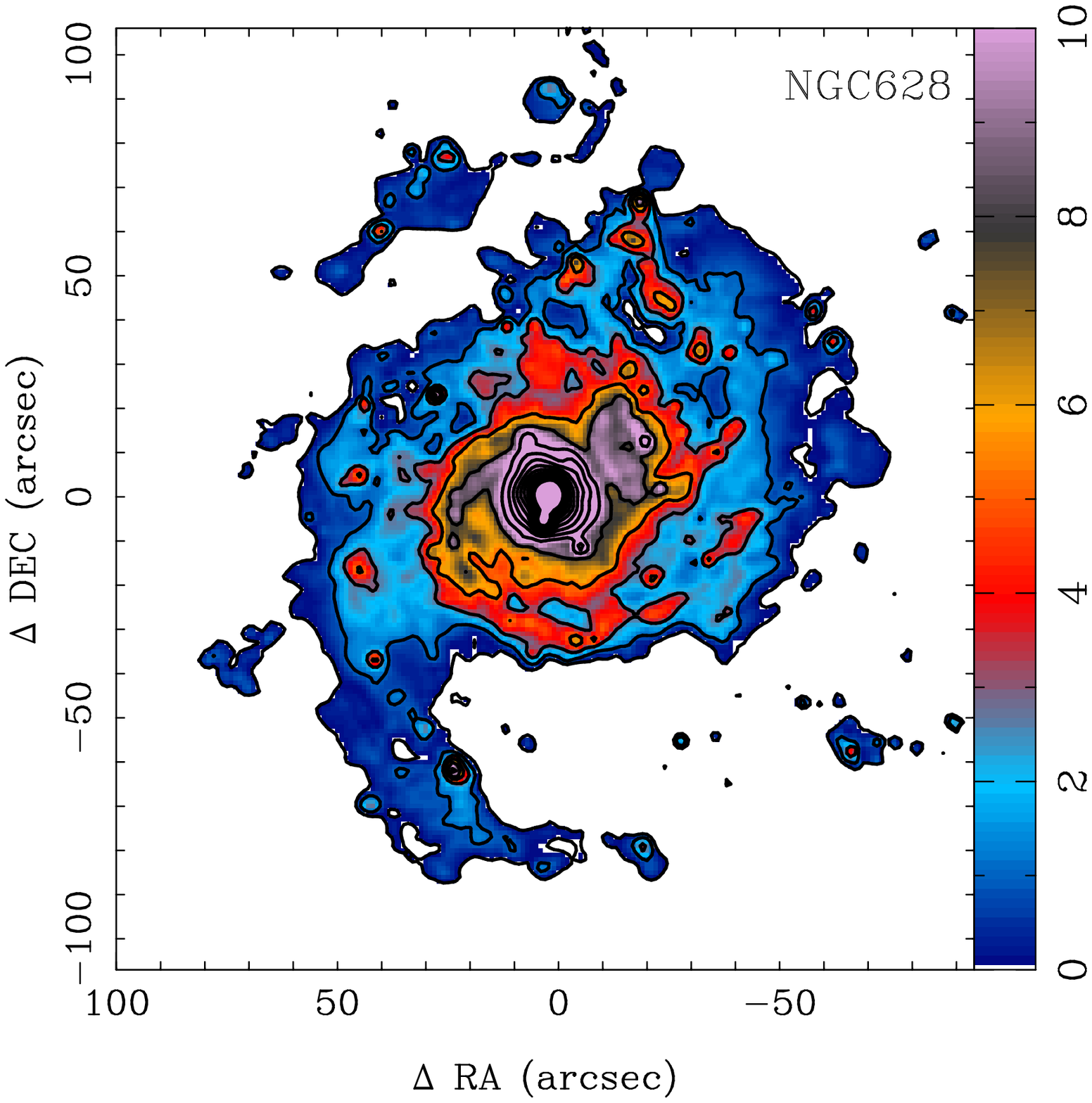}~
  \includegraphics[width=0.32\linewidth,clip=true,angle=-90]{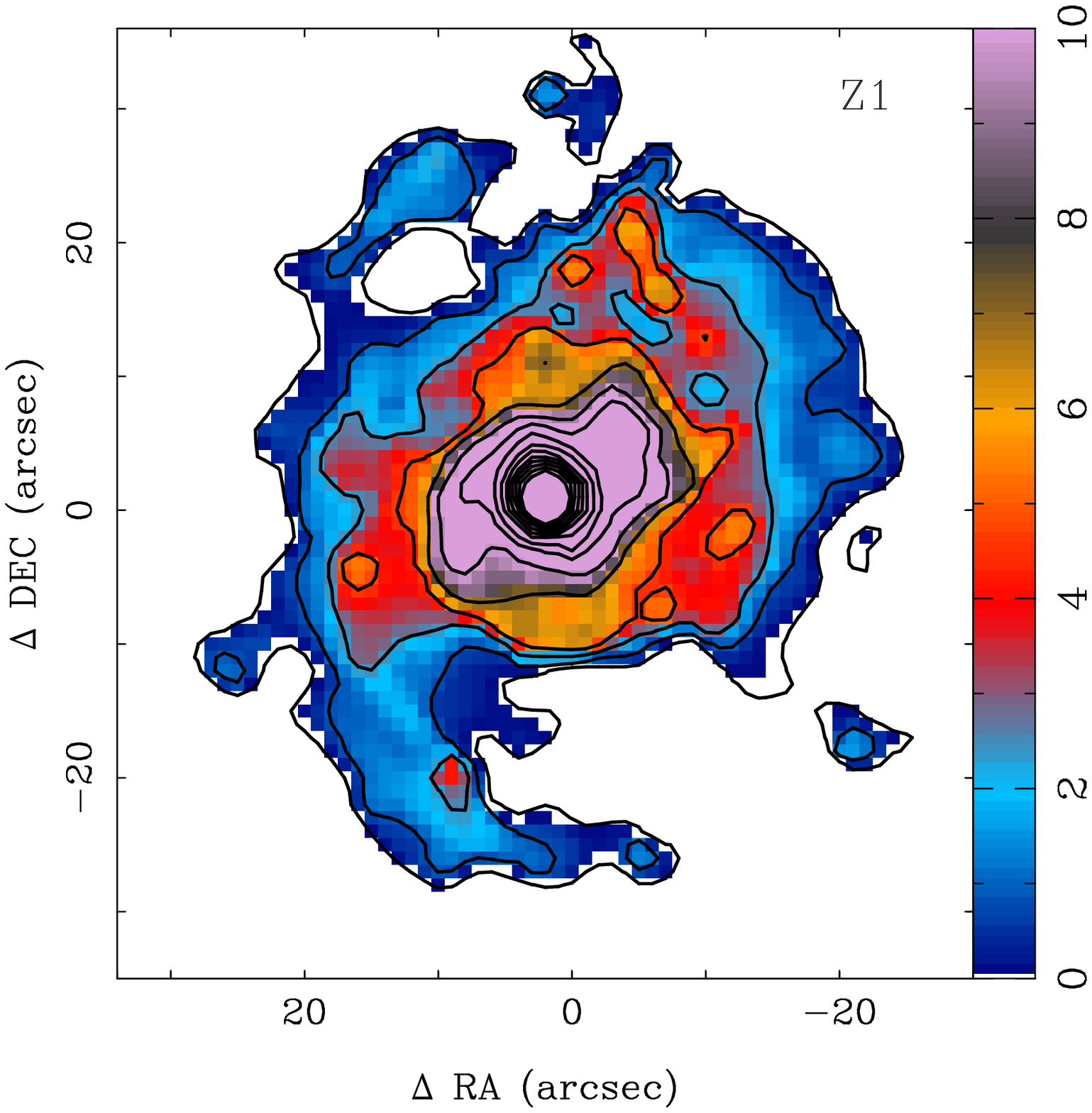}~
  \includegraphics[width=0.32\linewidth,clip=true,angle=-90]{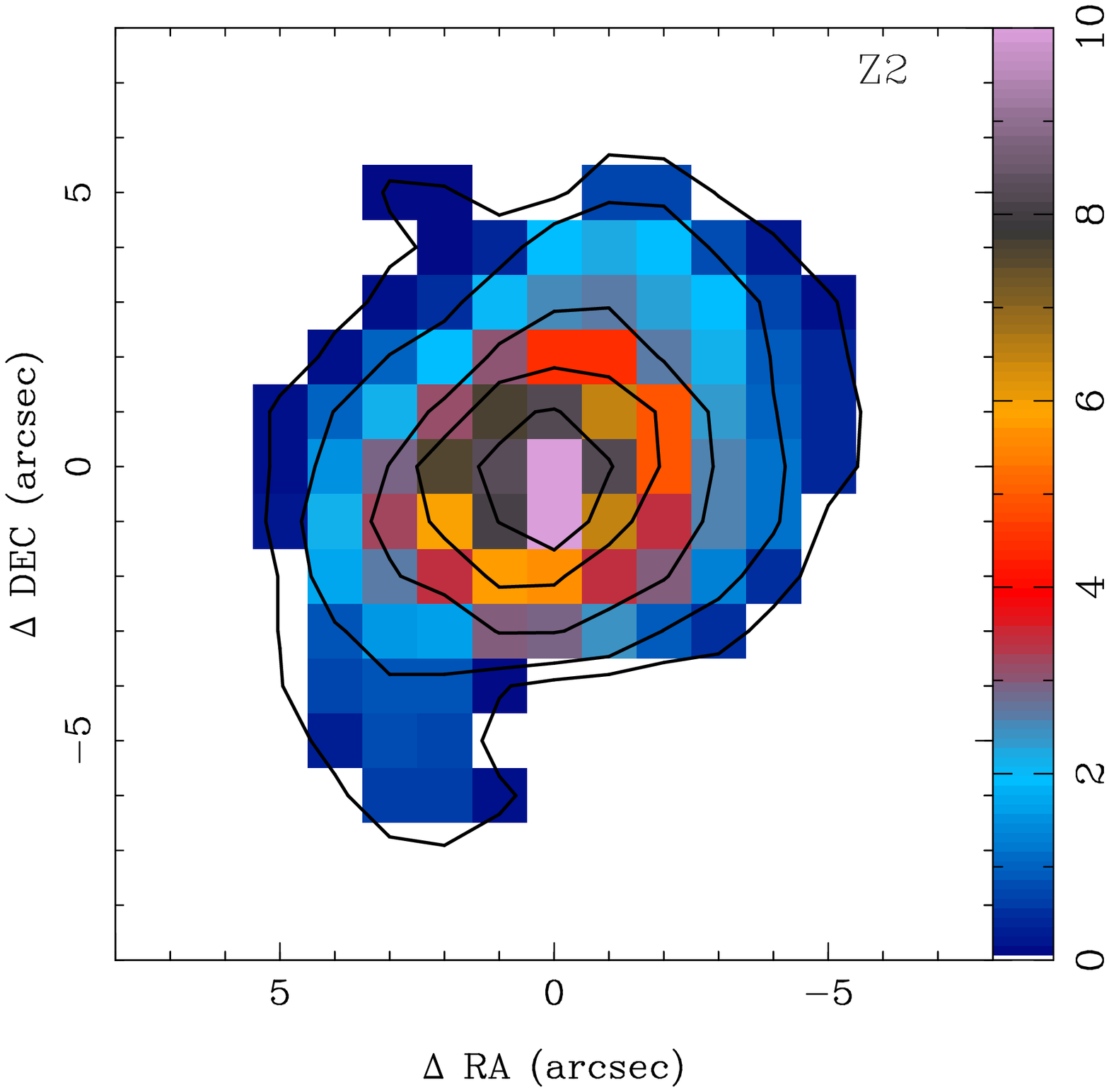}

  \includegraphics[width=0.30\linewidth,clip=true,angle=-90]{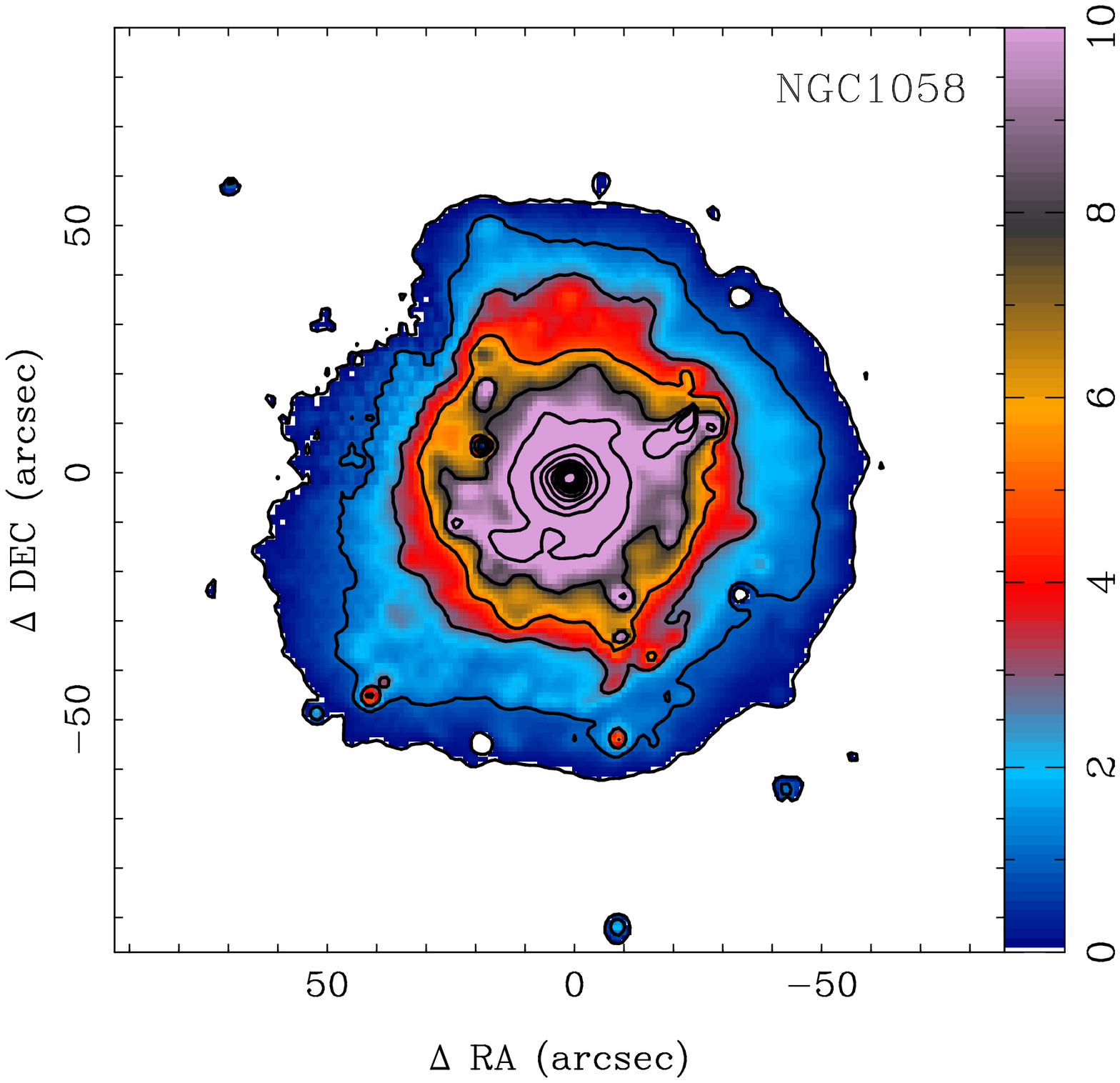}~
  \includegraphics[width=0.30\linewidth,clip=true,angle=-90]{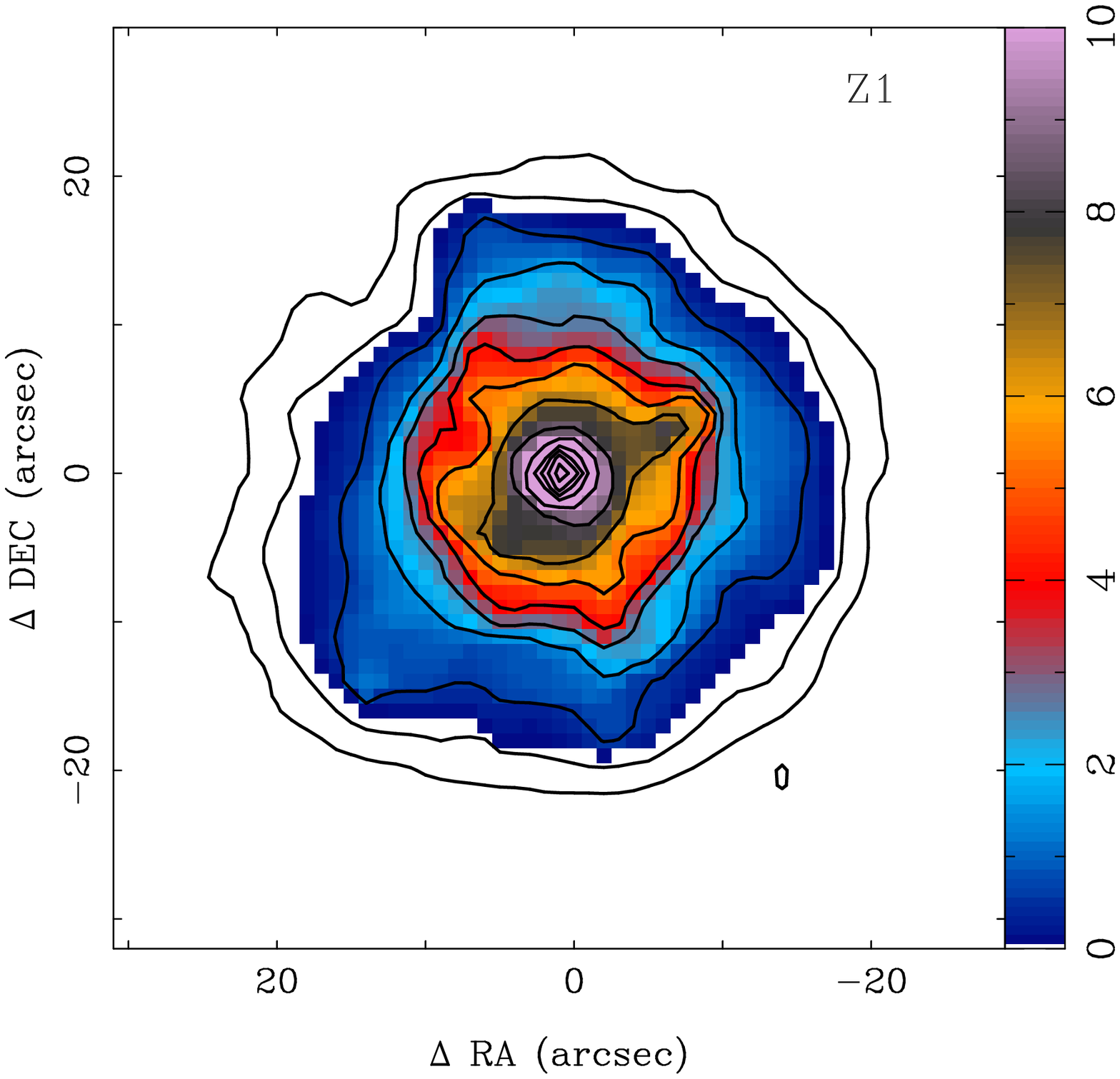}~
  \includegraphics[width=0.30\linewidth,clip=true,angle=-90]{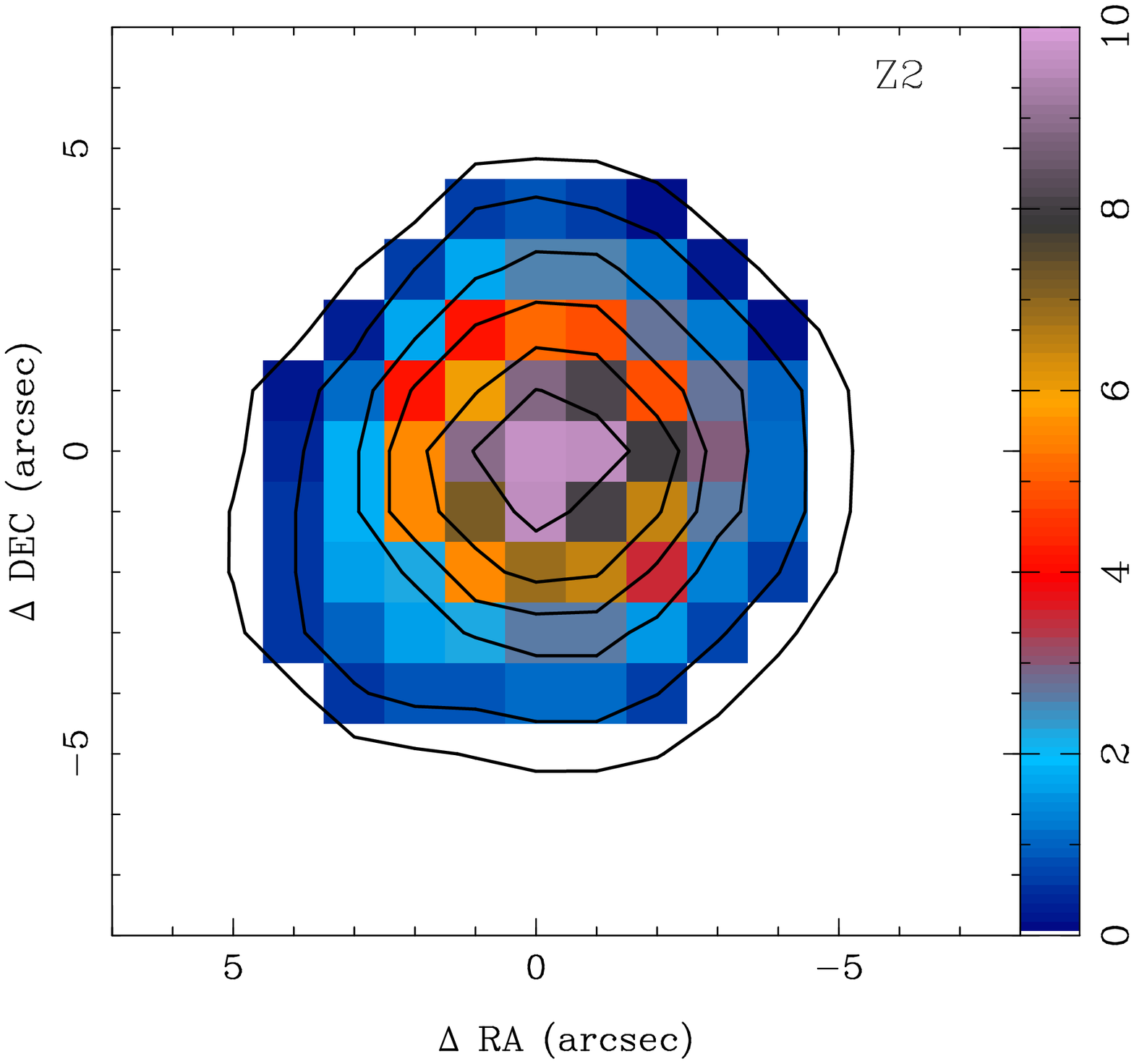}

  \includegraphics[width=0.30\linewidth,clip=true,angle=-90]{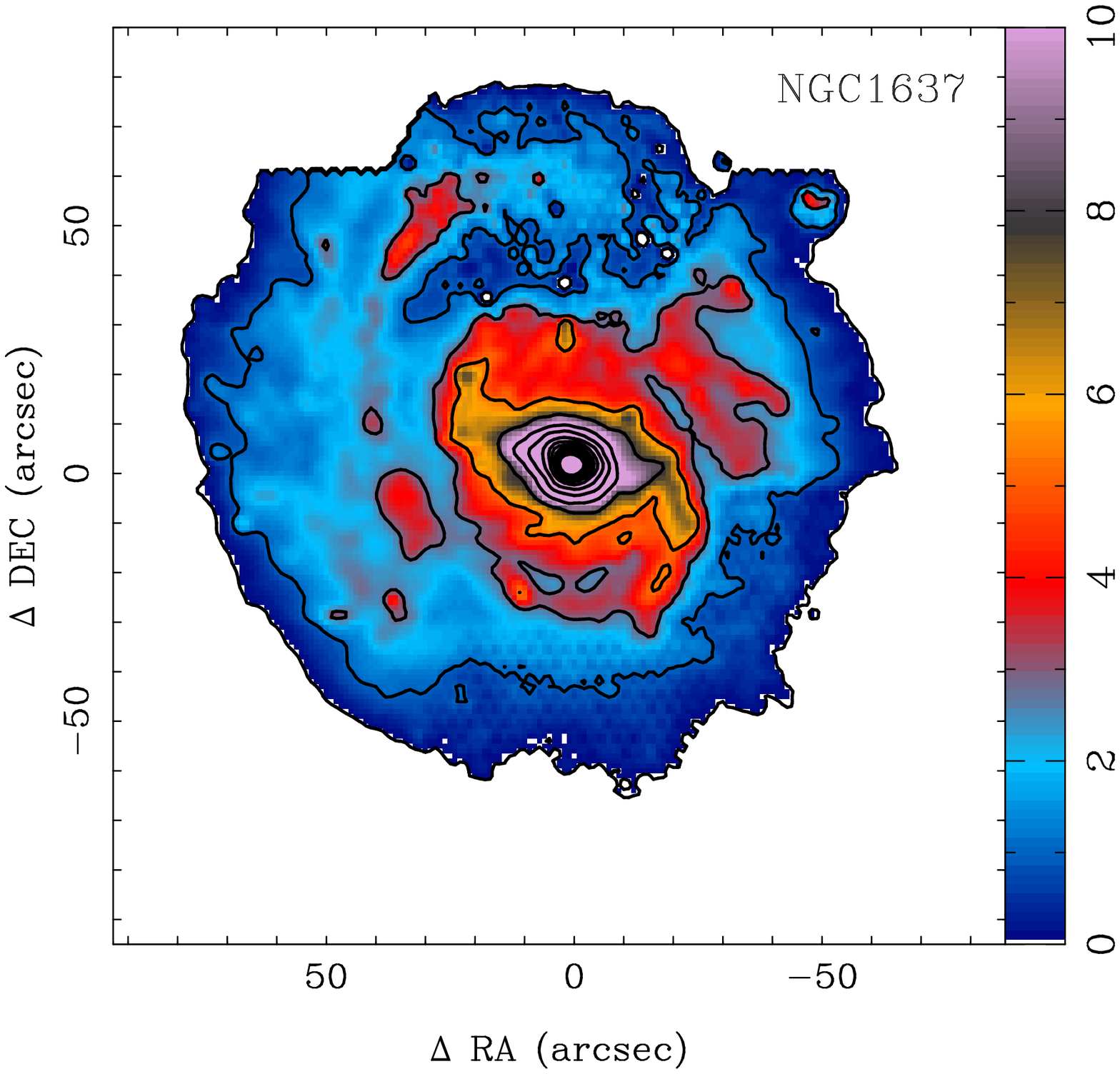}~
  \includegraphics[width=0.30\linewidth,clip=true,angle=-90]{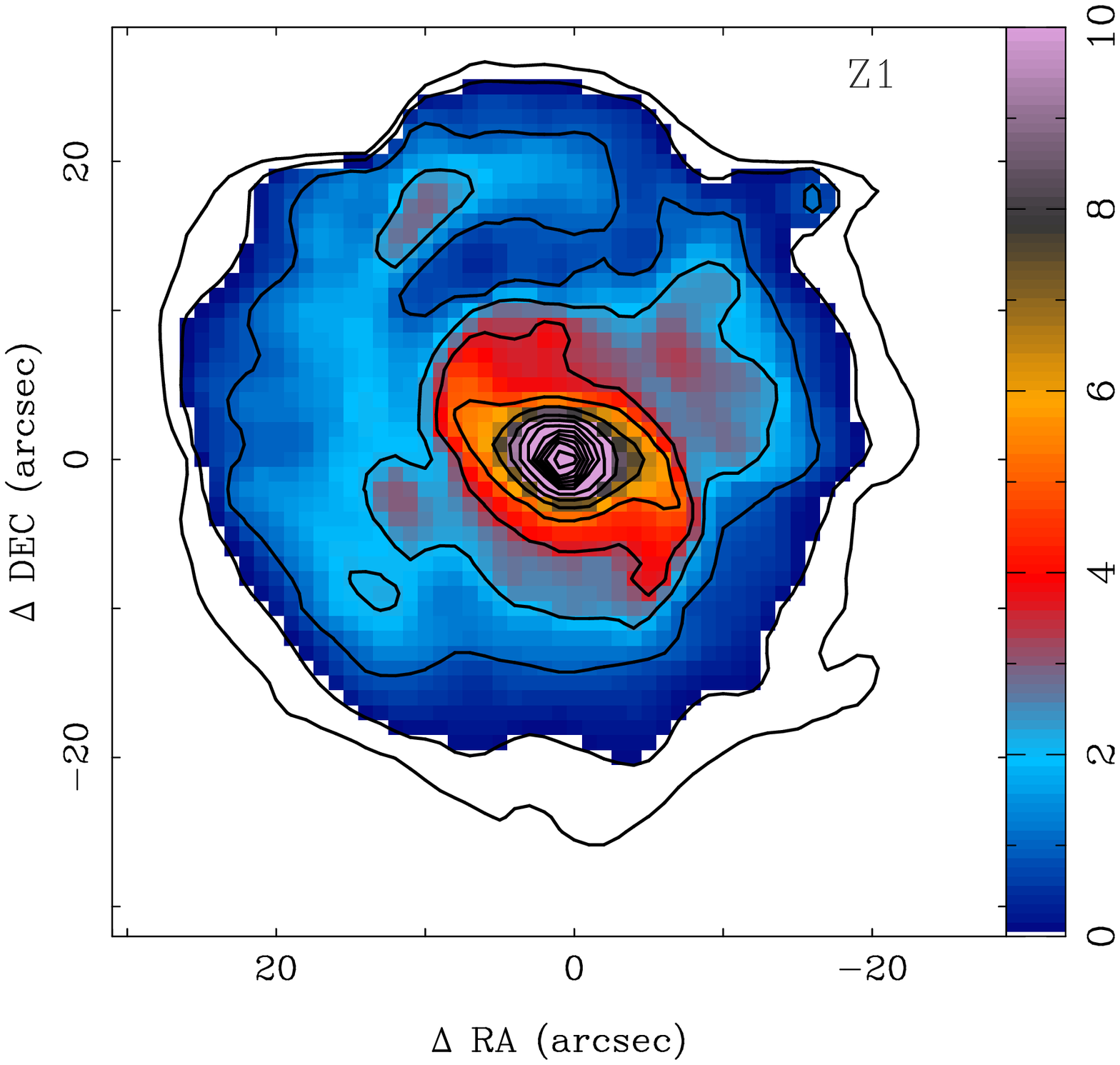}~
  \includegraphics[width=0.30\linewidth,clip=true,angle=-90]{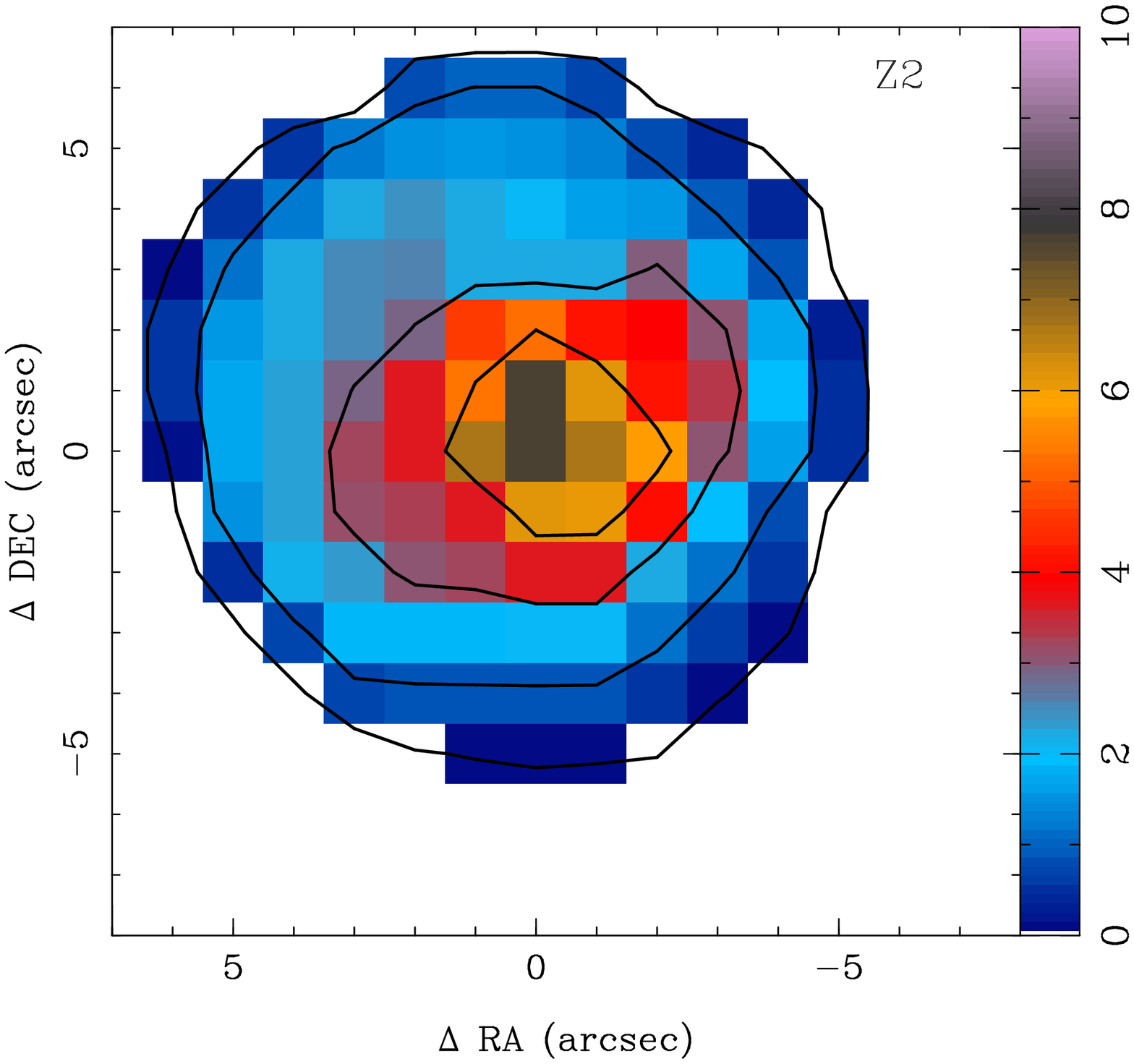}

  \includegraphics[width=0.37\linewidth,clip=true,angle=-90]{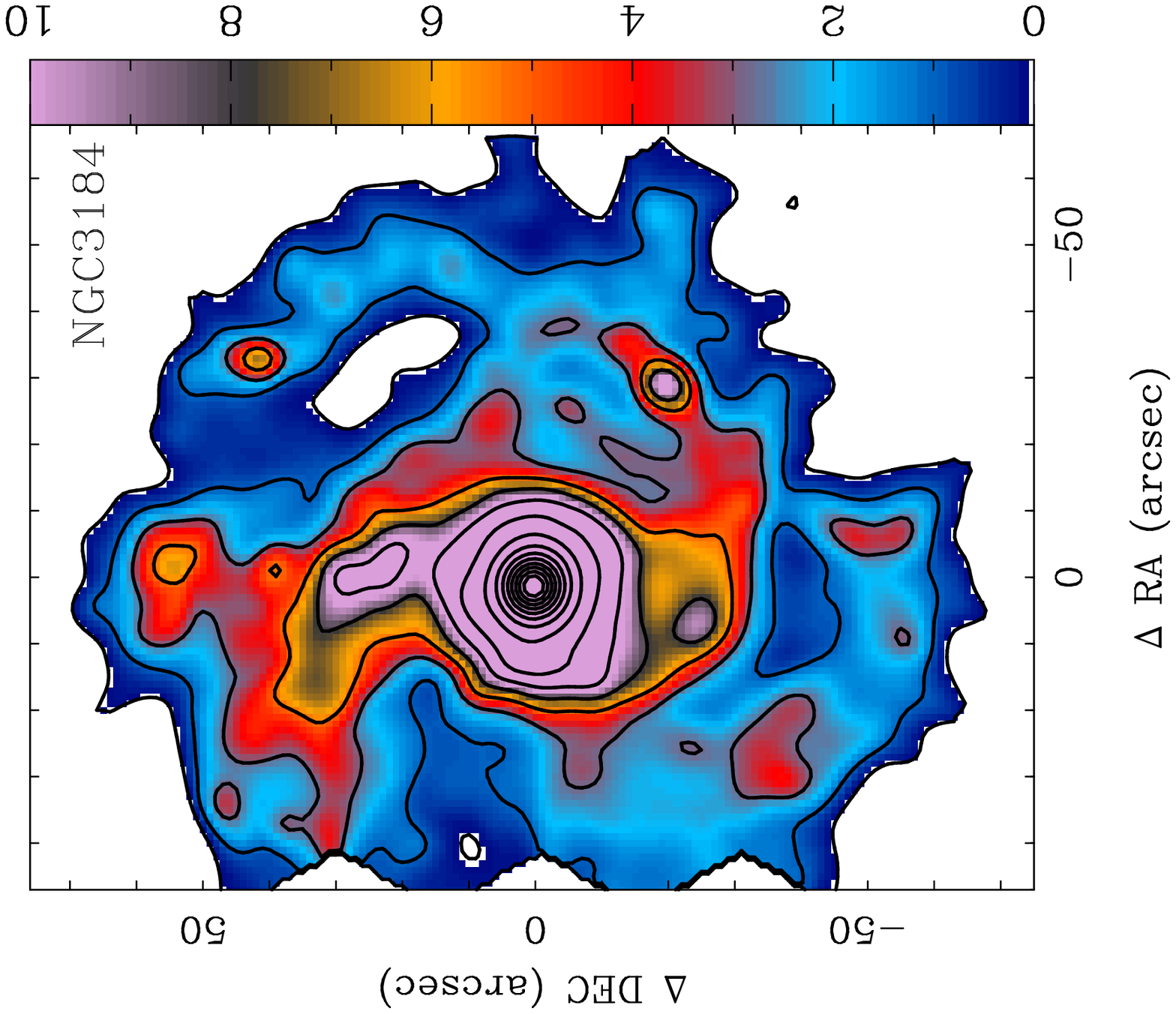}~
  \includegraphics[width=0.37\linewidth,clip=true,angle=-90]{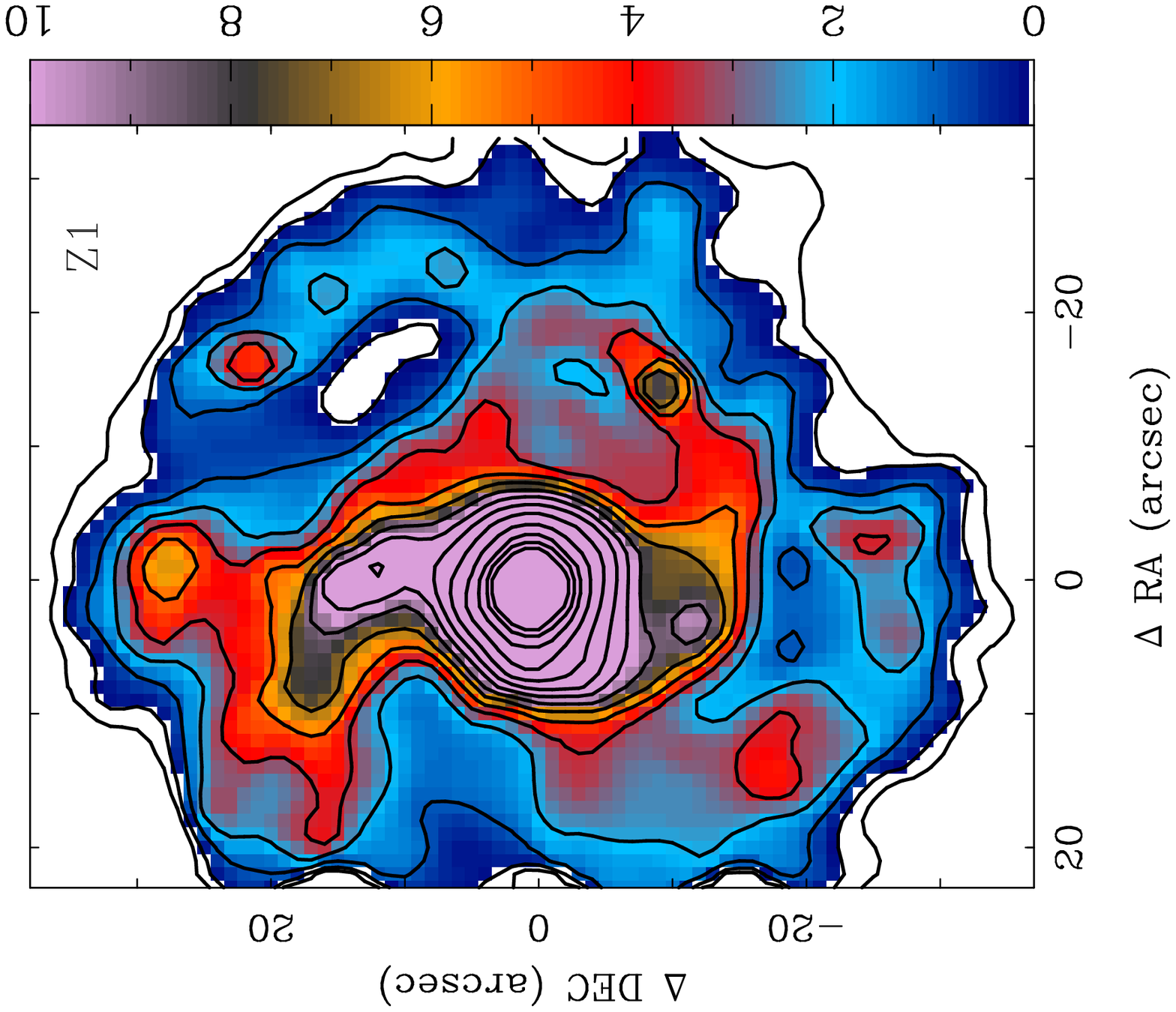}~
  \includegraphics[width=0.37\linewidth,clip=true,angle=-90]{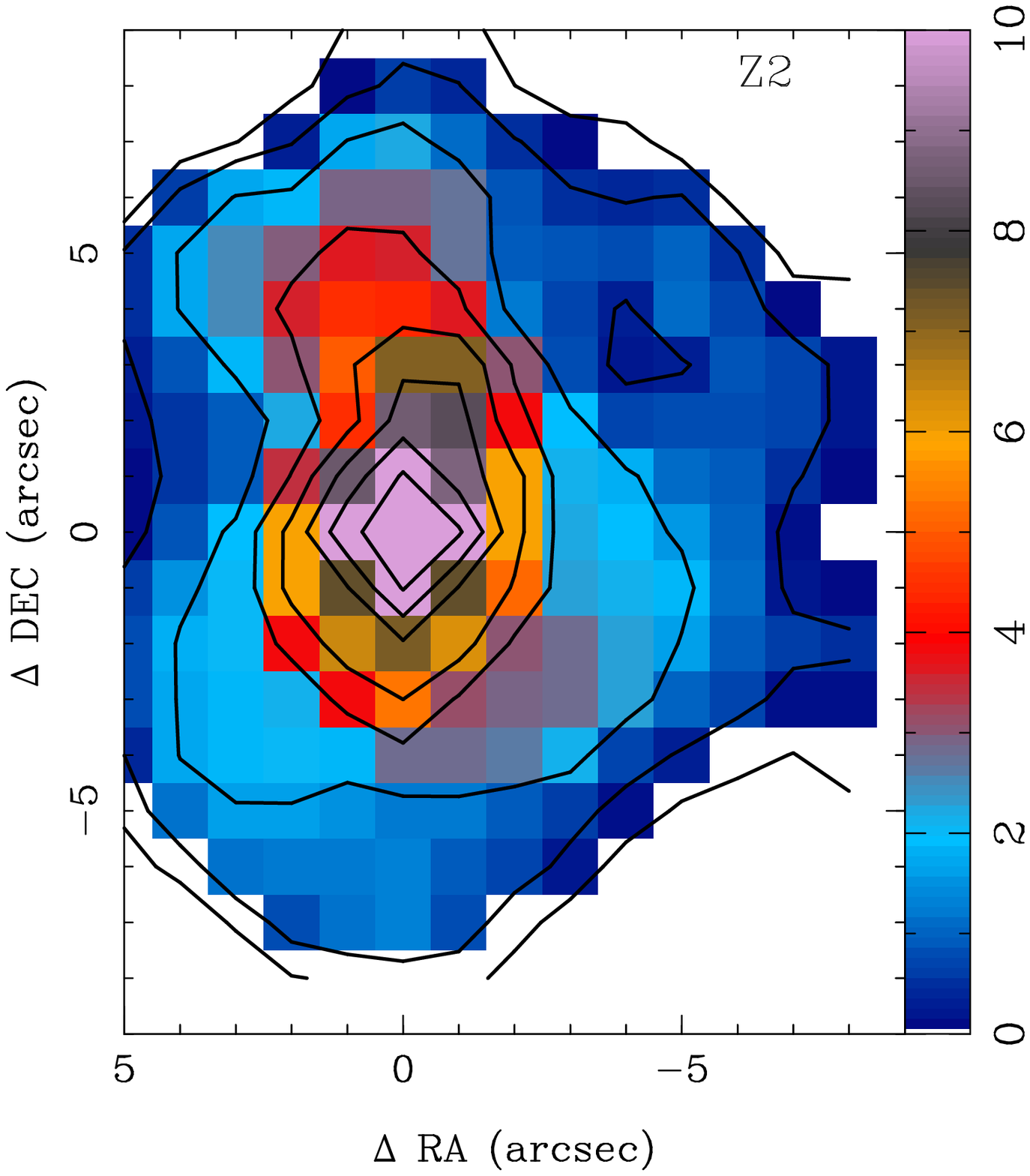}

  \caption{$V$-band images constructed from the datacubes. Flux scale in
    arbitrary units. {\it Left:} original PINGS galaxies. {\it Middle:} $Z1$
    galaxies. {\it Right:} $Z2$ galaxies. North is up, East is to the left.}
  \label{fig:simulation}
\end{figure*}

\begin{figure*}
  \ContinuedFloat
  \centering
  \includegraphics[width=0.28\linewidth,clip=true,angle=-90]{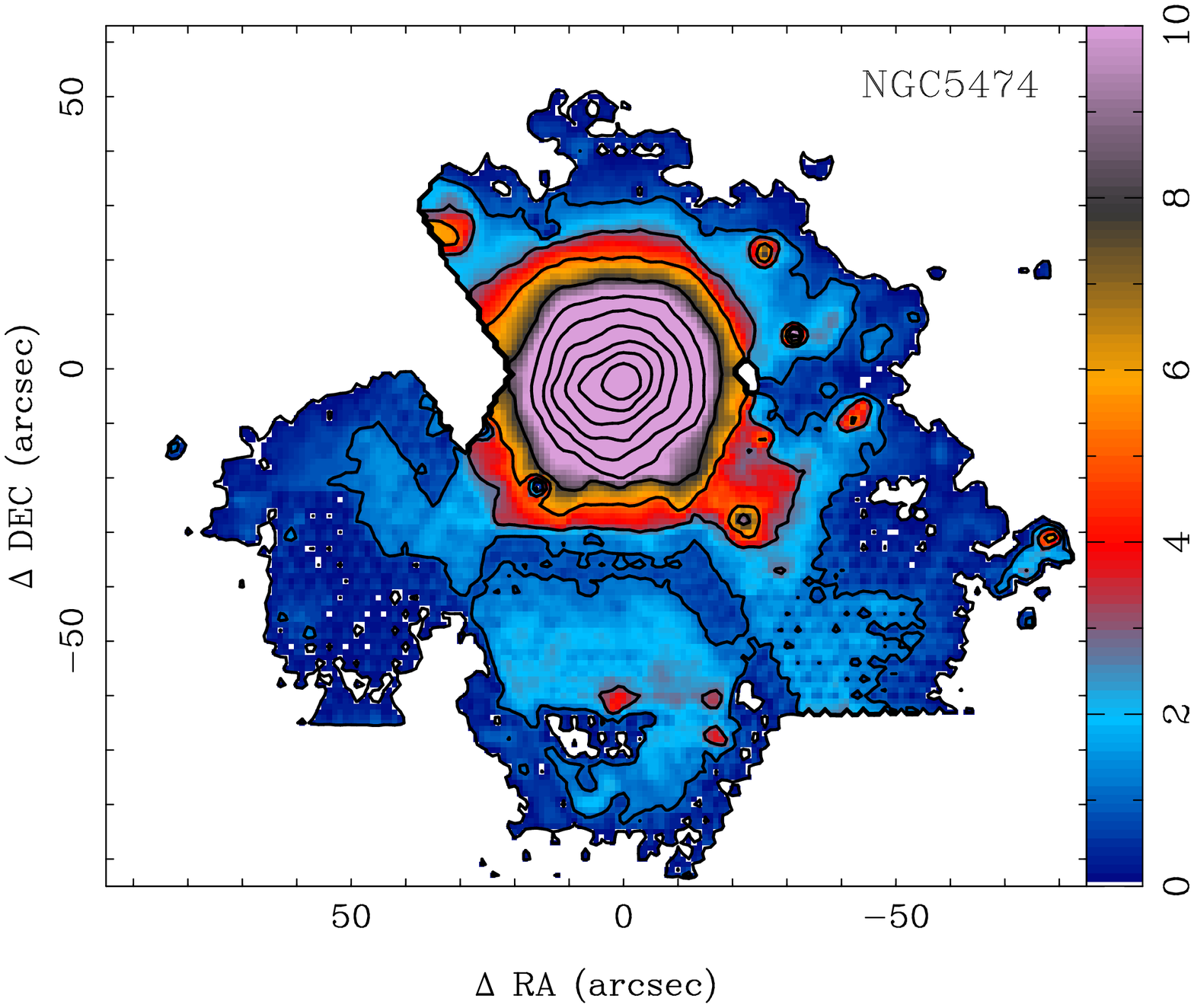}~
  \includegraphics[width=0.28\linewidth,clip=true,angle=-90]{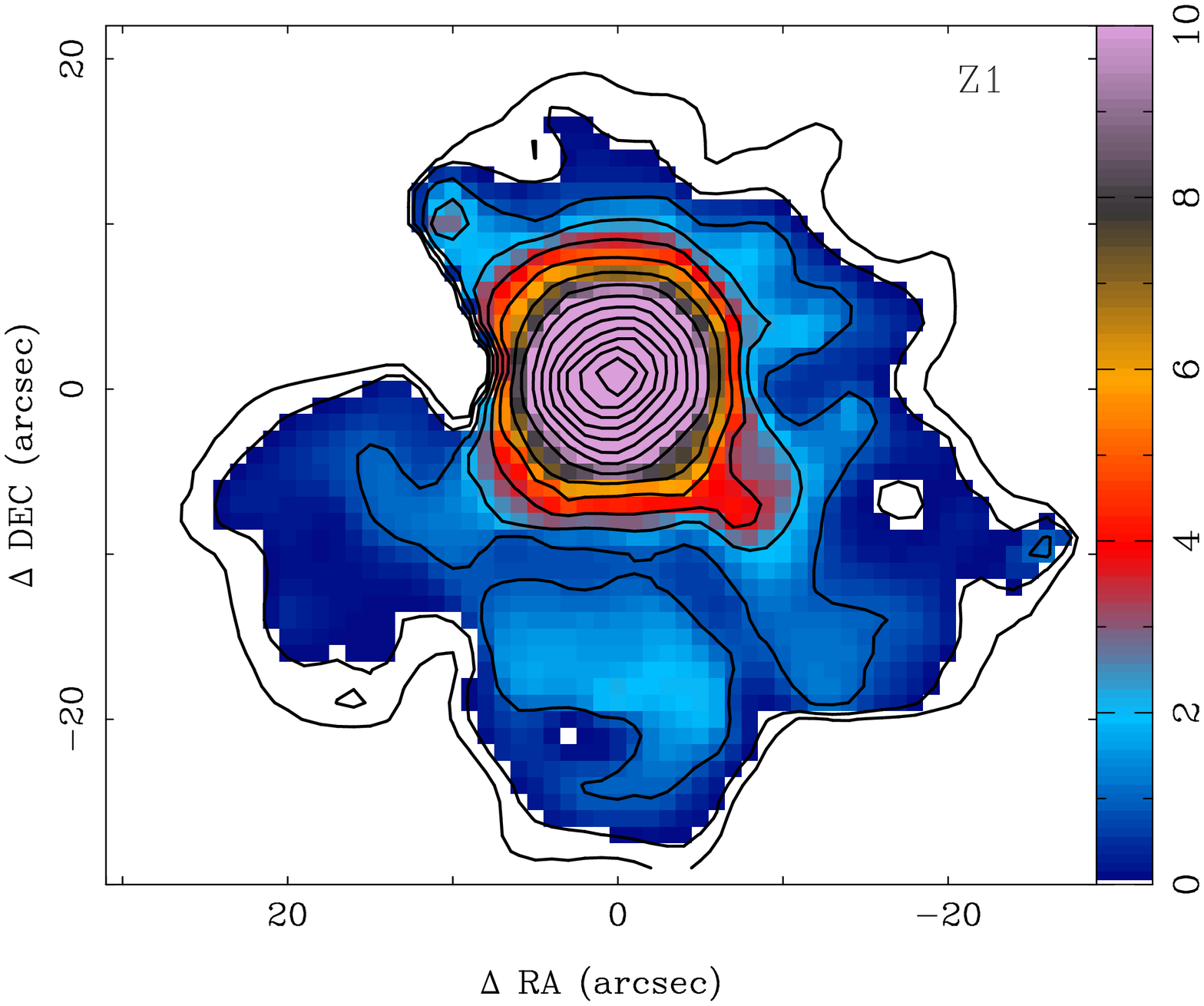}~
  \includegraphics[width=0.28\linewidth,clip=true,angle=-90]{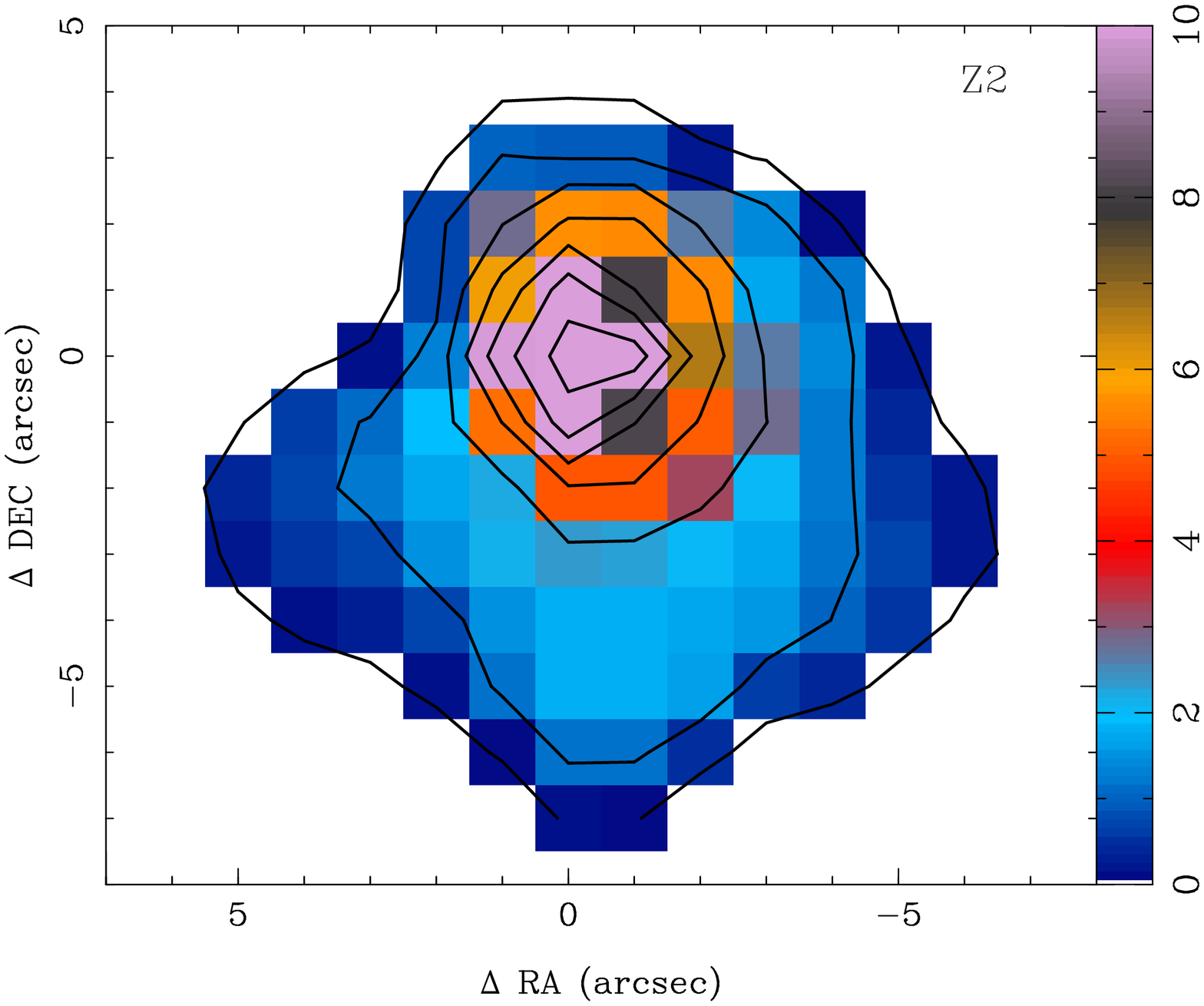}
  \caption{\it{Continued.}}
\end{figure*}

\subsection{Simulations}

We binned the PINGS cubes using R3D to match the $Z1$ (CALIFA) sampling, and then binned again these cubes to match the $Z2$ sampling. Given an original cube of size $N_x\times N_y$ and a bin factor $f$, we add all the spectra in a box of size $f\times f$ and generate a new cube of size  $N_x/f\times N_y/f$. The spatial sampling detailed above results in $f=3$ and $f=12$ for all galaxies accept for NGC 3184 where we used $f=2$ and $f=8$.  As we are not considering noise, nor cosmological dimming, we consider the mean flux of the binned pixels, rather than integrating. In the absence of noise this will be statistically identical, and makes direct comparisons much easier. 
Table \ref{tab:simu} gives the spatial sampling and fiducial redshift for each galaxy in the three surveys.  
To avoid binning artefacts such as border discontinuities, we applied a
two-dimensional Gaussian smoothing to the cubes after binning. This resembles
the effects of an interpolation kernel passed over fibre-based (dithered or
not) IFS observations. We have also tested convolution of the data before
binning, more appropriate for regularly gridded observations, such as SINFONI,
VIMOS, MUSE or GMOS.  In all tests we obtained qualitatively similar, although
quantitatively different, results. The differences do not affect the
conclusions of this work in any way. Fig. \ref{fig:simulation} shows $V$-band
images of the original and simulated galaxies. The images are constructed by
convolving a $Johnson\, V-band$ filter with the datacubes.

\section{Analysis}\label{analysis}

The aim of this paper is to understand how observed properties of \hh regions, in particular radial trends and spatial distribution, are affected by spatial resolution degradation. A complete analysis of the behaviour of derived properties and impact on physical conclusions will be the subject of forthcoming papers. 

We performed a set of standard analyses of increasing complexity: (i) reconstruction of broad-band images from the IFS
datacubes; (ii) creation of emission line flux maps; (iii) measurement of more complex
spectroscopic properties, such as line ratios, chemical abundance tracers and
ionisation indicators, following the procedures described in \cite{Sanchez:2012a,Sanchez:2012b}.

To analyse the information contained in the cubes, we used FIT3D to separate
the underlying stellar continuum from the emission lines in each spectrum,
following  the process described in detail in \cite{2011MNRAS.410..313S} and
RO10. FIT3D routines fit the underlying stellar population combining linearly
a set of stellar templates within a multi-SSP model. A simple SSP template
grid was adopted, consisting of three ages (0.09, 1.00 and 17.78 Gyr) and two
metallicities (Z$\sim0.0004$ and $0.03$). The SSP templates were taken from
the MILES project \citep{2010MNRAS.404.1639V}. The metallicities and ages
cover the largest ranges possible within the MILES library. The oldest stellar
population was selected to reproduce the reddest possible underlying stellar
population, mostly due to larger metallicities than the one considered in our
simplified model, although it is clearly older than the accepted cosmological
age of the Universe. Our youngest stellar population is the 2nd youngest in
the MILES library for which both metallicities are available. No appreciable
difference was found between using this or the youngest SSP ($\sim80$
Myr). Finally, we selected an average age stellar population, of $\sim1$ Gyr,
required to reproduce the intermediate-to-blue stellar populations, and to
produce more reliable corrections of the underlying stellar absorption.

We study commonly used indicators of gaseous and stellar
  properties. In the case of the gas emission, these are the H$\alpha$ flux,
 the \oiii$\lambda$5007/H$\beta$ vs. \nii $\lambda$6583/H$\alpha$
 BPT diagnostic diagram \citep{1981PASP...93....5B,1987ApJS...63..295V} and the
 radial oxygen abundance gradient derived from the widely-used O3N2 strong-line indicator 
\citep{2004MNRAS.348L..59P,2013arXiv1307.5316M}. For the stellar component we
decided to show the 4000 \AA\, break D$_{4000}$ (i.e. the break strength at
4000 \AA\, defined as the ratio of the average flux densities in the narrow
bands $4000-4100$ \AA\ and $3850-3950$ \AA\ , \citealt{1999ApJ...527...54B})
as it is the most commonly used proxy for mean stellar age.

\begin{figure*}
  \centering
   \includegraphics[width=0.32\linewidth,clip=true,bb=80 0 570 720,angle=-90]{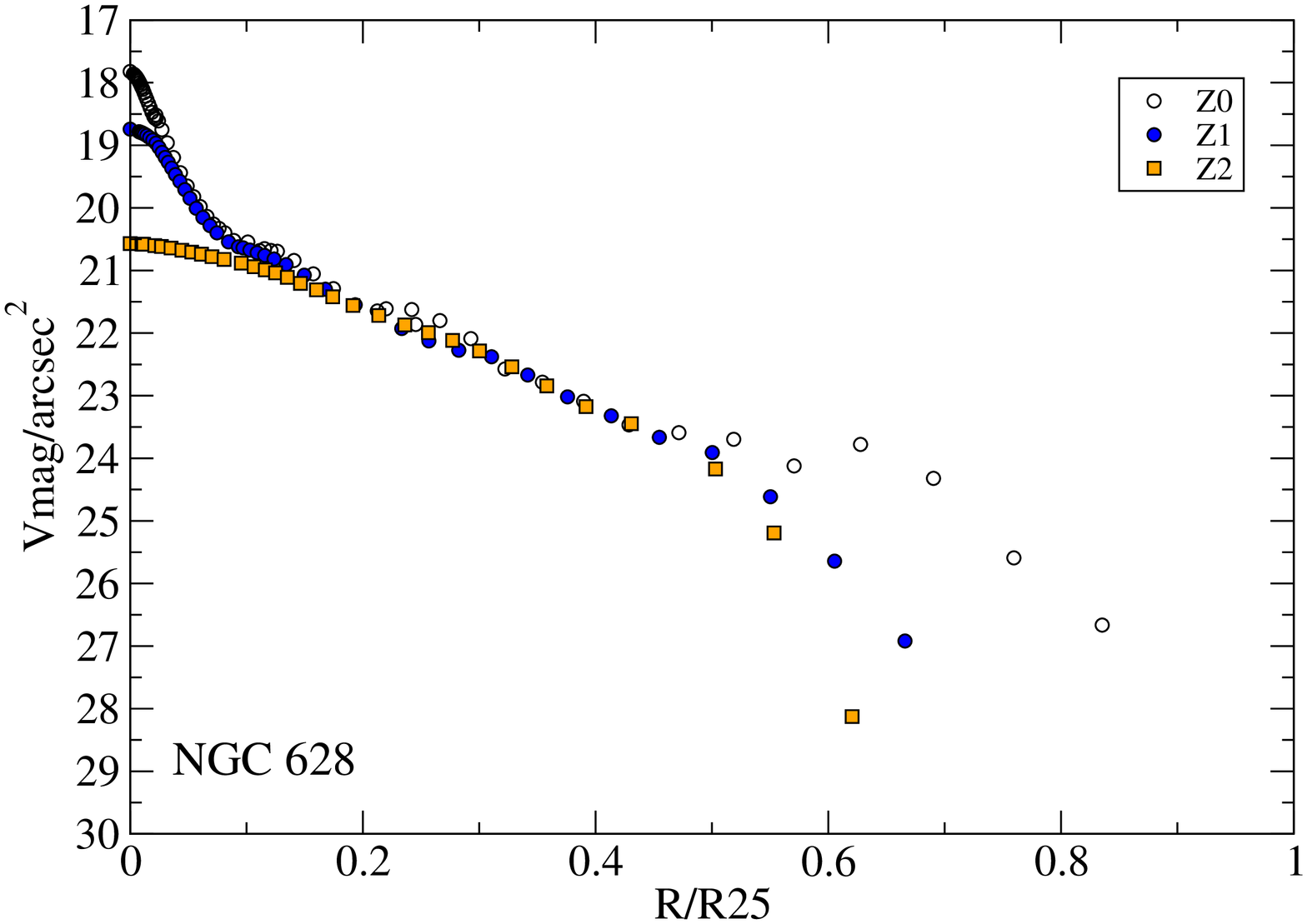} ~
   \includegraphics[width=0.32\linewidth,clip=true,bb=80 0 570 720,angle=-90]{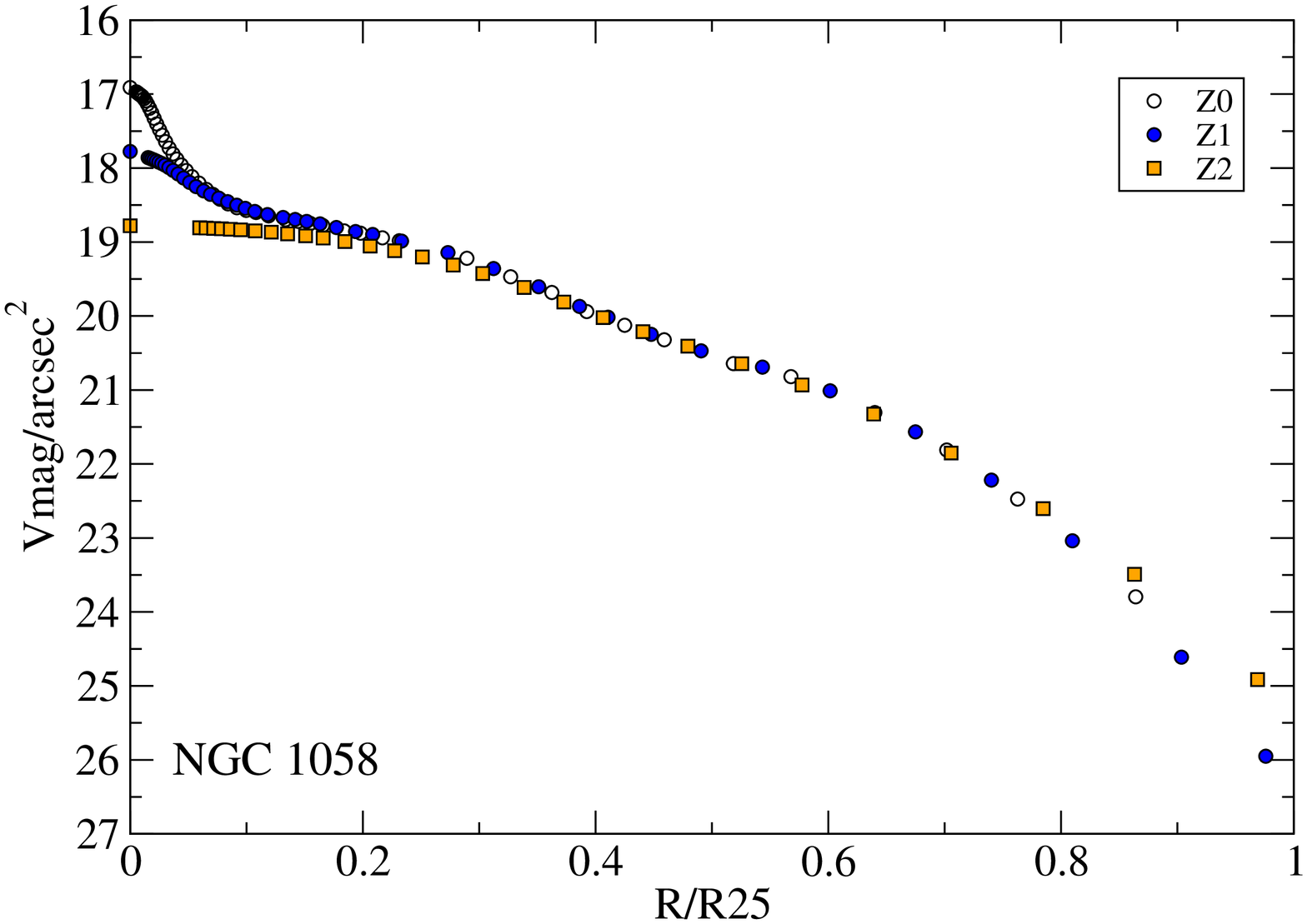} 
   \includegraphics[width=0.32\linewidth,clip=true,bb=80 0 570 720,angle=-90]{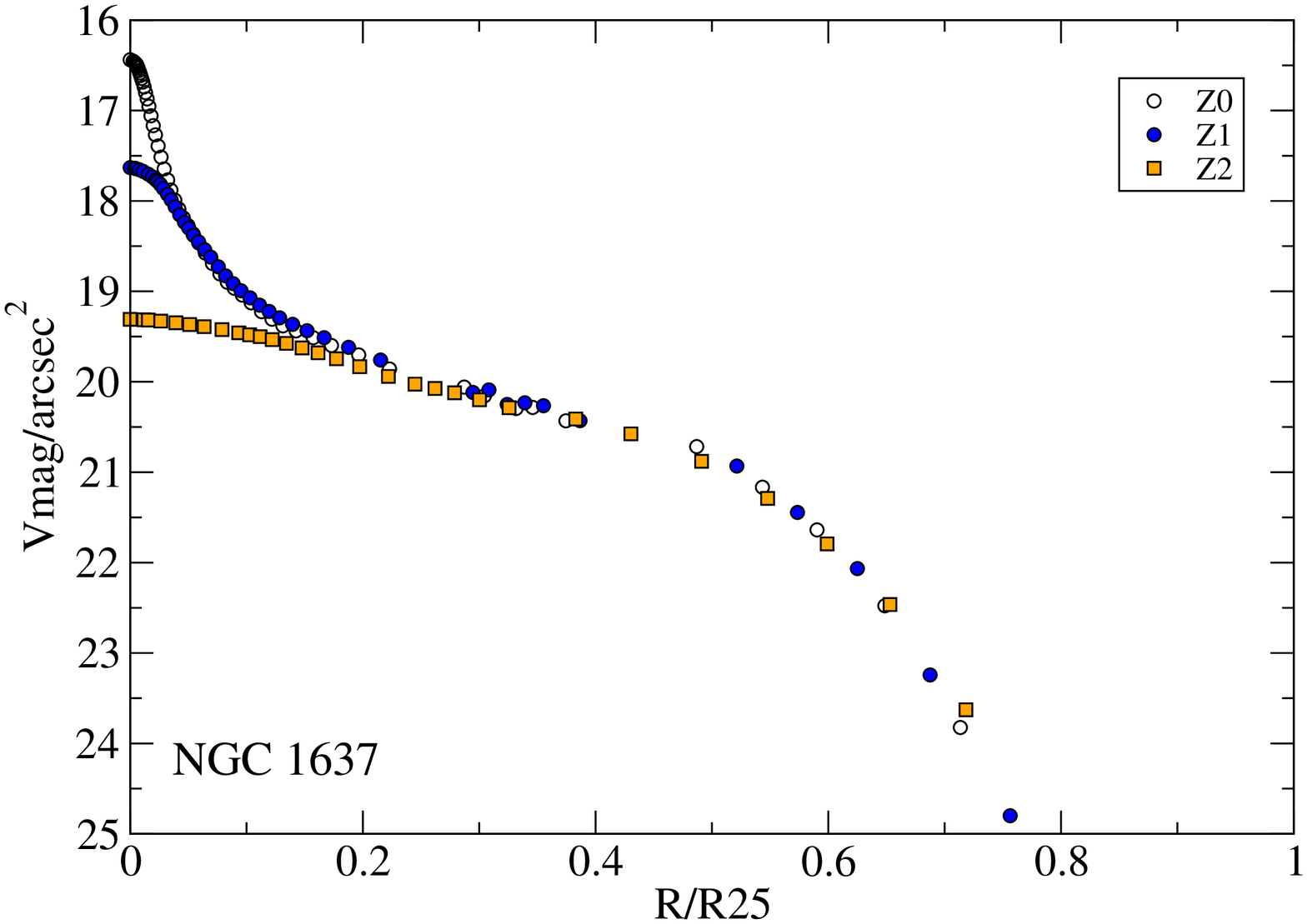} ~
   \includegraphics[width=0.32\linewidth,clip=true,bb=80 0 570 720,angle=-90]{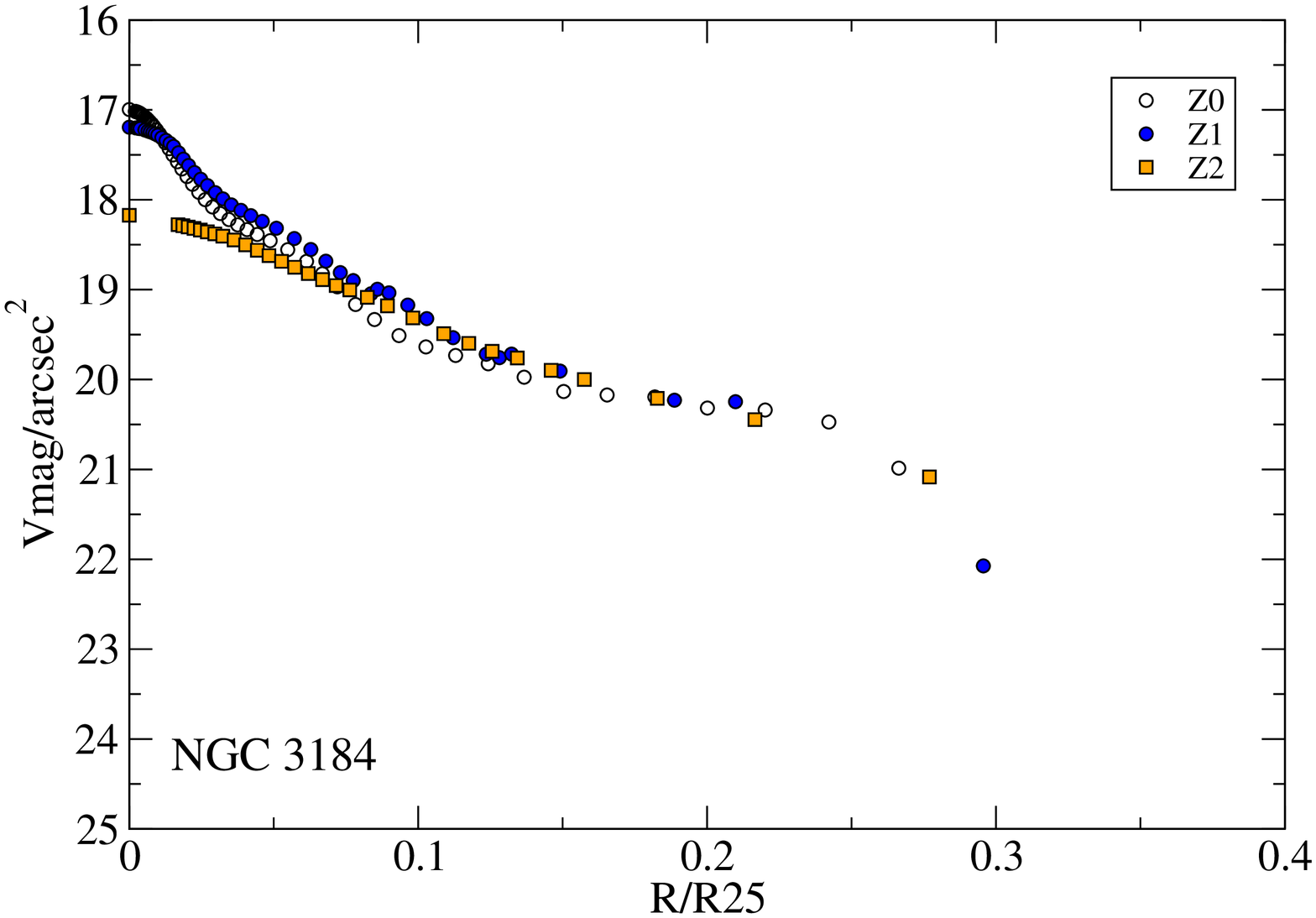} 
   \caption{Surface brightness profiles of the original PINGS galaxies and the
     simulated ones, constructed fitting ellipses to $V$-band images extracted
     from the datacubes. The profiles were scaled in the y-axis  to make them
     coincide at R$_{max}/2$  (maximum measured radius divided by 2).}
   \label{fig:profiles}
\end{figure*}

All radial profiles are plotted in units of R$_{25}$ obtained from the RC3 catalogue \citep{1992yCat.7137....0D} and scaled to each redshift regime. Surface brightness profiles for each galaxy  (Fig. \ref{fig:profiles}) were constructed using the IRAF\footnote{IRAF is distributed by the National Optical Astronomy Observatories, which are operated by the Association of Universities for
Research in Astronomy, Inc., under cooperative agreement with
the National Science Foundation.} task {\it ellipse} and then scaled in the y-axis  to make them coincide at R$_{max}/2$ (maximum measured radius divided by 2), in order to analyse the profile differences between galaxies. 
For the ellipse fitting we used $V$-band images extracted from the datacubes as explained before. 

All the above analyses (fitting and subtracting the underlying stellar
continuum and fitting of the emission lines) were performed independently on
each datacube, i.e both the original and simulated ones.

\subsection{Morphology}\label{sec:morpho}

From Fig. \ref{fig:simulation} we can see that at the representative redshift
of CALIFA ($Z1$, middle panel) we are able to qualitatively reproduce  most
morphological features. In all cases we can trace the spiral arms, rings and
bulge extension seen in the original images. Finer detail, such as individual
regions or spots, are smeared out on the $Z1$ case, so qualitatively we can
see that the analysis of the distribution of some features along the arms or
small regions will be limited by the resolution effect.

In the $Z2$ regime (right panel), the situation is totally different. Except
in the case of NGC 628, where, as a grand-design galaxy, the spiral arms are
very well defined and still traceable to high redshift, in all other galaxies
the identification of spiral signatures is hampered. Individual regions are
not easily detected and the separation between the bulge and the disk becomes
subtle. The latter effect can also be appreciated from the surface brightness
profiles in Fig.  \ref{fig:profiles}. In all cases, the primary effect of
spatial resolution loss is that the profiles become flatter in the centre. In
the $Z1$ case the central component, though smoothed, can be detected as an
excess of light above the inward extrapolation of the disk. That is not the
case for the $Z2$ profiles where the inner regions are shallow and a single
component fit to the data seems to be the best option, although a slight
disc/bulge differentiation could be seen in the radial plots.

\subsection{Ionized gas component}

\subsubsection{\hh region distribution}

\begin{figure*}
  \centering
  \subfloat{\epsfig{file=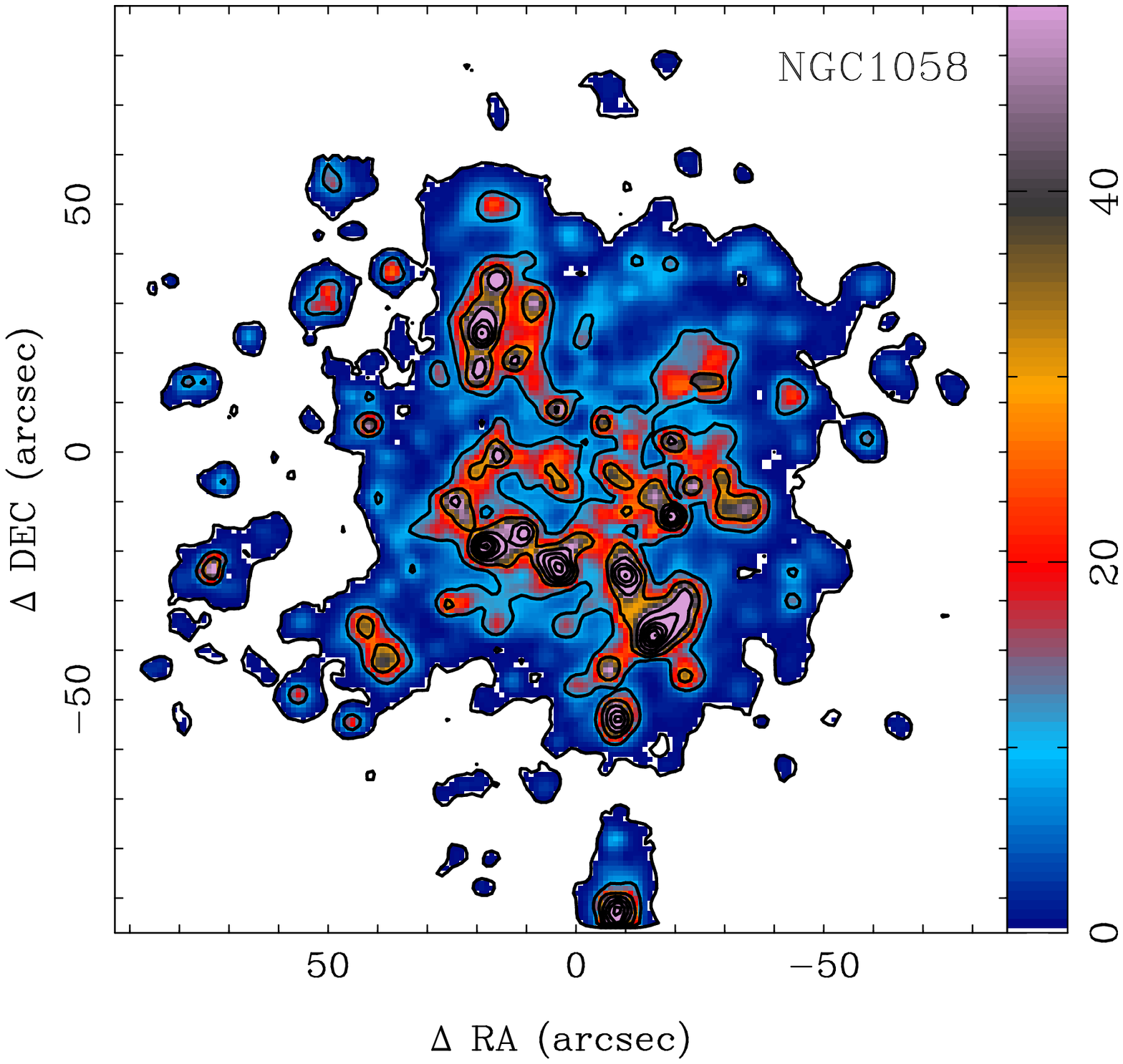 ,width=0.3\linewidth,clip=,angle=-90}}
  \subfloat{\epsfig{file=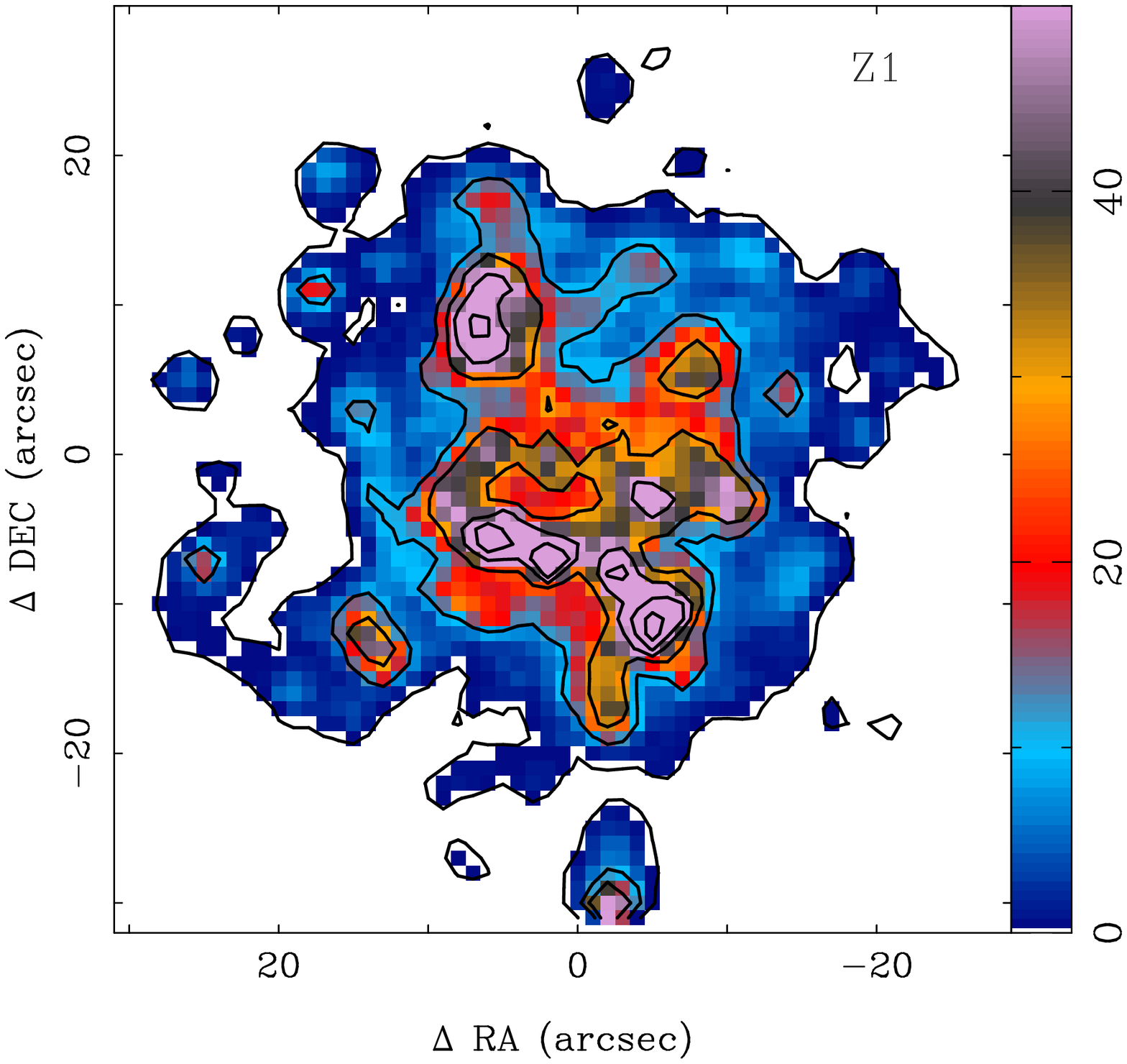 ,width=0.3\linewidth,clip=,angle=-90}}  
  \subfloat{\epsfig{file=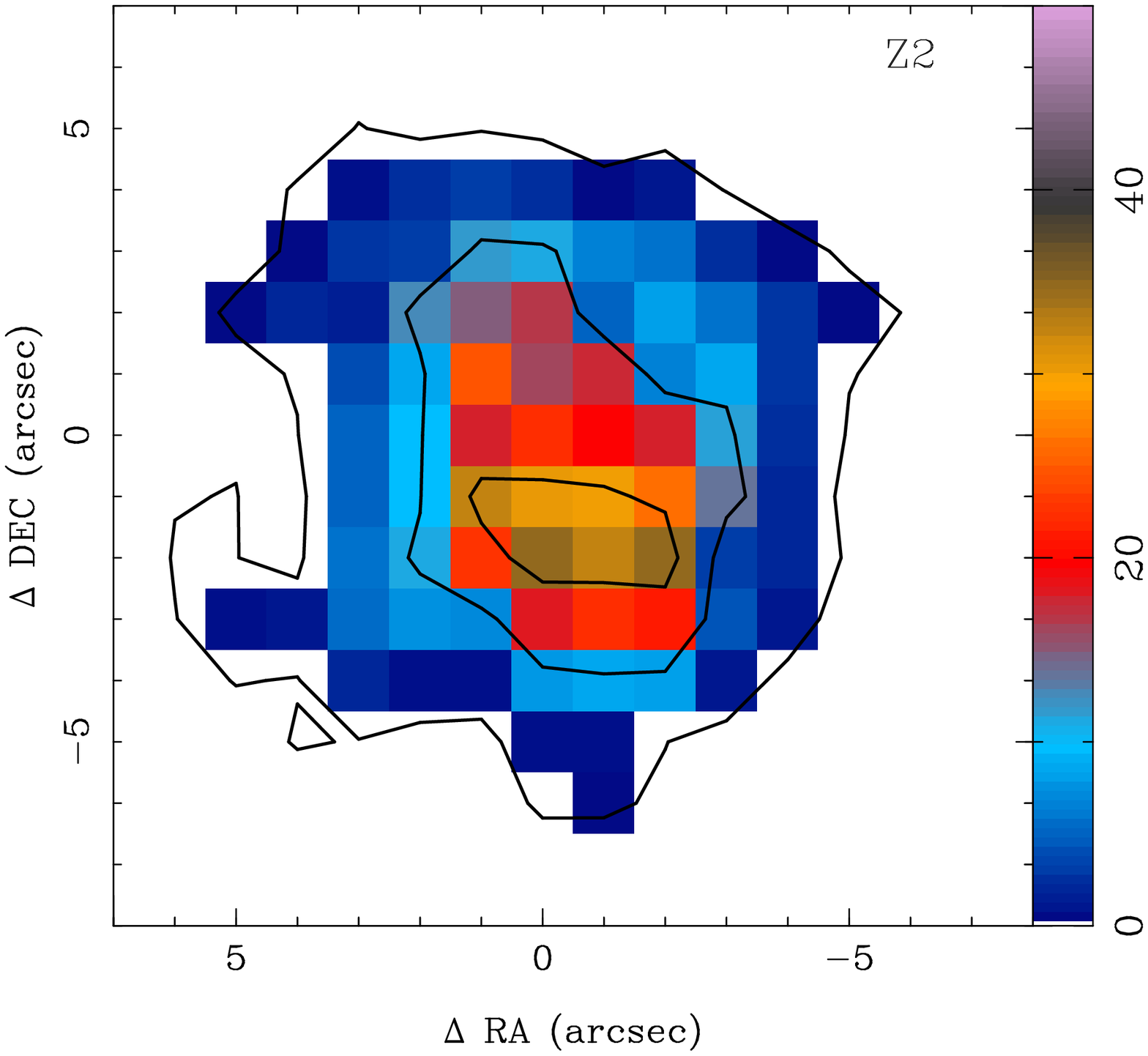 ,width=0.3\linewidth,clip=,angle=-90}}

  \subfloat{\epsfig{file=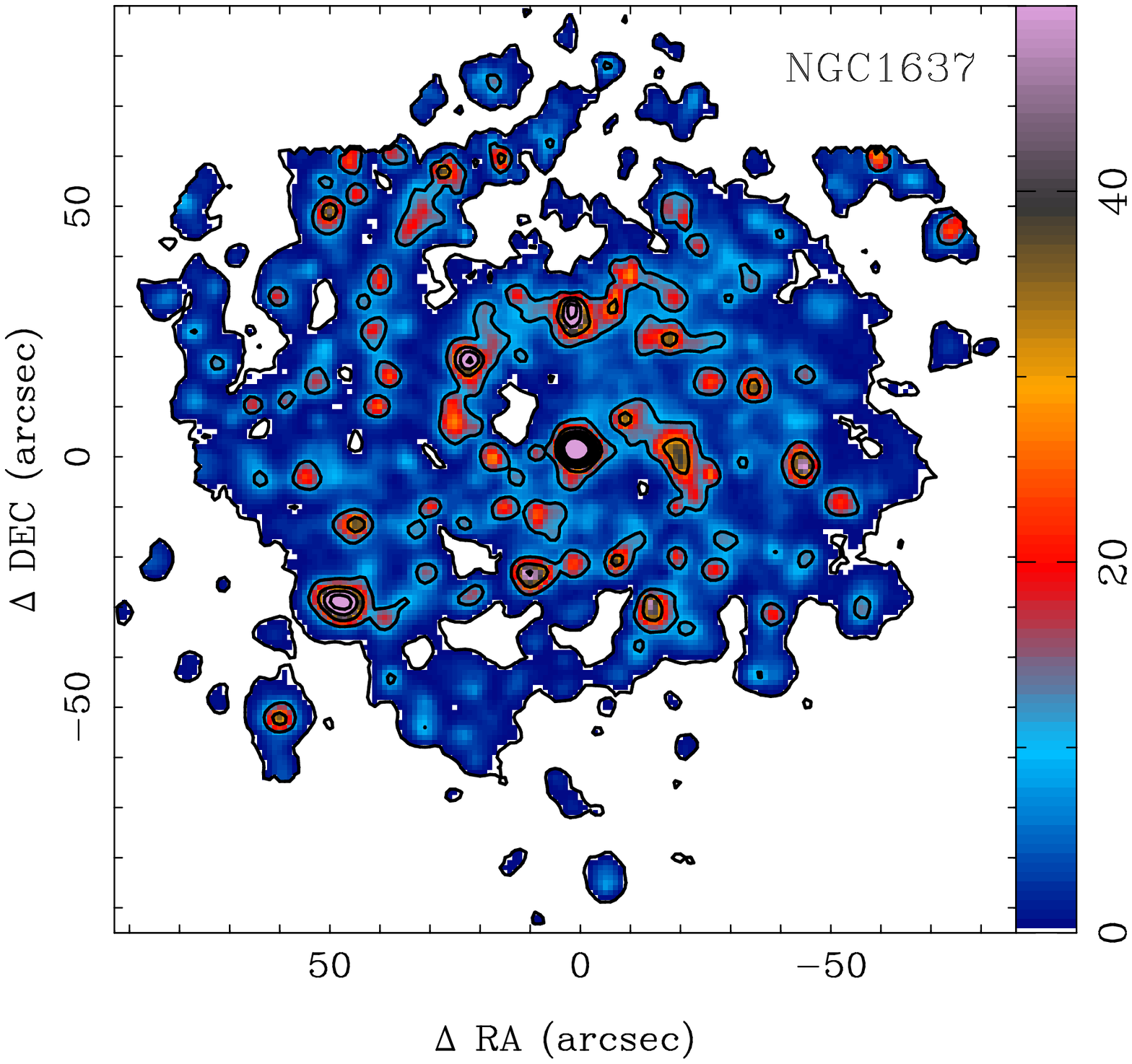 ,width=0.3\linewidth,clip=,angle=-90}}
  \subfloat{\epsfig{file=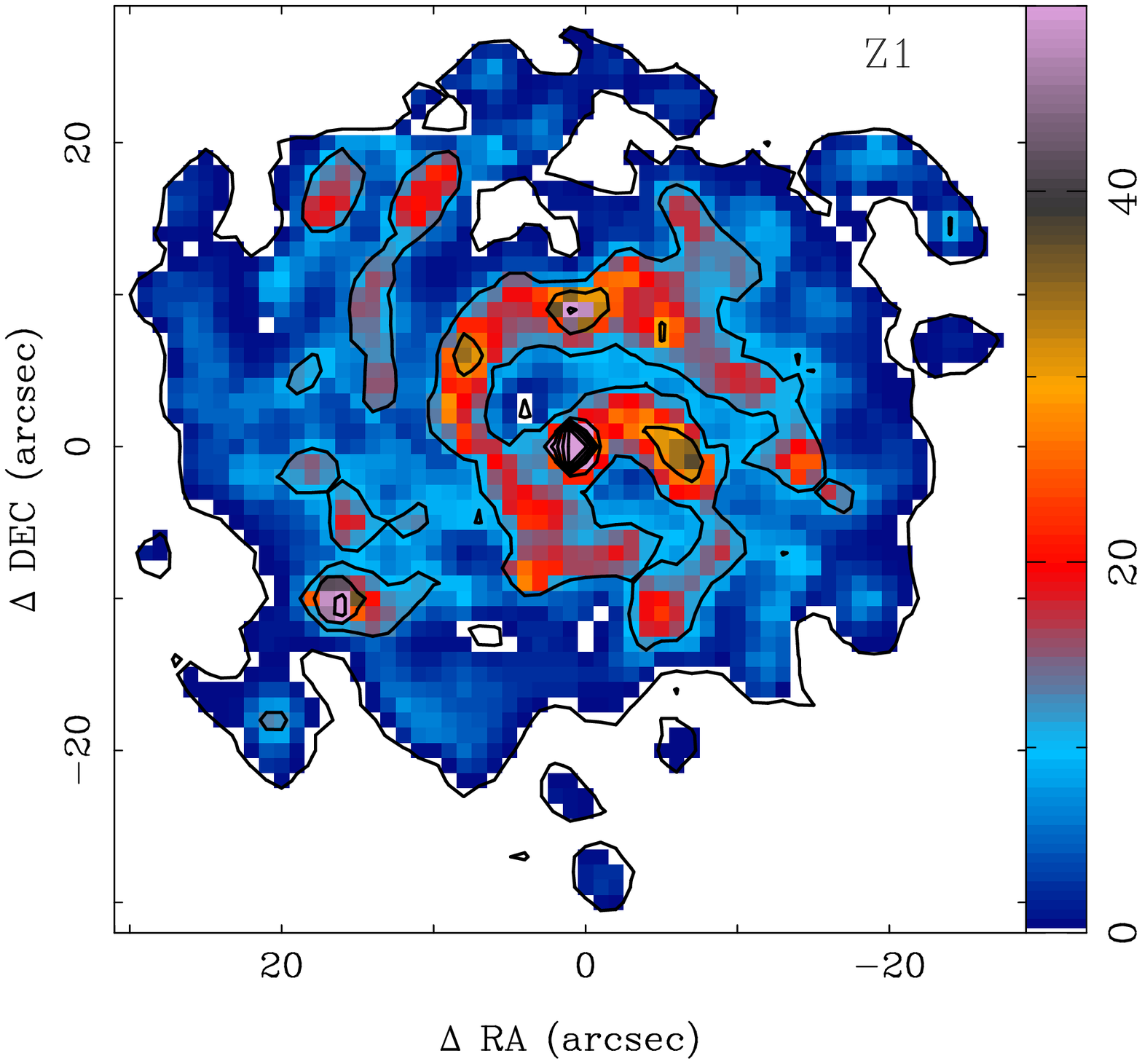 ,width=0.3\linewidth,clip=,angle=-90}}  
  \subfloat{\epsfig{file=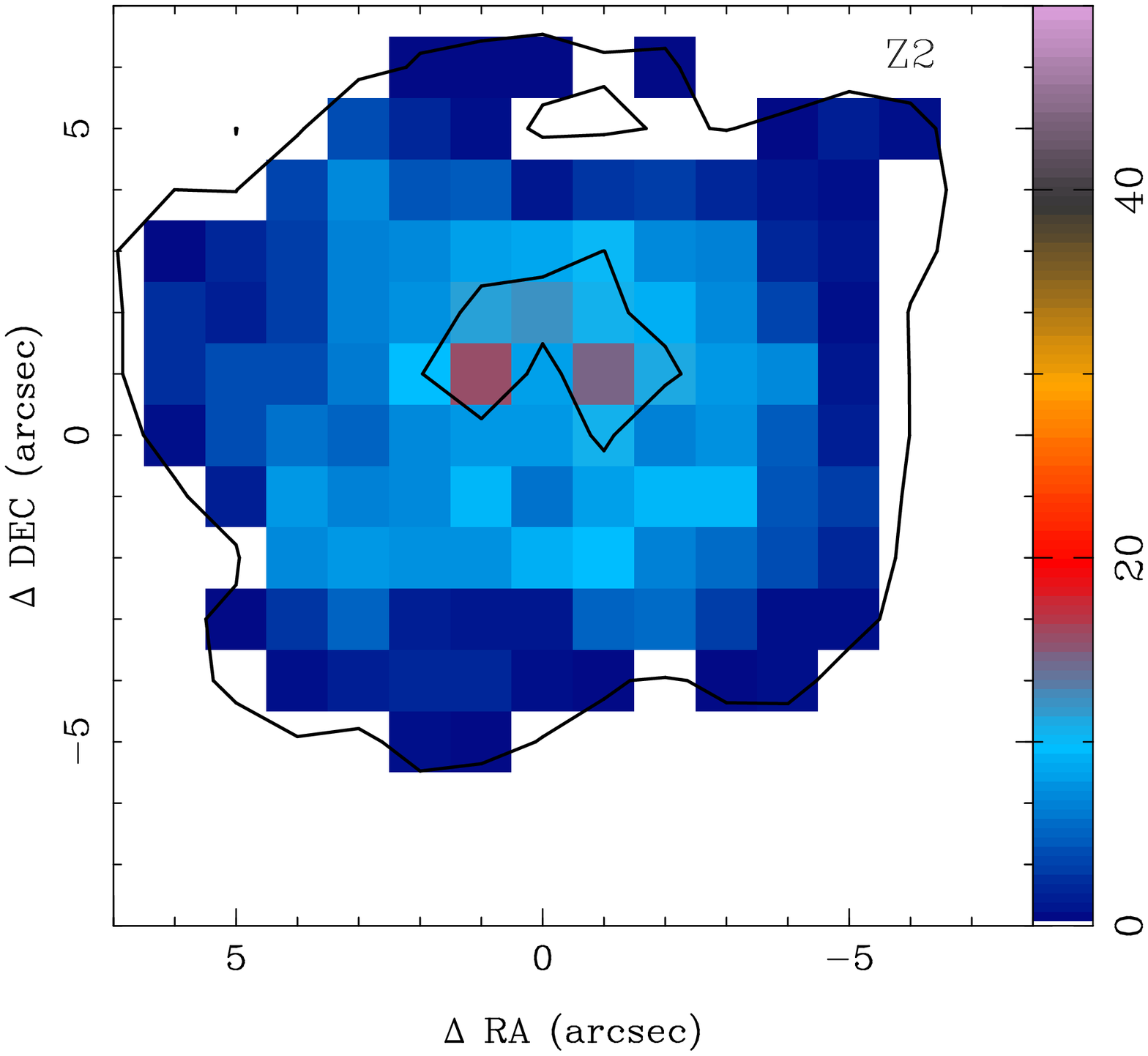 ,width=0.3\linewidth,clip=,angle=-90}}

  \subfloat{\epsfig{file=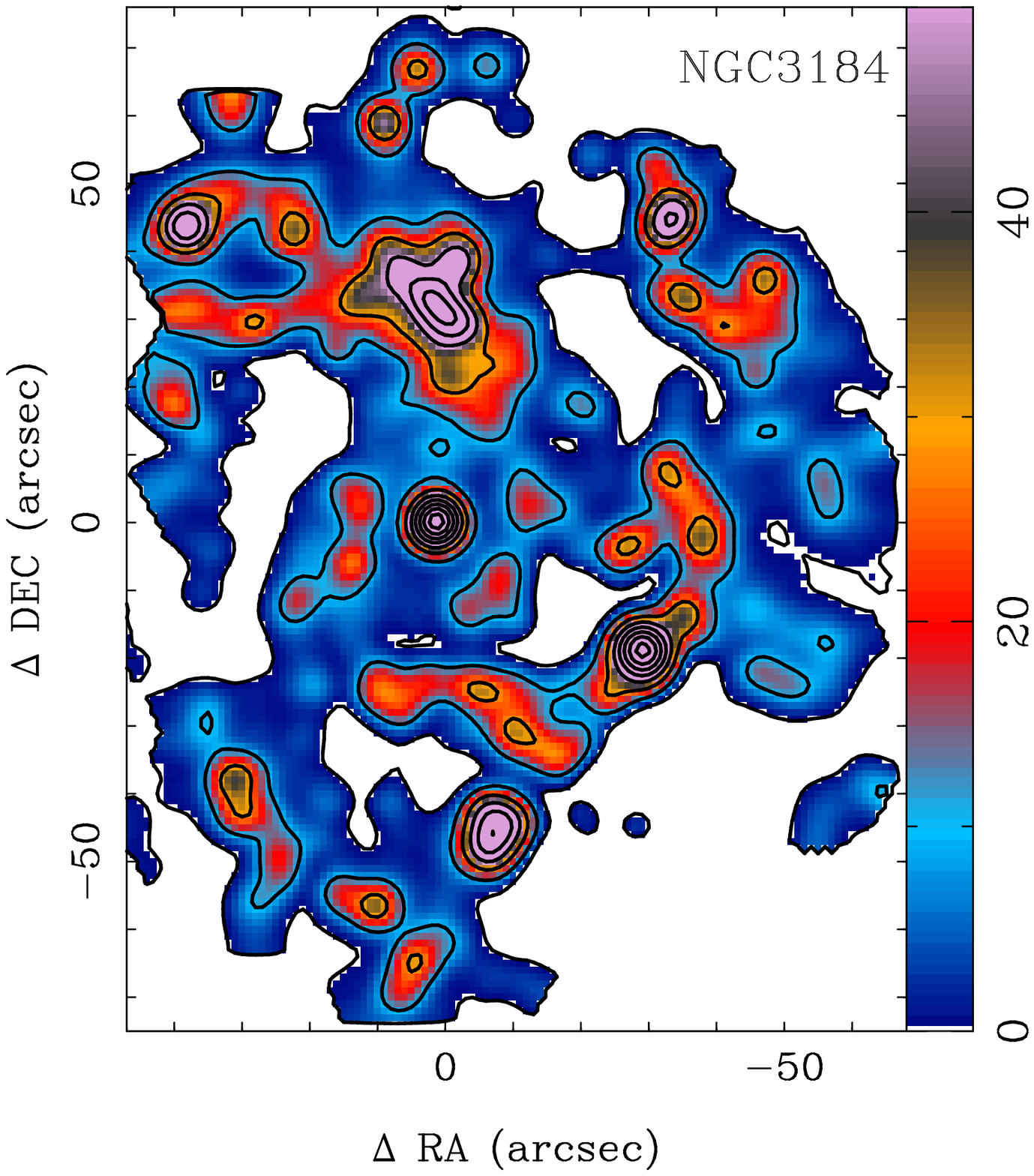 ,width=0.36\linewidth,clip=,angle=-90}}
  \subfloat{\epsfig{file=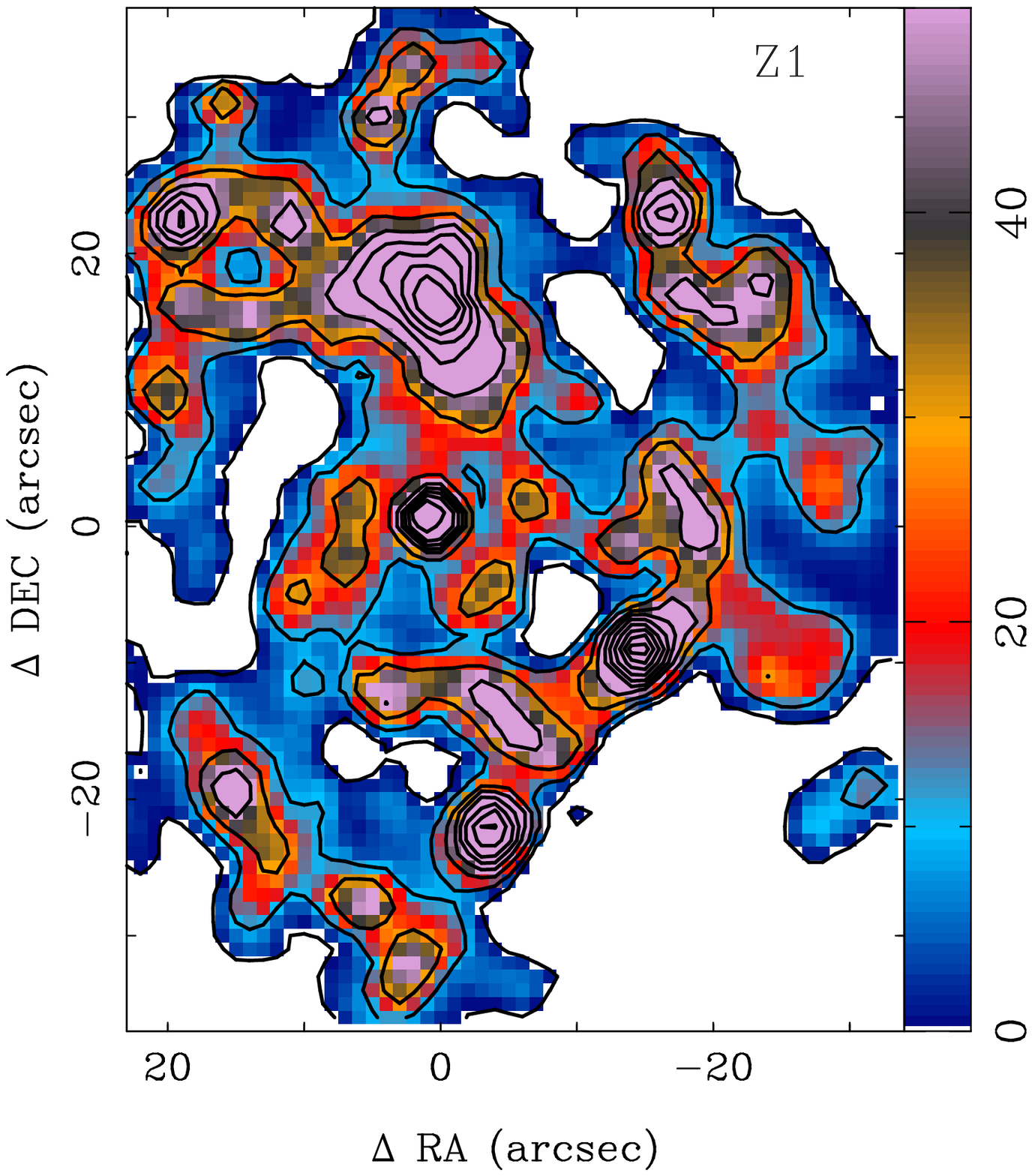 ,width=0.36\linewidth,clip=,angle=-90}}  
  \subfloat{\epsfig{file=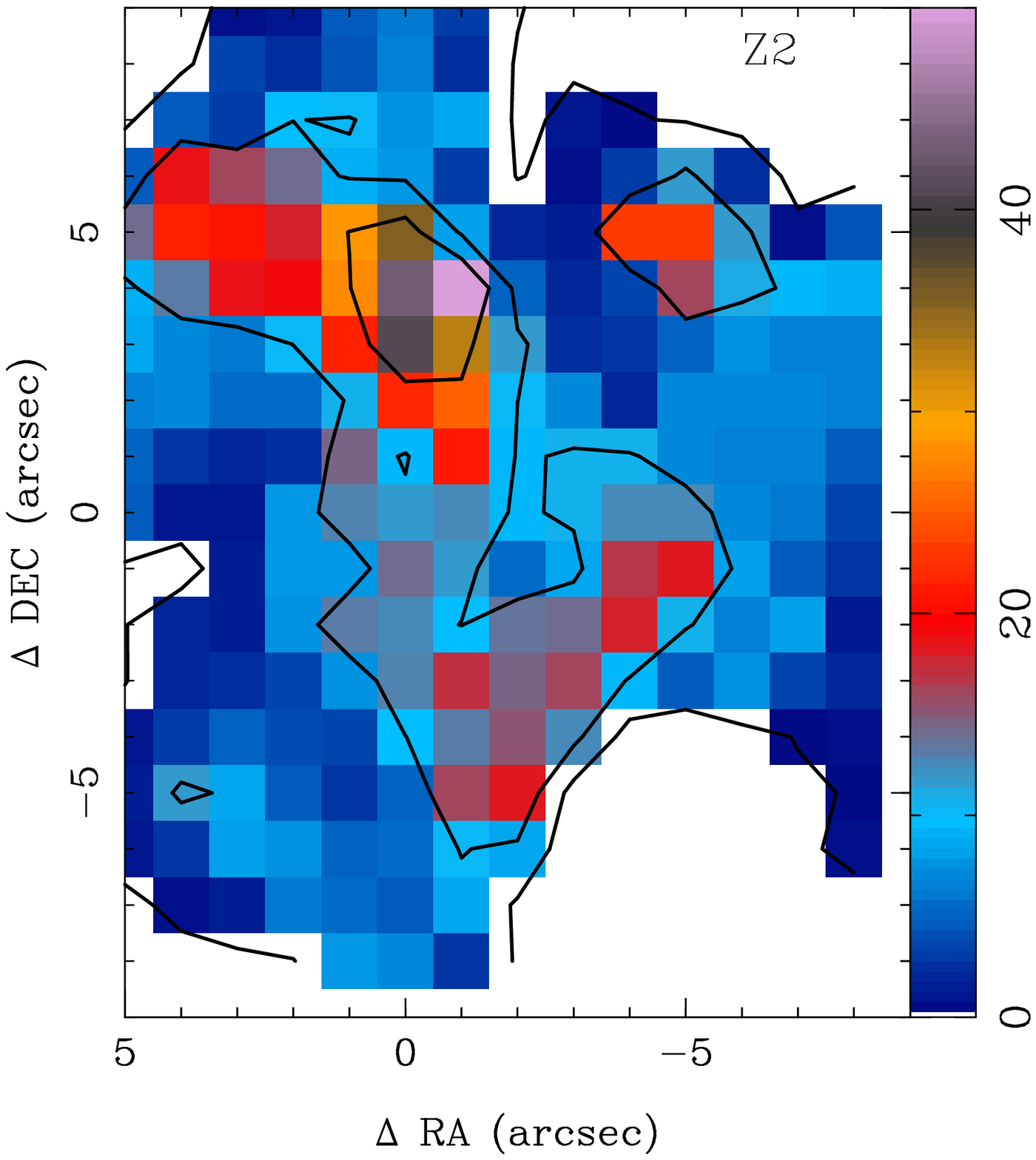 ,width=0.36\linewidth,clip=,angle=-90}}

  \subfloat{\epsfig{file=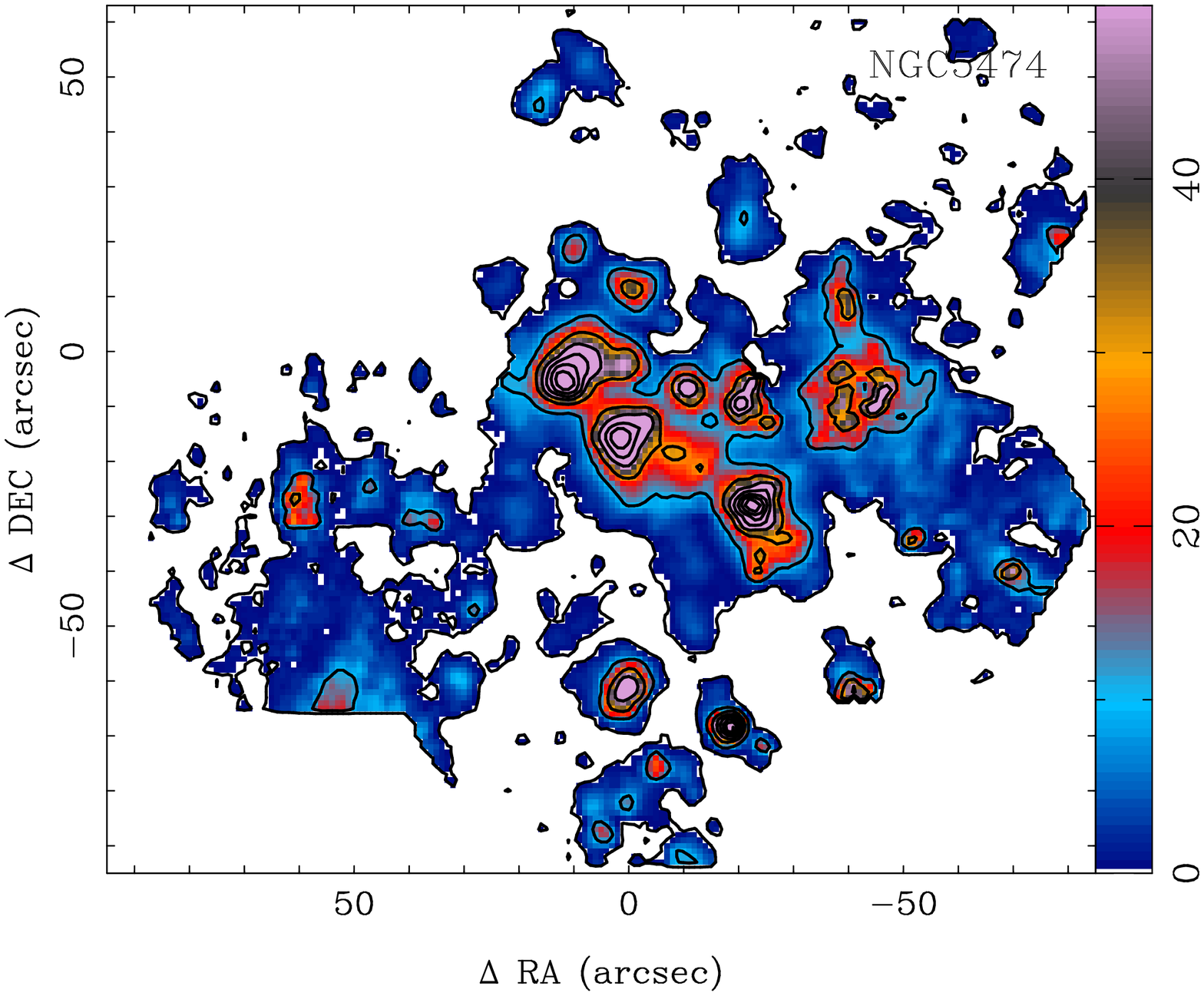 ,width=0.26\linewidth,clip=,angle=-90}}
  \subfloat{\epsfig{file=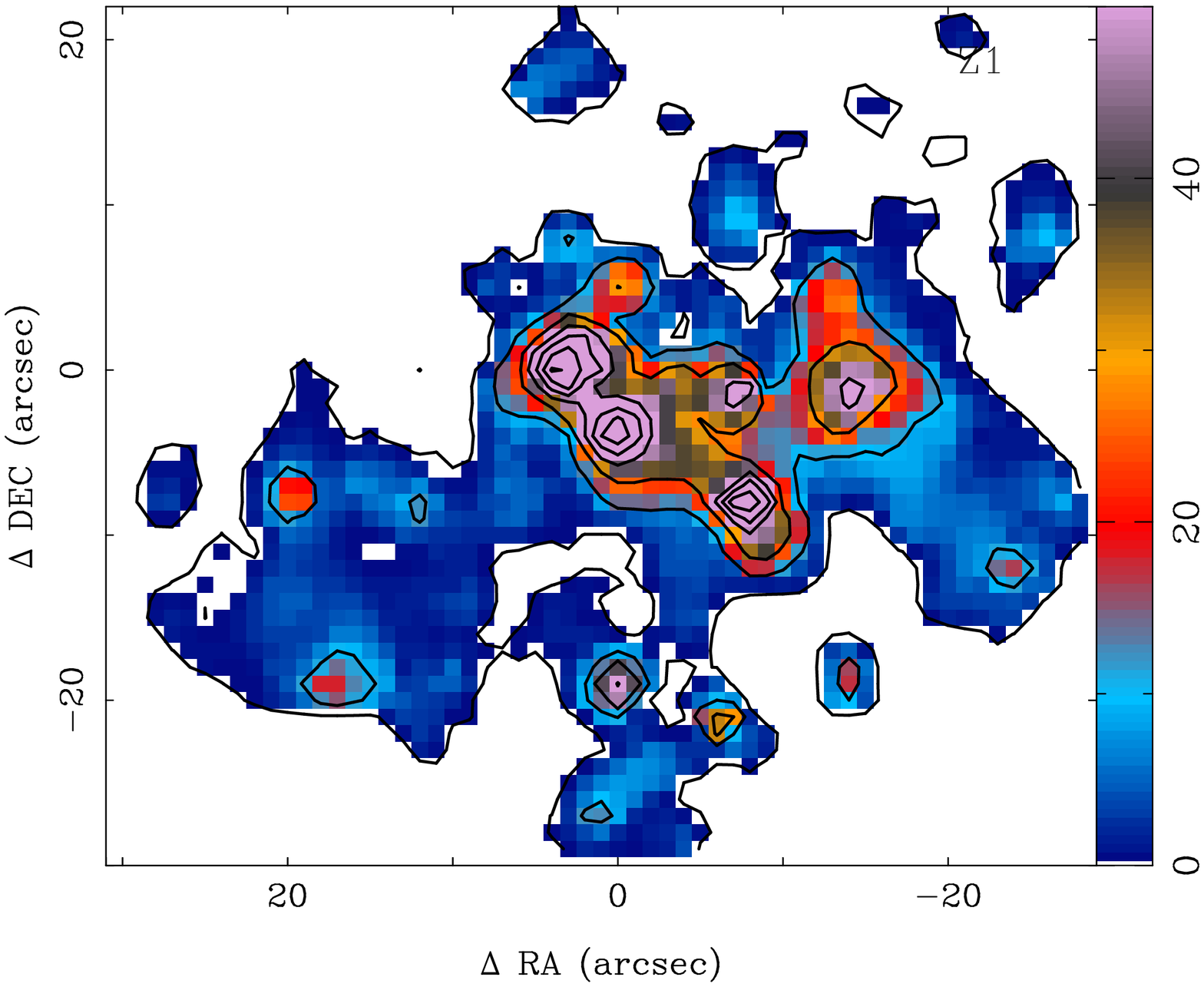 ,width=0.26\linewidth,clip=,angle=-90}}  
  \subfloat{\epsfig{file=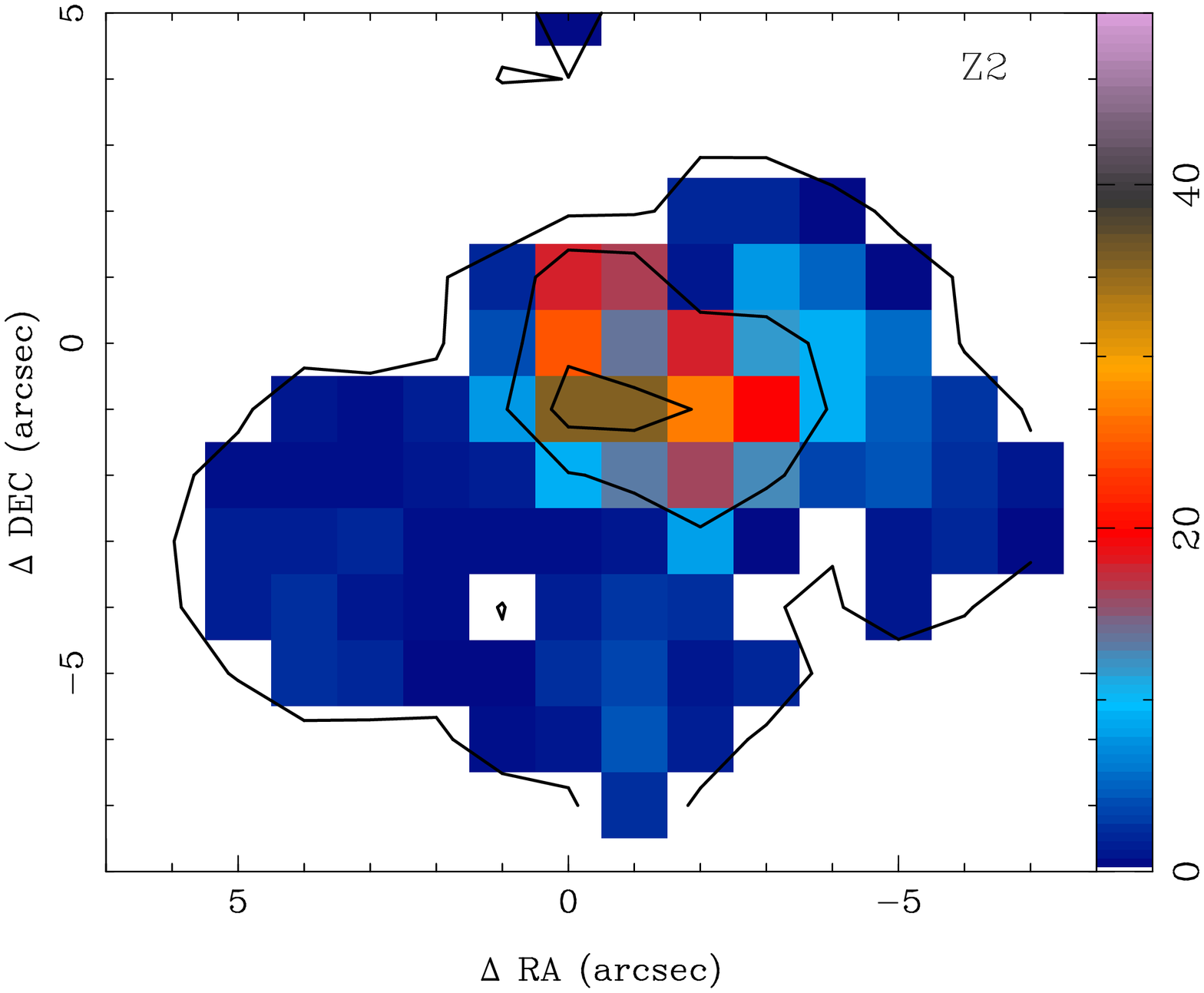 ,width=0.26\linewidth,clip=,angle=-90}}

  \caption{H$\alpha$ images extracted from the cubes. Flux is in units of
    \funits. $Z1$ and $Z2$ are scaled to $Z0$ flux. {\it Left:} original $Z0$
    galaxies. {\it Middle:} $Z1$ galaxies. {\it Right:} $Z2$ galaxies. North is
    up, East is to the left. The H$\alpha$ map of NGC 628 is displayed in
    Fig. \ref{fig:HII_explorer}.}
  \label{fig:Halfa}
\end{figure*}

\begin{figure*}
  \centering
  \subfloat{\epsfig{file=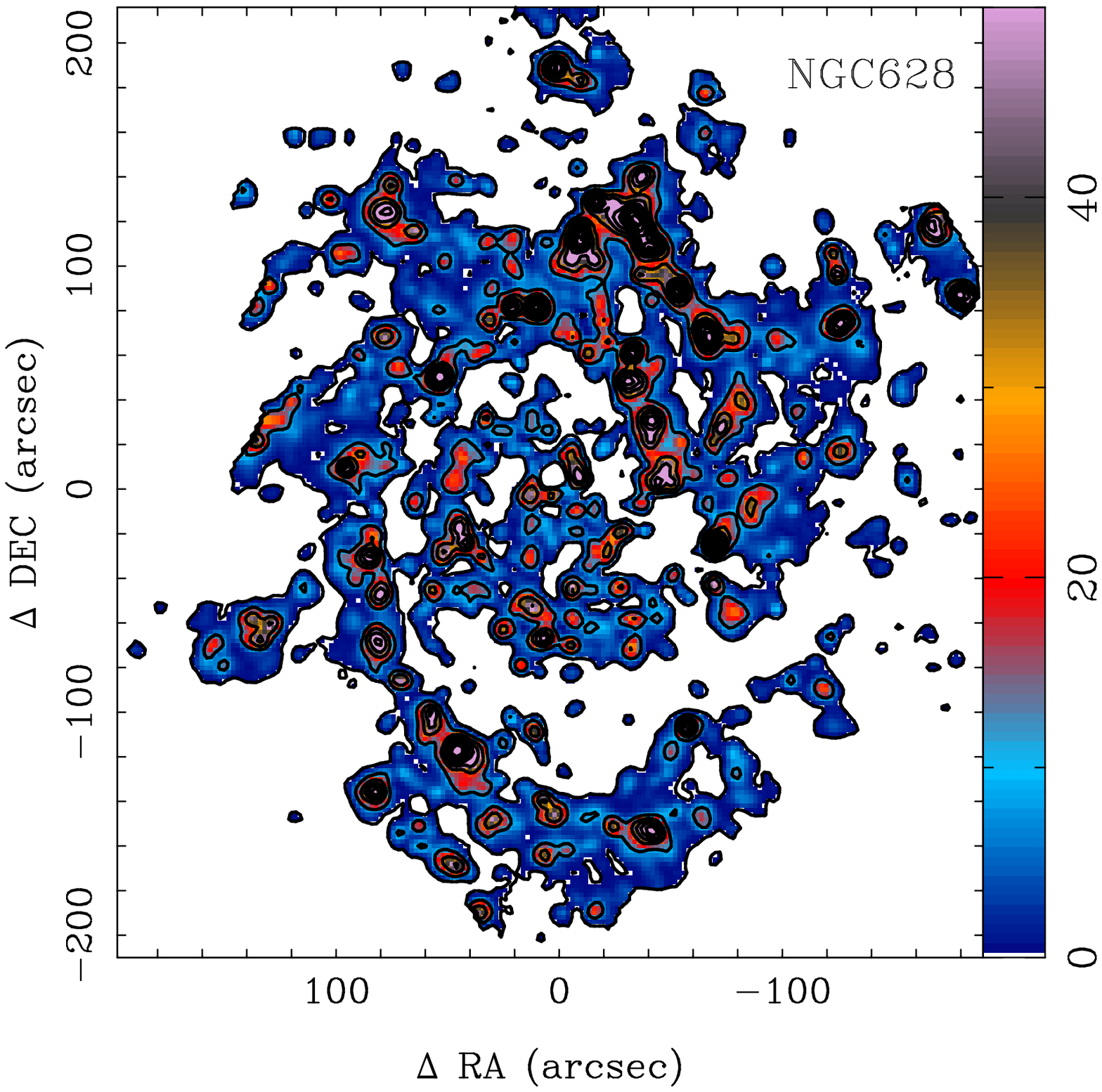,width=0.31\linewidth,clip=,angle=-90}}~
  \subfloat{\epsfig{file=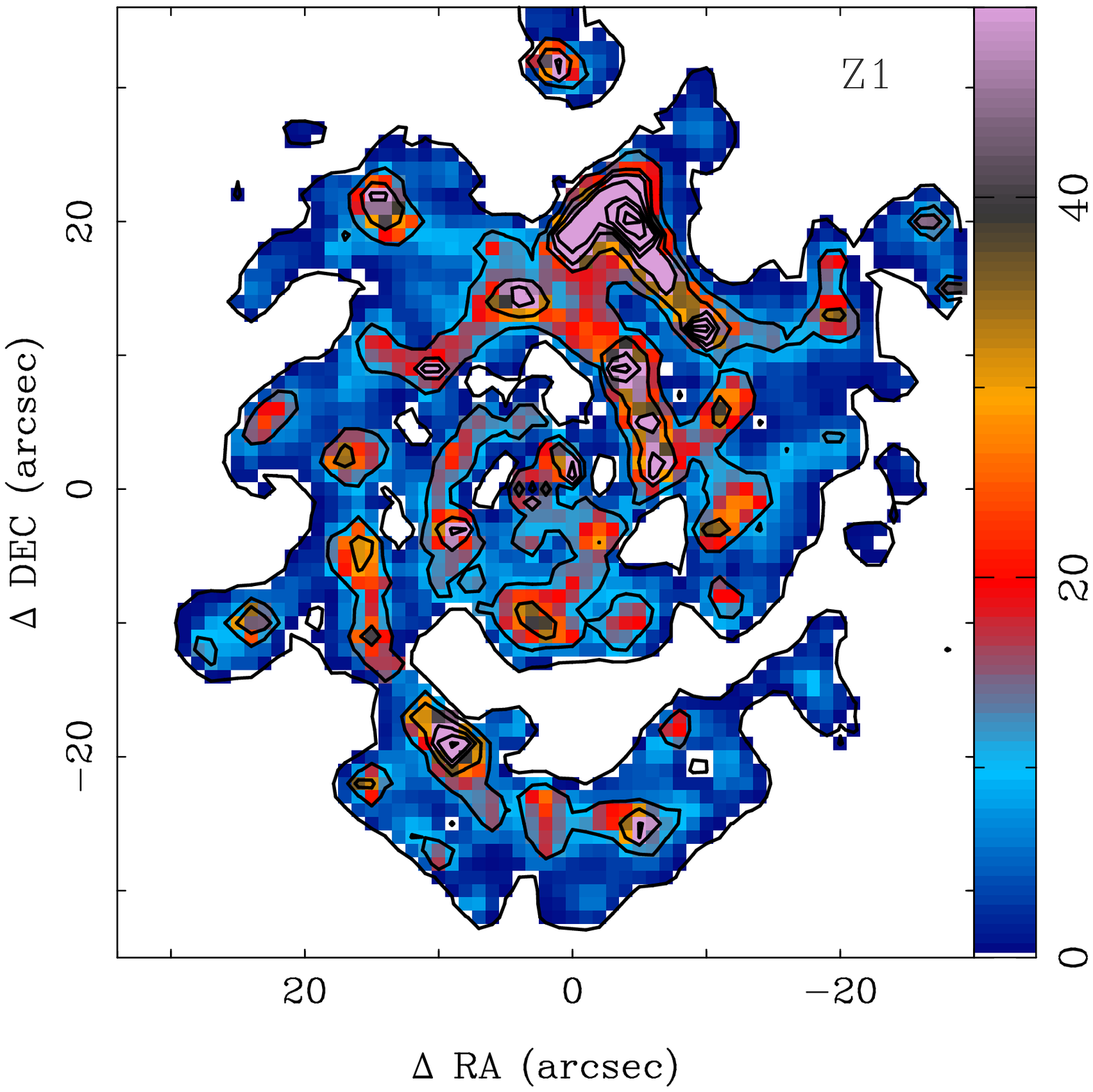,width=0.31\linewidth,clip=,angle=-90}}~ 
  \subfloat{\epsfig{file=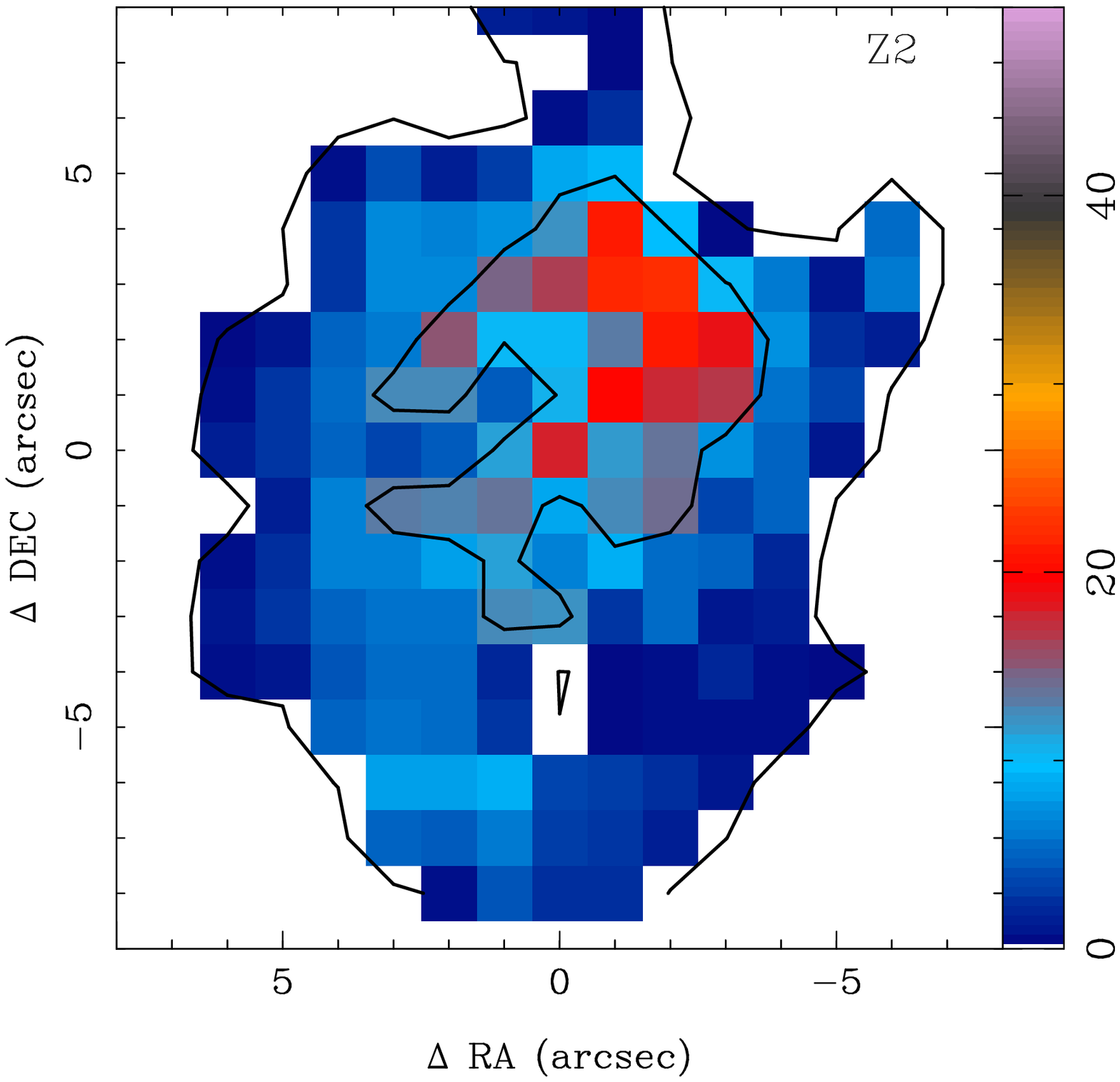,width=0.31\linewidth,clip=,angle=-90}} \\
  
  \subfloat{\epsfig{file= 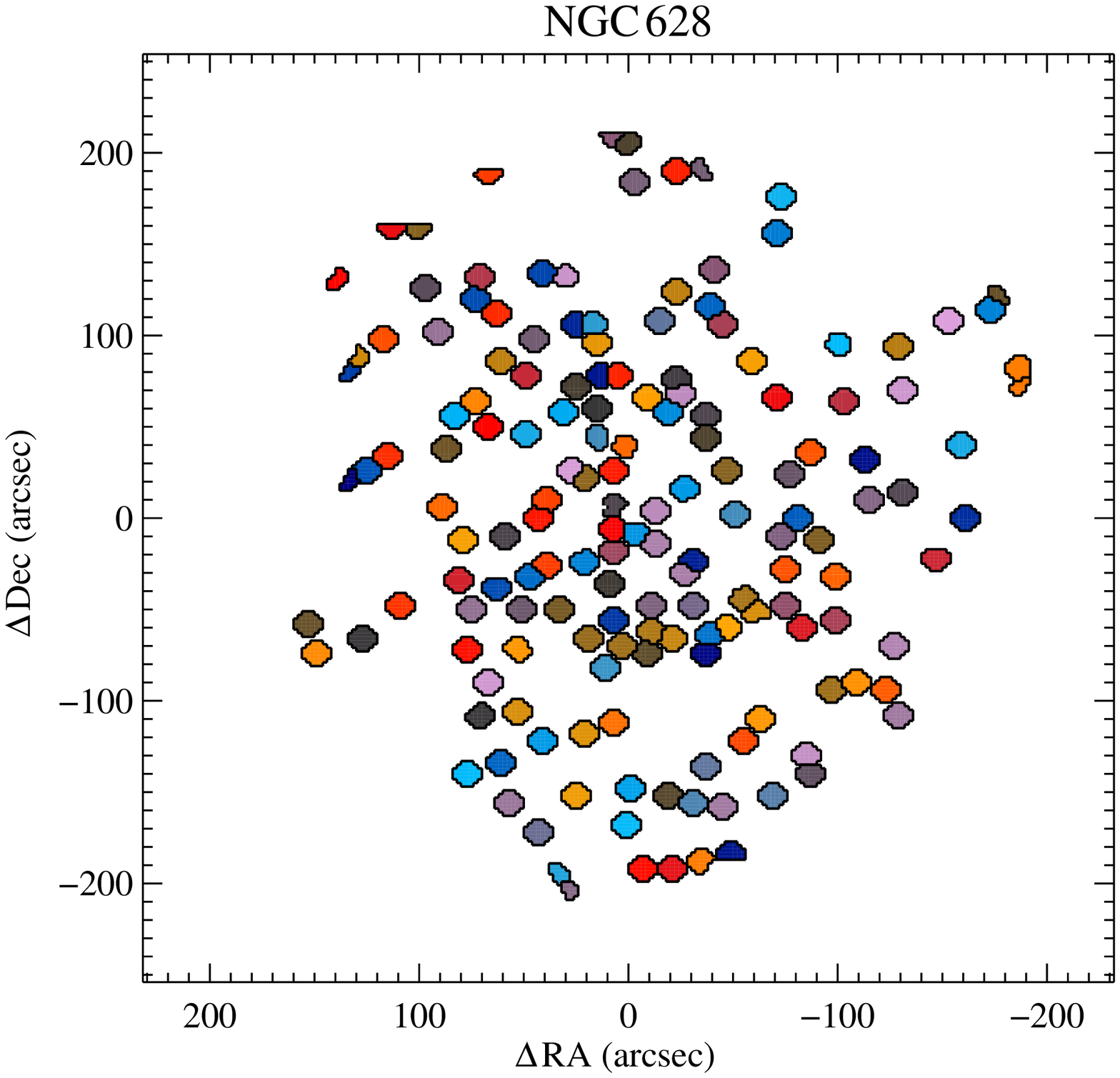,width=0.32\linewidth,clip=,angle=0}}  
  \subfloat{\epsfig{file= 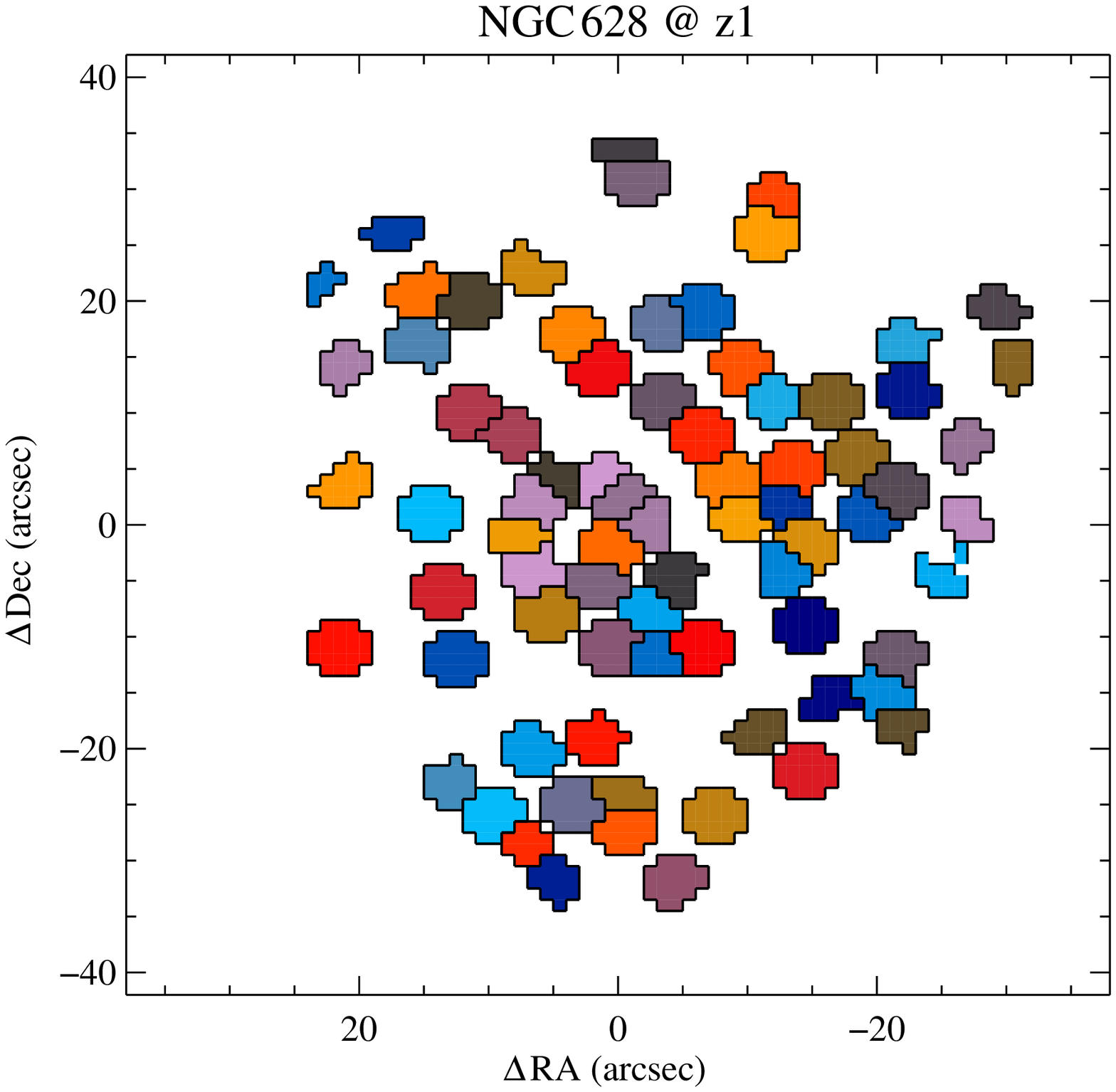,width=0.32\linewidth,clip=,angle=0}}
  \subfloat{\epsfig{file= 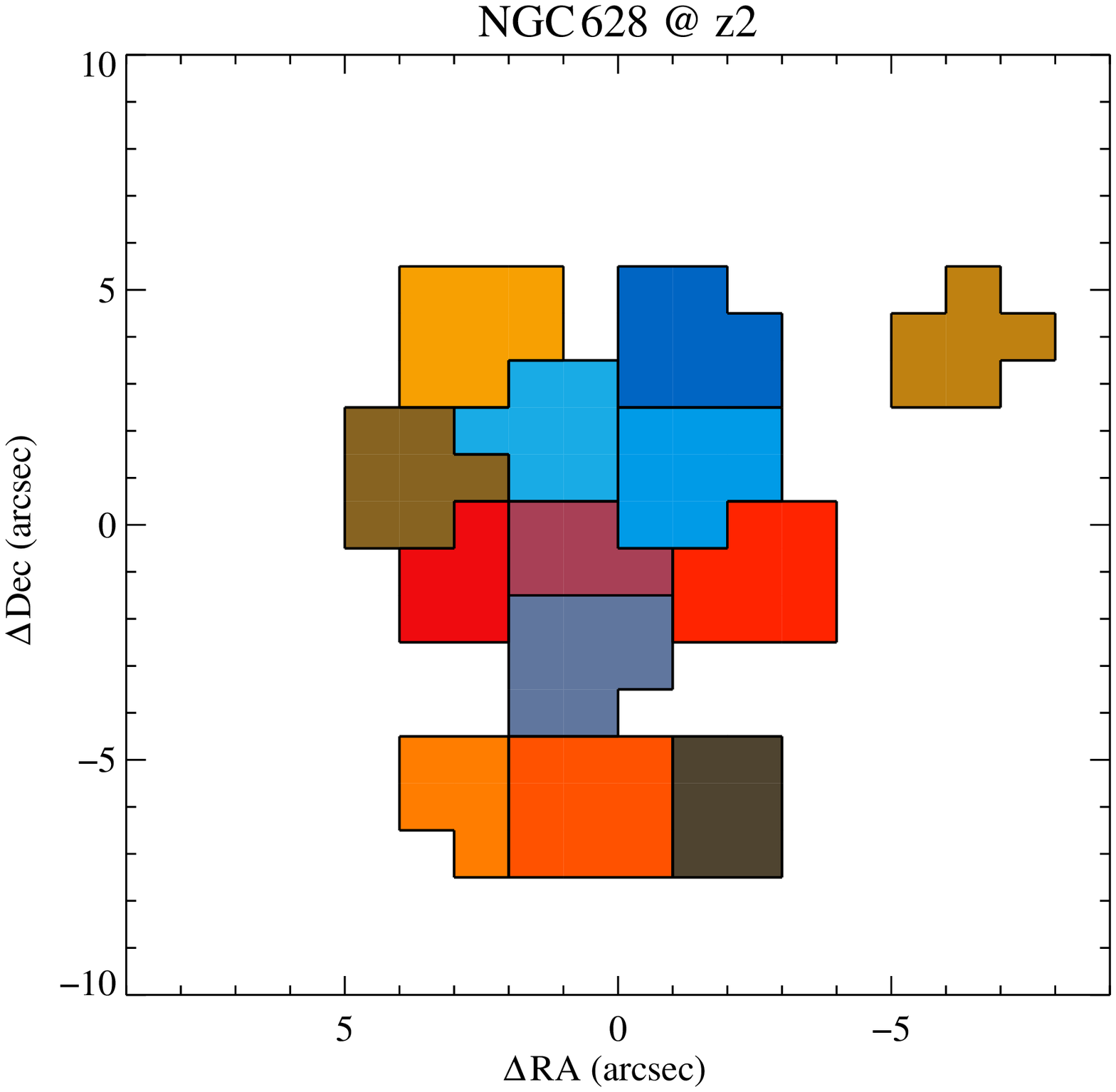,width=0.32\linewidth,clip=,angle=0}}
  \caption{\hh regions detection showing the application of the {\sc
      HIIexplorer} software to NGC~628. The panels display the segmentation
    maps generated by the software. {\it Left:} original $Z0$ galaxies. {\it
      Middle:} $Z1$ galaxies. {\it Right:} $Z2$ galaxies. North is up, East is
    to the left. These segmentation maps are later used for the extraction of
    the \hh region spectra. }
  \label{fig:HII_explorer}
\end{figure*}

To extract the  H$\alpha$ intensity maps for all the cubes (see
  Fig. \ref{fig:Halfa} and \ref{fig:HII_explorer}) we co-add the flux
intensity within a square-shaped simulated filter centred at the wavelength of
H$\alpha$ (6563 \AA) shifted to the observed frame considering the redshift of
the object, with a width of 60 \AA. The adjacent continuum for each pixel was
estimated averaging the flux intensity within two bands on both sides of the
centre, separated 100 \AA\  from it and with a width of 60 \AA. This continuum
intensity is then subtracted from the H$\alpha$ intensity to derive a
continuum-subtracted emission line map. It is worth noting that this H$\alpha$
intensity map is contaminated with the adjacent [N II] emission lines, and it
is not corrected for Balmer absorption of the underlying stellar population.

Figure \ref{fig:Halfa} show the resulting maps from the original and degraded
datacubes. The $Z1$ maps show the same distribution of \hh regions tracing the
spiral arms as in the $Z0$ case. Because of the coarse resolution, the \hh
regions found in the $Z1$ case are complexes of several PINGS regions but
without a severe alteration of H$\alpha$ flux local maxima position. On the
contrary, as can be expected from the global morphology noted in the previous
section, the H$\alpha$ 2D distribution in the $Z2$ simulation, presents a
strong difference as many \hh complexes are averaged into one single H$\alpha$
aggregation. Again, no traceable spiral arms are visible for the $Z2$ regime,
and even in the NGC 628 case, in which the spiral structure is perfectly
traced by the \hh complexes at $Z0$ and $Z1$, the $Z2$ galaxy 
presents a morphology that at a first sight it is not easily associated
  with a spiral structure. In any case, it is worth noting that the
  morphological analysis presented here is meant to show the effects of
  resolution in the morphology, in the understanding that any spectroscopic survey
  will have an imaging follow up to derive the morphology with better
  resolution and that kind of work will not be done from the line maps
  (although this might not be the case for all high-redshift studies, since
  some of them relies only in the IFS data).

We used the {\sc HIIexplorer} software \citep{Sanchez:2012b}  to segregate and perform the corresponding spectral extraction of all the \hh regions detected on each map. The {\sc HIIexplorer} works assuming: (a) \hh regions are peaky/isolated structures with strong ionized gas emission clearly above the continuum level emission and the average ionized gas emission across the galaxy, and 
(b) \hh regions have a typical physical size of about a hundred or few hundreds of parsecs (e.g. \citealt{1997ApJS..108..199G}, \citealt{2011ApJ...731...91L}, \citealt{2003AJ....126.2317O}), which corresponds to a typical projected size of a few arcsec at the distance of the observed galaxies.
The algorithm requires a line emission map, with the same world-coordinate system (WCS) and resolution as the input datacube (preferentially an H$\alpha$ emission line map). {\sc HIIexplorer} outputs a segmentation FITS file describing the pixels associated to each H II region map that should be used for the extraction of the spectra. Each region will have a single spectrum generated by co-adding all the spectra indicated by the segmentation map. In Fig. \ref{fig:HII_explorer} we can see, as an example, the case of NGC 628 in the three redshift regimes. The number of \hh regions detected in each regime immediately indicates the effect of resolution degradation, i.e. as we go to higher redshift we are considering as one region the contribution from all \hh regions smaller than the resolution element. Even in the $Z0$ case, the real physical size of an \hh region is below the spatial resolution: in all cases, when we refer to \hh regions we are in fact considering \hh complexes or aggregates. Table \ref{tab:regions} shows the number of detected regions, with good enough S/N to perform the analysis, following the criteria from \cite{Sanchez:2012b}, for all the galaxies in the present study. For example, from 331 regions detected in NGC 1058, the $Z2$ regime identifies only 7. For the $Z1$ case, the average factor of under-detection is $\sim3$, and the difference between $Z1$ and the $Z2$ regime $\sim5$. Also, as the resolution becomes coarser, the difficulty in delimiting the \hh regions implies that we are combining both diffuse gas and \hh region emission. It is important to mention that this is always the case if one is considering  spaxel-to-spaxel analysis instead of  individual or aggregations of \hh regions. The latter is required to minimize the
effect of contamination due to diffuse gas. If \hh regions cannot be 
segregated, it is not guaranteed that the dominant ionized emission is
due to photoionization from young massive stars and therefore, analysis as the
abundance derivations are not reliable. This is probably the case for the
  $Z2$ case where, due to the low number of pixels, it is difficult to
  identify peaky/isolated structures as required by the {\sc HIIexplorer}. As
  will be shown later for the abundance determination, although for the $Z2$
  case the conditions are not easily fulfilled, the comparison between the
  spaxel-spaxel and the {\sc HIIexplorer} analysis indicates that the former
  has a greater misinterpreting possibility, even considering a S/N
  threshold. As already mentioned in \cite{Sanchez:2012b}, the spatial
sampling and resolution of the data presented in this work is not optimal for the derivation of additive properties of individual \hh regions. 
Therefore these data are not optimal for the study of the H$\alpha$ luminosity function or the characteristic optical
extension of \hh regions, to mention just a couple of quantities. Although in every analysis of this kind one has to bear in mind that the measurements could correspond to an \hh region complex. 

When we reach the higher redshift regime the situation gets more
complicated. We are adding together regions belonging to very different
galactic components, considering that we are not distinguishing disc, arms and
bulge, but, in any case, it will depend on the morphology of the studied
galaxy.

\begin{figure}[h]
  \begin{minipage}[b]{0.7\linewidth}
    \centering
    \includegraphics[width=\textwidth, angle=-90]{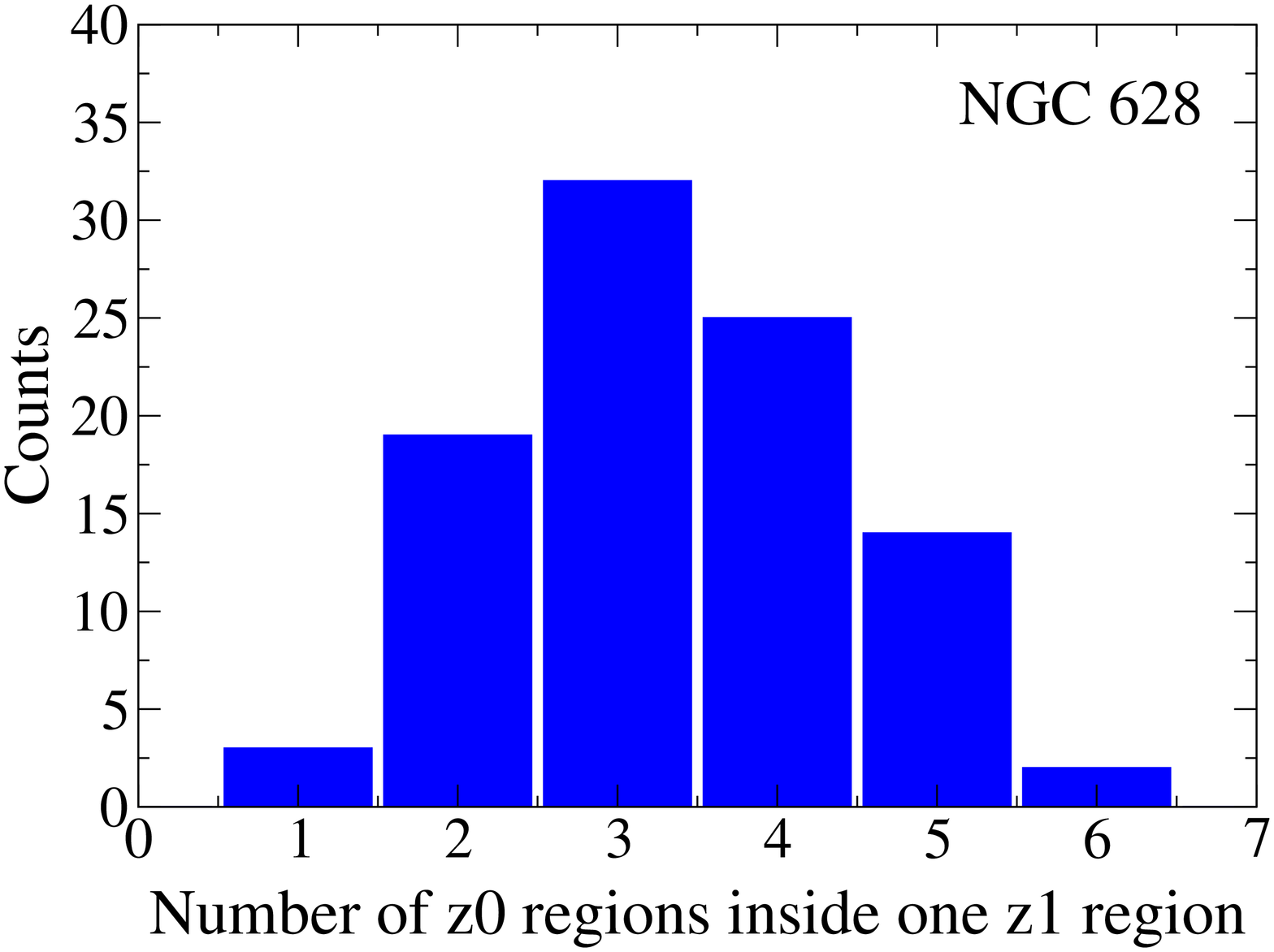}
  \end{minipage}
  \begin{minipage}[b]{0.7\linewidth}
    \includegraphics[width=\textwidth, angle=-90]{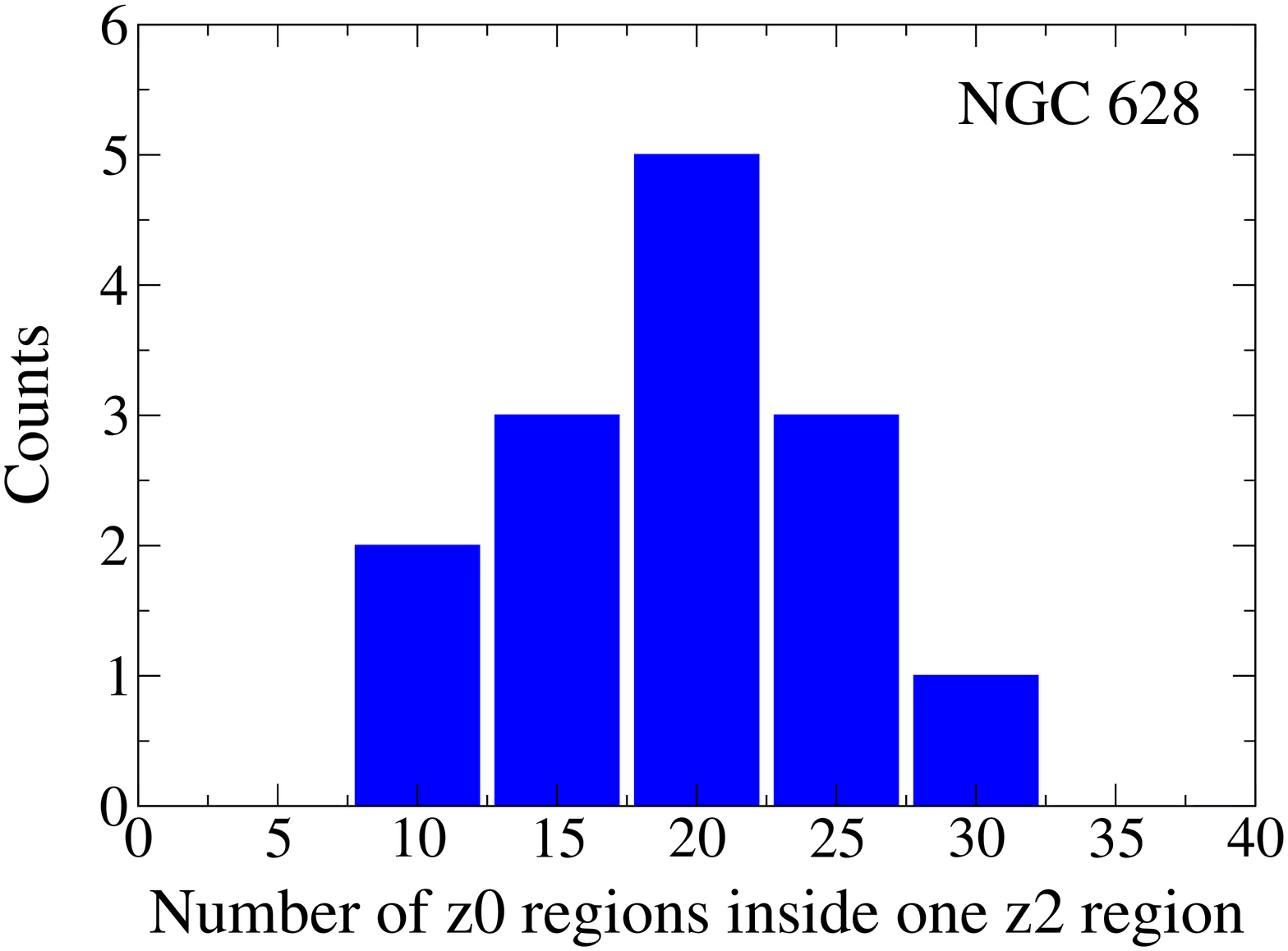}
  \end{minipage}
  \begin{minipage}[b]{0.7\linewidth}
    \includegraphics[width=\textwidth, angle=-90]{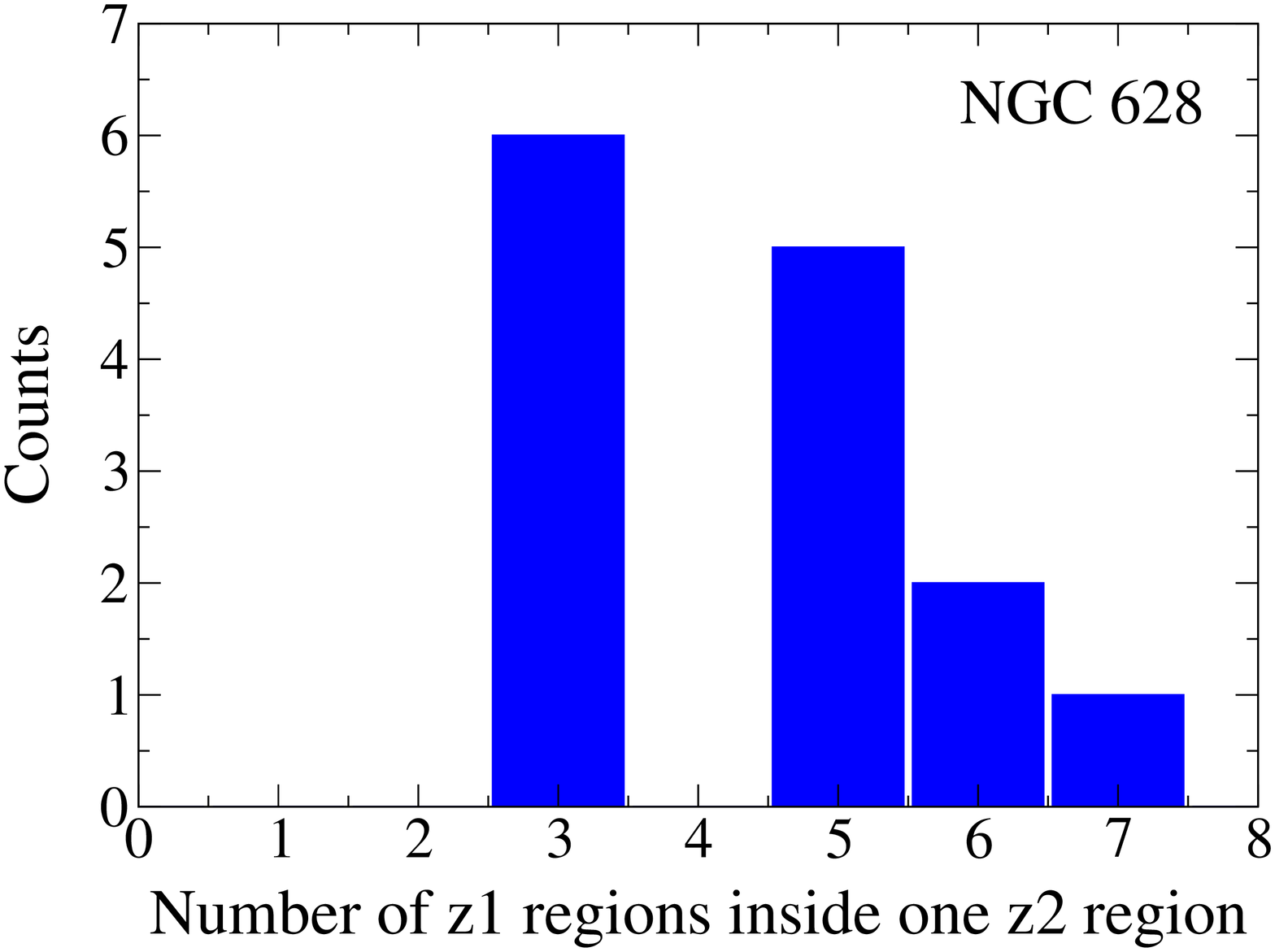}
  \end{minipage}
  \caption{Correspondence between number of regions on each regime for NGC
    628, i.e. the number of regions lying inside an individual region of the
    other redshift case. \textit{Upper panel:} correspondence between $Z0$ and
    $Z1$. \textit{Middle panel:} correspondence between $Z0$ and
    $Z2$. \textit{Lower panel:} correspondence between $Z1$ and $Z2$.}
  \label{fig:numbers}
\end{figure}

To explore further the \hh region limits mentioned above, we identified all
the regions from the higher resolution datacube that are detected as an
individual region at lower resolution. In this way we can analyse the real
source (or sources) of a particular measurement, and study how it is affected
by the spatial binning. Fig. \ref{fig:numbers} shows an example for NGC628.
The upper panel of this figure displays an histogram of the number of $Z0$
regions inside a $Z1$ one. This distribution indicates that, on average, each
$Z1$ detected \hh region is built of 3 smaller $Z0$ regions. For the higher
redshift case, each individual region is, in reality, an agglomerate of around
15 to 25 $Z0$ regions, while it is built by the contribution of $\sim4$ $Z1$
regions. Table \ref{tab:regionaverage} summarize the distribution of the
  number of regions in terms of median and standard deviation for all the
  sample.

\begin{table}[ht]
  \begin{minipage}[t]{\linewidth}
    \caption{Number of \hh regions detected for each galaxy}             
    \label{tab:regions}    
    \begin{center}
      \begin{tabular}{cccccc}
        \hline
        \hline
        Galaxy &  \multicolumn{3}{c}{Number of \hh regions at $z$} & $Z0/Z1$ & $Z0/Z2$  \\
        &  Z0 &  Z1 & Z2  \\
        \hline
        
        NGC628 &  286  &  77 &  14 & 3.7 & 20.4 \\
        \hline
        NGC1058 & 331 &  67 &  7  & 4.9 & 47.2 \\
        \hline
        NGC1637 & 297 &  95 &  14  & 3.1 & 21.2 \\
        \hline
        NGC3184 & 169 &  66 &  17  & 2.5 & 9.9 \\
        \hline
        NGC5474 & 122 &  47 &  10  & 2.5 & 12.2 \\
        \hline
        \hline
      \end{tabular}\end{center}
  \end{minipage}
\end{table}

\begin{table}
  \begin{minipage}[t]{\linewidth}
    \caption{Distribution of number of regions corresponding to different regimes.}             
    \label{tab:regionaverage}    
    \begin{center}
      \begin{tabular}{cccc}
        \hline
        \hline
        Galaxy & $Z0$ in $Z1$ & $Z0$ in $Z2$ & $Z1$ in $Z2$  \\
        \hline
        NGC628 &  $3\pm1$  &  $20\pm5$ &  $4\pm1$ \\
        \hline
        NGC1058 & $3\pm2$ &  $19\pm9$ &  $5\pm2$  \\
        \hline
        NGC1637 & $2\pm1$ &  $14\pm5$ &  $4\pm1$ \\
        \hline
        NGC3184 & $1\pm1$ &  $4\pm3$ &  $2\pm1$   \\
        \hline
        NGC5474 & $2\pm1$ &  $8\pm4$ &  $3\pm1$  \\
        \hline
        \hline
      \end{tabular}\end{center}
  \end{minipage}
\end{table}

\subsubsection{Diagnostic diagrams}

\begin{figure*}[ht]
   \centering
   \includegraphics[width=0.37\linewidth,clip=true,bb=40 0 590 710,angle=-90]{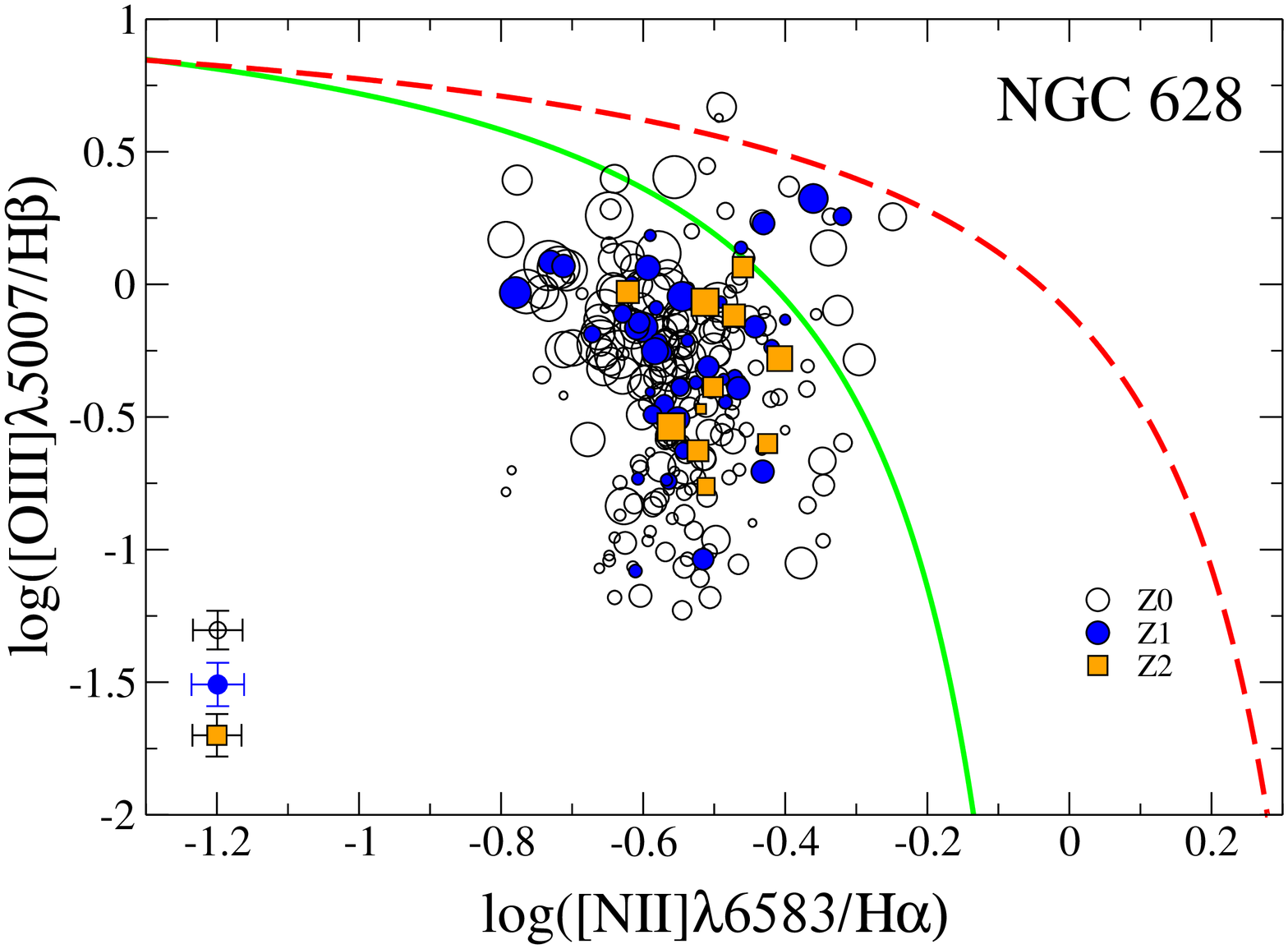} ~
   \includegraphics[width=0.37\linewidth,clip=true,bb=40 0 590 710,angle=-90]{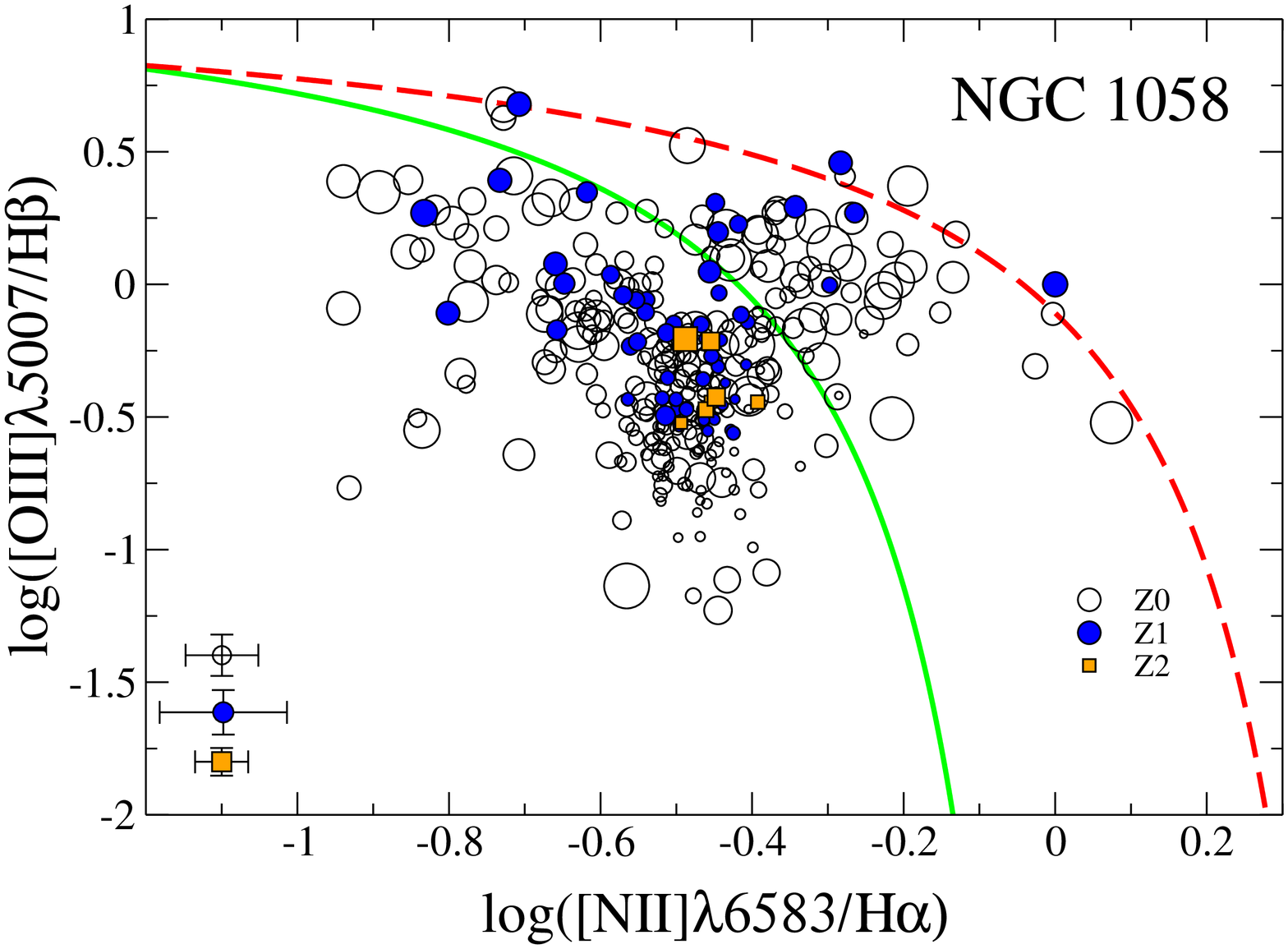}\vspace{0.5cm}
   \includegraphics[width=0.37\linewidth,clip=true,bb=40 0 590 710,angle=-90]{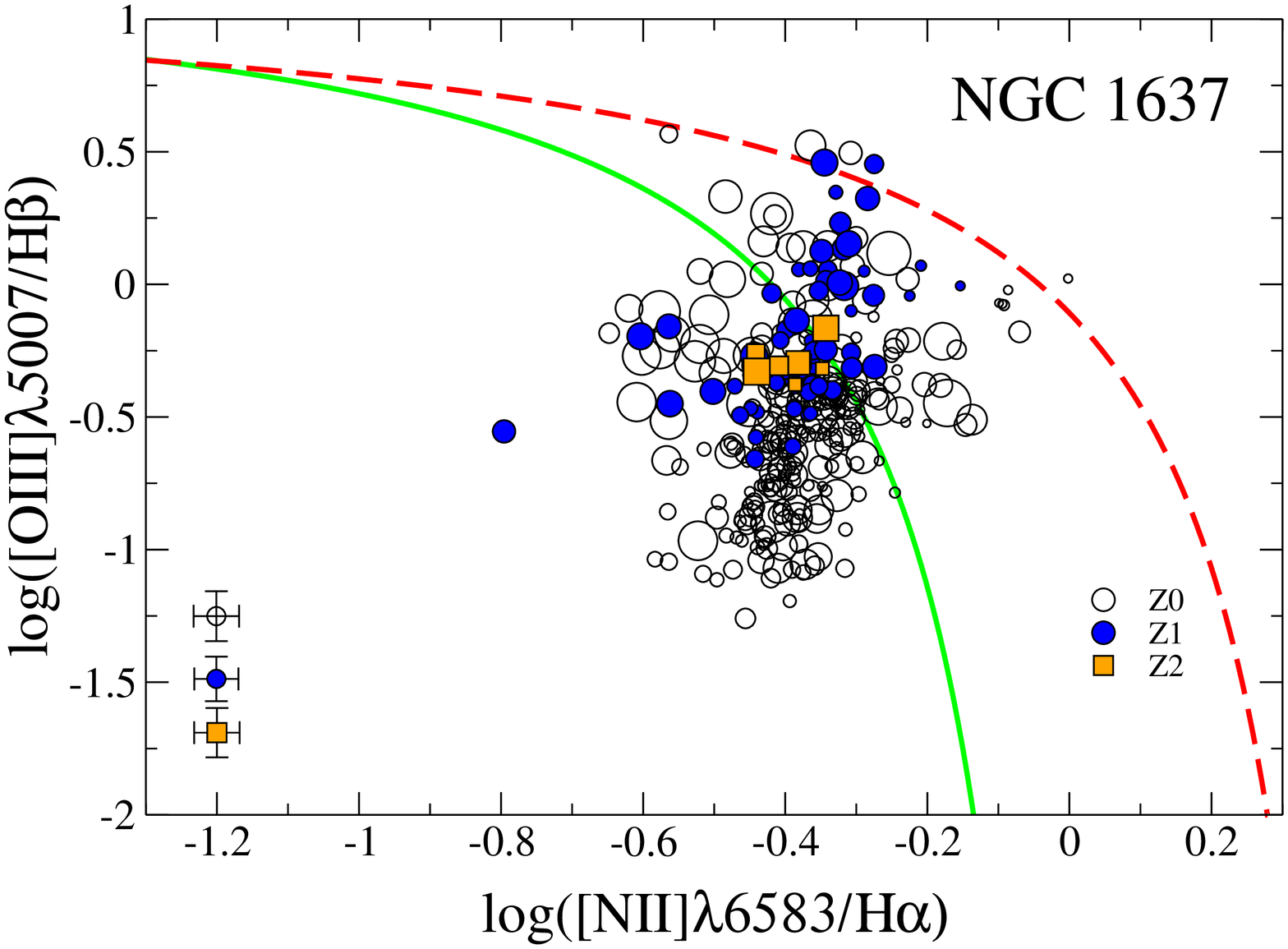}~
   \includegraphics[width=0.37\linewidth,clip=true,bb=40 0 590 710,angle=-90]{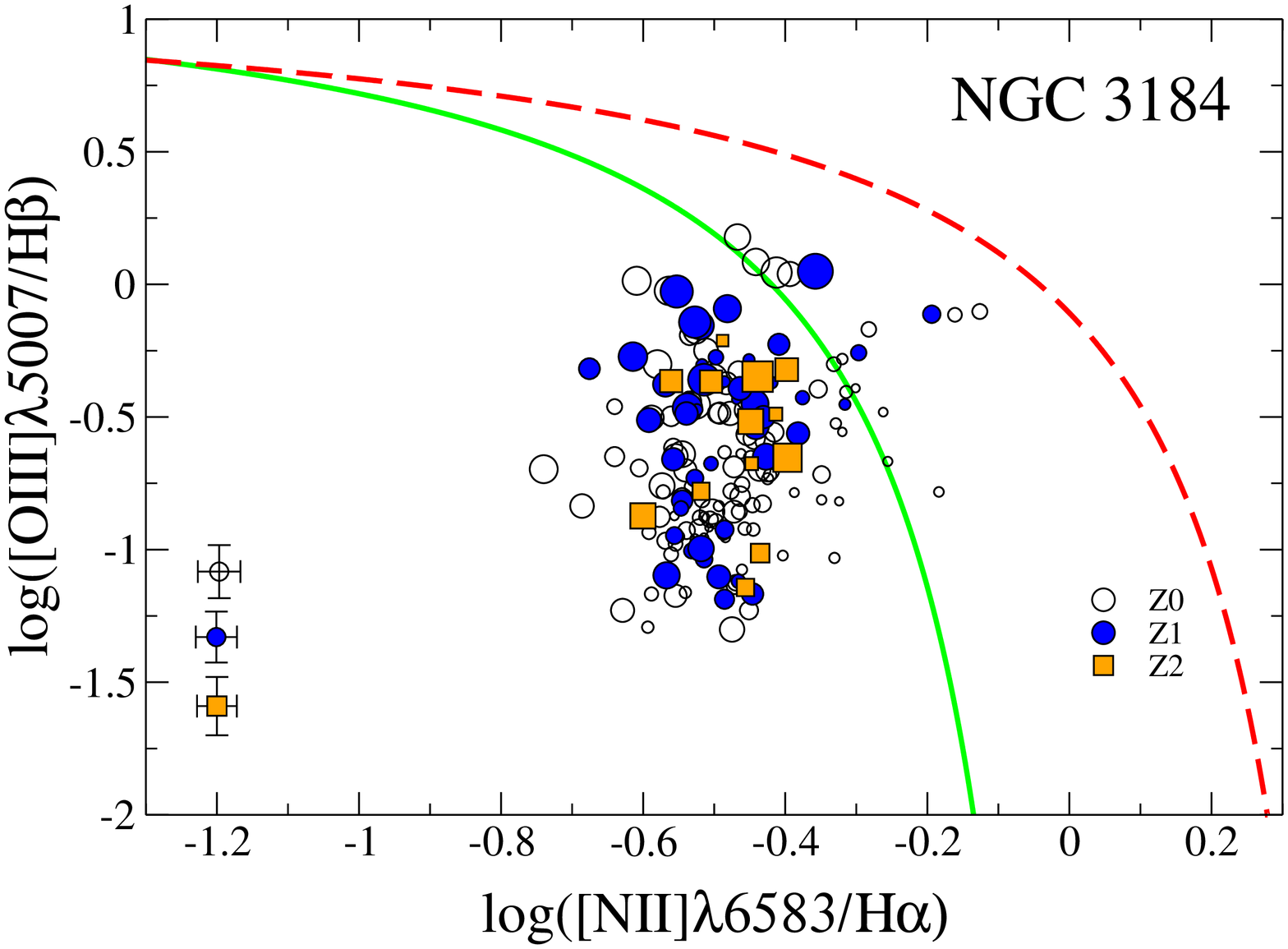}\vspace{0.5cm}
   \includegraphics[width=0.37\linewidth,clip=true,bb=40 0 590 710,angle=-90]{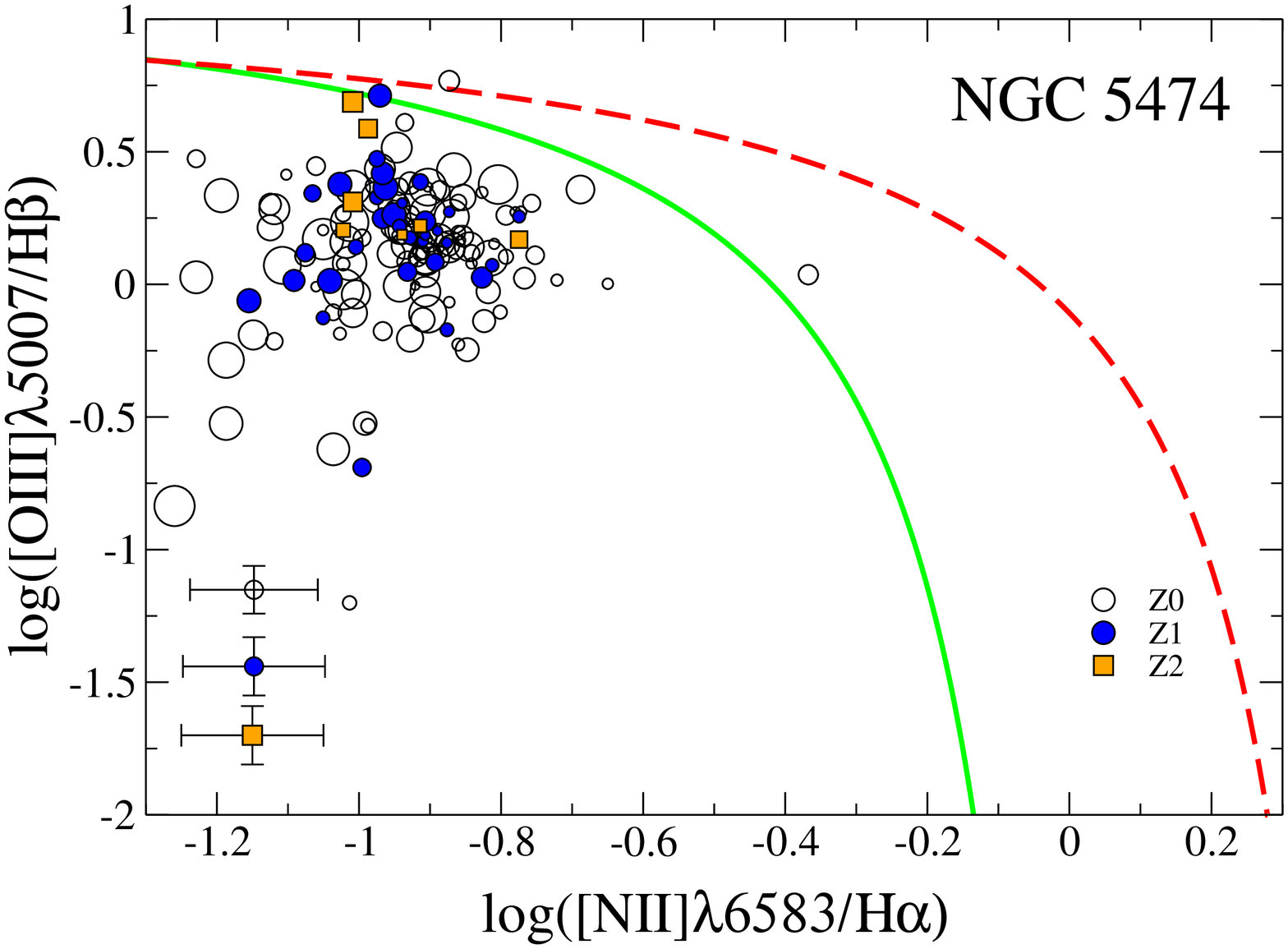}
   \caption{Classical BPT emission-line diagnostic diagrams
     \citep{1981PASP...93....5B,1987ApJS...63..295V} with the demarcation curve
     of  \cite{2001ApJ...556..121K}  (upper curve, red) and
     \cite{2003MNRAS.346.1055K} (lower curve, green) for each galaxy. Empty
     circles correspond to $Z0$, the blue circles to $Z1$ and the orange
     squares to $Z2$. Symbol sizes are proportional to galactocentric distance,
     i.e., the larger symbols are the farthest from the centre.}
   \label{fig:BPT}
\end{figure*}

We can use the \oiii\lam5007/\hb\, ratio, together with  \nii\lam6583/\ha, to
construct the classical BPT diagram for all the galaxies, as shown in
Fig. \ref{fig:BPT}. These plots allow us to see how the spatial information
loss is affecting our understanding and characterisation of the ionisation
sources at higher redshift. It is clear that when the idea is to separate,
e.g., AGN activity from star-formation, one has to consider the degree of
contamination due to the area subtended by the spectroscopic aperture. In the
present study, where we are dealing with \hh regions, we have the opportunity
to map the way that this resolution degradation alters the position of each
individual region on the BPT diagram. We have also plotted the
\cite{2001ApJ...556..121K}  (upper red dashed-line) demarcation curve, often
invoked to distinguish between star-forming regions (below the red line), and
other sources of ionisation, such as AGN/shocks/post-AGB stars (above the
line) and the \cite{2003MNRAS.346.1055K} (lower curve, green) to indicate the
so-called composite zone. In all the plots, empty circles correspond to the
$Z0$ case, while full blue circles and orange squares correspond to $Z1$ and
$Z2$ respectively. We also plot symbol size proportional to the galactocentric
distance, i.e., the larger the symbol, the farthest from the galaxy centre.

This distribution of regions in the BTP diagram is expected since star-formation is the dominant ionising source
in all the selected galaxies. As described in \citet[e.g.][]{2008ApJ...681.1183K}  the
objects lying in the composite region do not require an
additional ionising source {\it per se}, in the case of a strong
nitrogen enrichment. Other ionisation sources,
like shocks, AGB star and AGN photoionization, mixed with pure star-formation,
can also populate this region. Thus, the intermediate region only indicates that
their nature is uncertain, not that the ionising source is definitely different than star-formation. However, for the particular scope of this article,
we have preferred a conservative approach and do not consider them in
any further discussion.

\clearpage

\begin{figure}
  \centering
  \includegraphics[width=0.75\columnwidth,clip=true,bb=40 0 590 710,angle=-90]{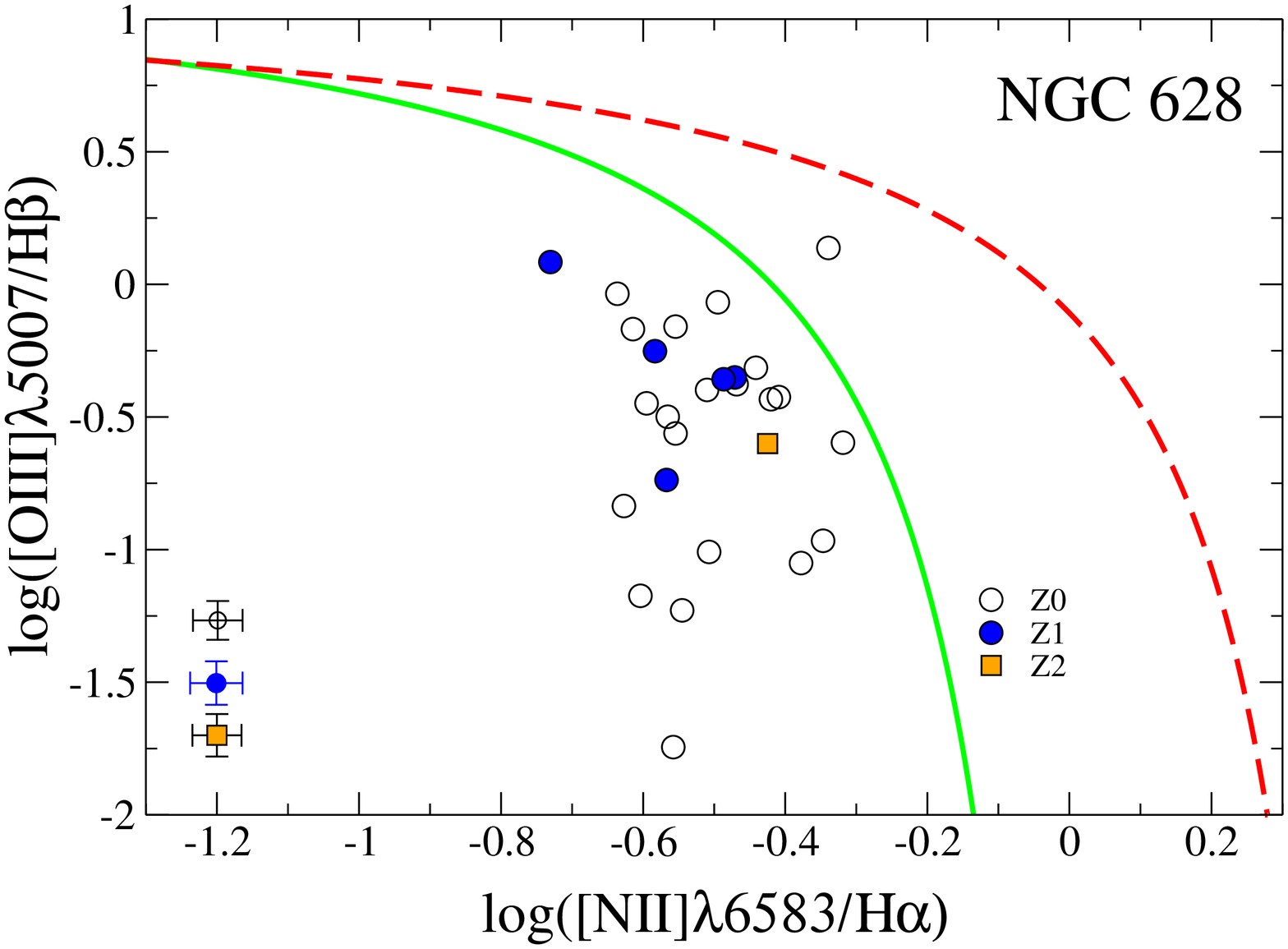}
  \caption{Same as Fig.\ref{fig:BPT} but for one $Z2$ region and its
    corresponding $Z0$ and $Z1$ region content. Empty circles correspond to
    $Z0$, the blue circles to $Z1$ and the orange squares to $Z2$.}
  \label{fig:BPTindividual}
\end{figure}

In the $Z1$ case of NGC 628 (blue circles), the effect of spatial resolution
degradations is a collapse of the regions into the most populated zone of the
$Z0$ plot (empty circles), and then again, for the $Z2$ regime, all regions
are more concentrated in the ``core'' of the lower redshift case. In the $Z2$
case we do not see composite regions, with all lying  in the star formation
zone. The same behaviour is displayed in the BPT diagrams of the other
galaxies, except for NGC 1637 with several $Z2$ points in the composite
area. NGC 1058 displays the strongest collapse
effect. Fig. \ref{fig:BPTindividual} shows the behaviour of one individual
$Z2$ region of NGC628 and its constituents from the other redshift
regimes. The underlying reason of the shrinking in the occupied parameter
  space with increasing Z could be because the ratios are
  luminosity-weighted. If two regions have $R1 = a1/b1$ and $R2 = a2/b2$, then
  the combined ratio $R = (a1+a2) / b = R1*(b1/b) + R2*(b2/b)$, where $b =
  b1+b2$. In the case where $b1>>b2 (b1<<b2)$, then $R \sim R1 (R2)$. When
  $b1\sim b2$, then $R \sim 0.5*(R1+R2)$. $R$ is always between $R1$ and
  $R2$. Therefore, summing multiple \hh regions and measuring line ratios mean
  taking a luminosity-weighted mean of the ratios.

In all the $Z1$ simulated galaxies the approximate shape of the two-branches
structure, visible in the original $Z0$ galaxies is acceptably
discerned. Thus, we suggest that $Z1$ can recover the global trend of \hh
regions in the BPT diagram while $Z2$ only reflects the denser part of the
distribution in the lower redshift case.

\subsubsection{Line ratios and abundances}\label{sec:abund}

\begin{figure}[ht]
  \centering
  \includegraphics[width=0.72\linewidth,clip=true,bb=60 0 560 720,angle=-90]{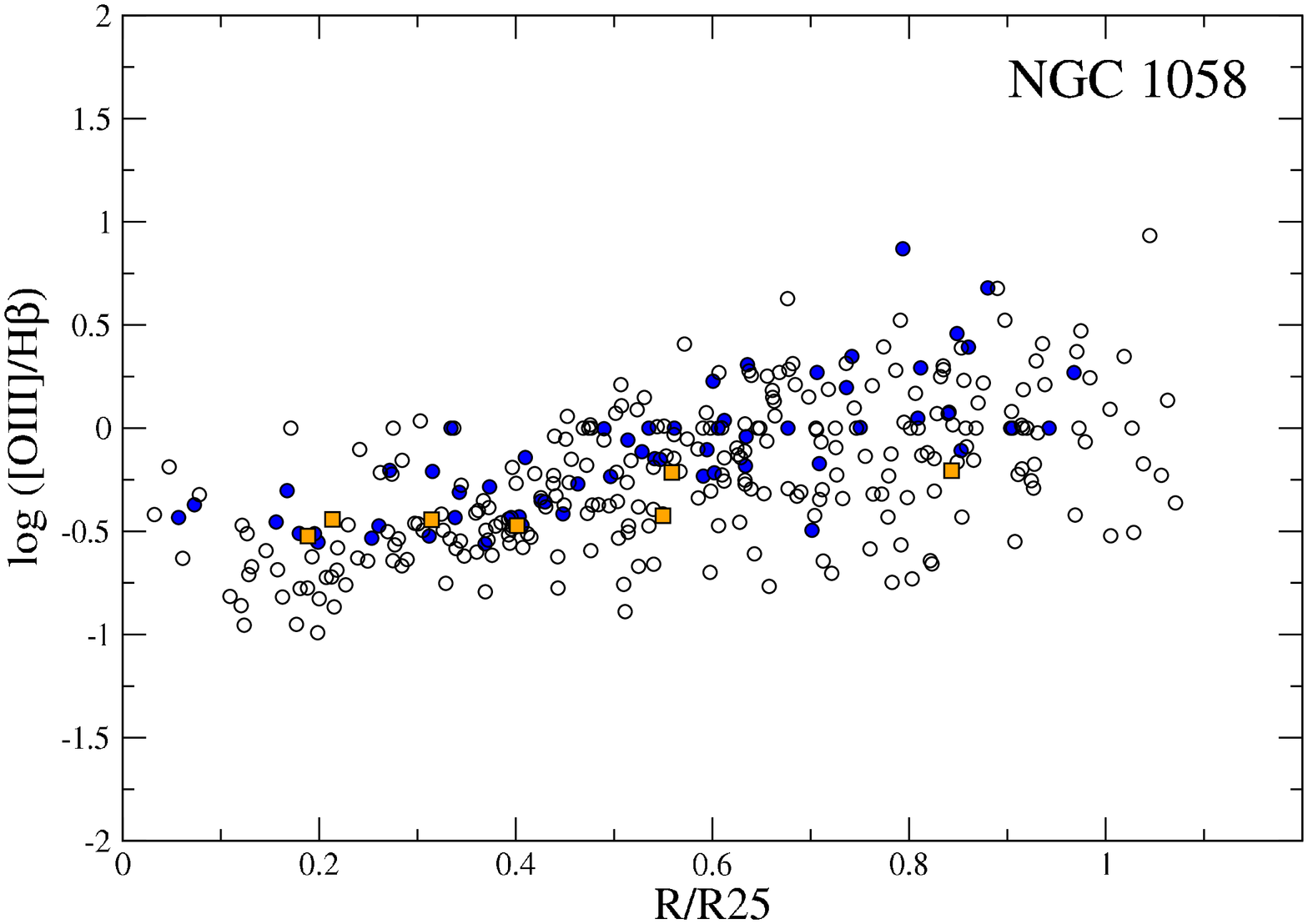}  ~
  \includegraphics[width=0.72\linewidth,clip=true,bb=60 0 560 720,angle=-90]{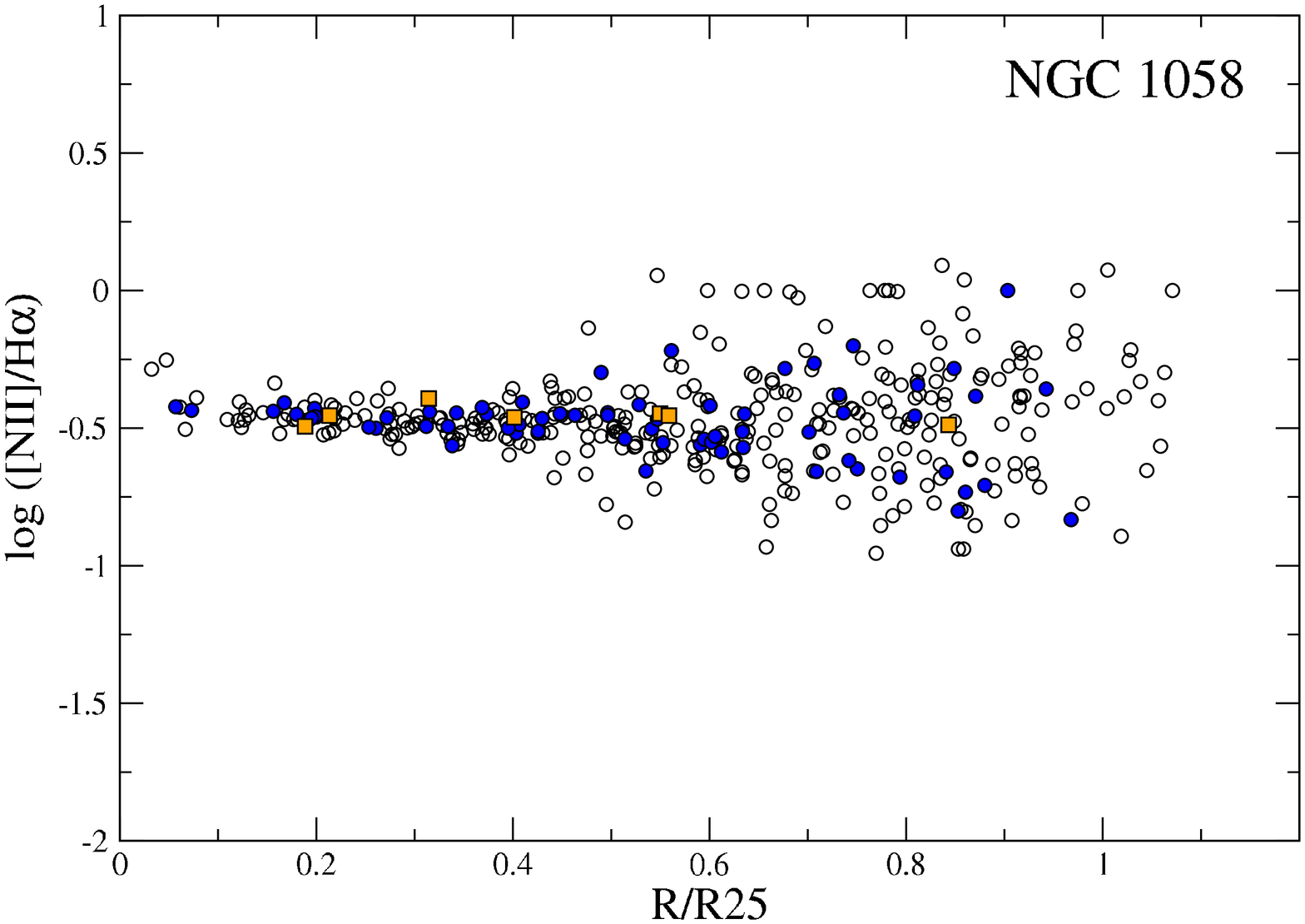}  
  \caption{\textit{Top}: radial distribution of \oiii/\hb\, for NGC
    1058. \textit{Bottom}: radial distribution of the \nii/\ha\, ratio
    for the same galaxy. Empty circles correspond to $Z0$, the blue circles to
    $Z1$ and the orange squares to $Z2$.}
  \label{fig:ngc1058ratios}
\end{figure}

\begin{figure*}[ht]
  \centering
   \includegraphics[width=0.34\linewidth,clip=true,bb=80 0 570 710,angle=-90]{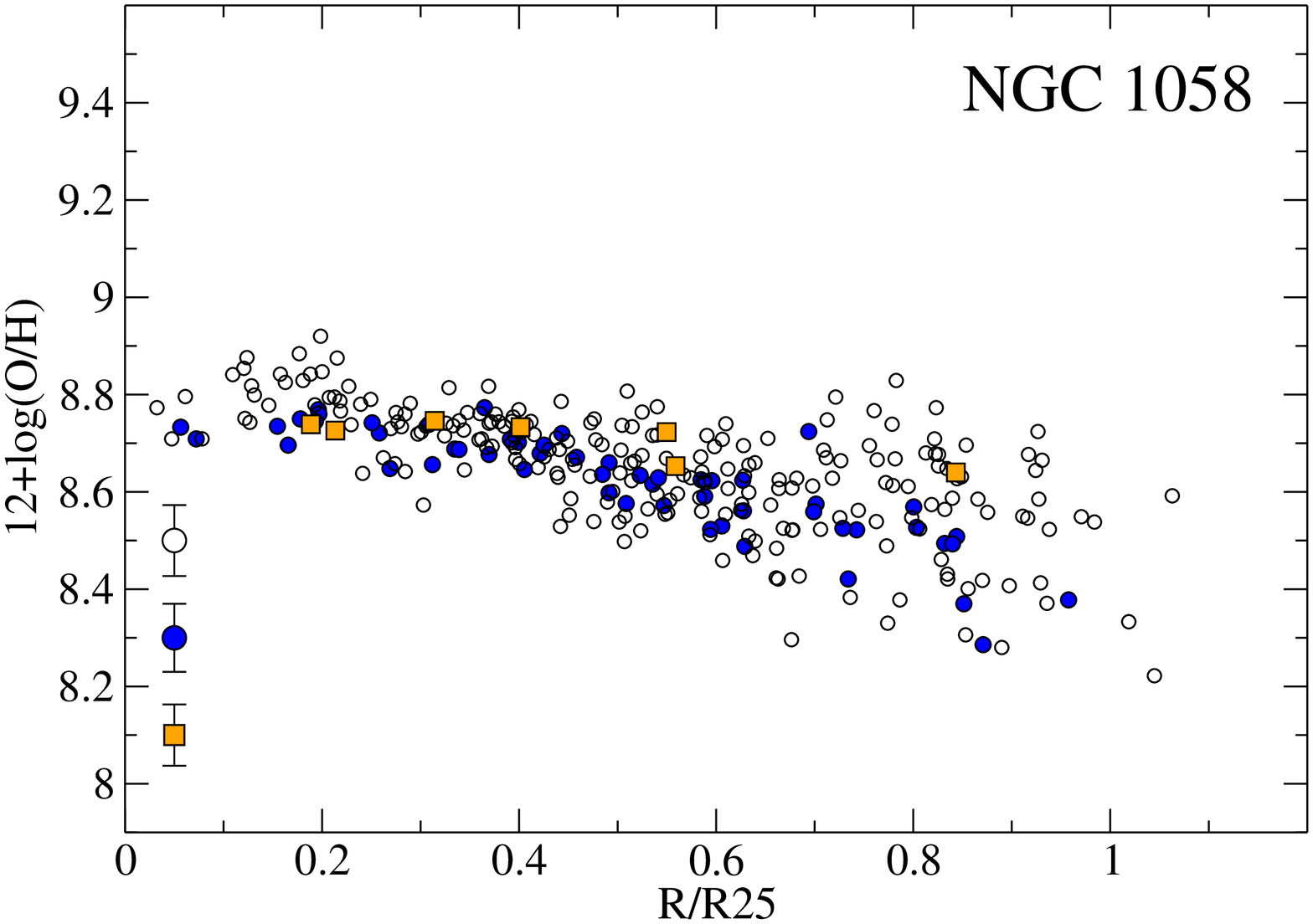} ~
   \includegraphics[width=0.34\linewidth,clip=true,bb=80 0 570 710,angle=-90]{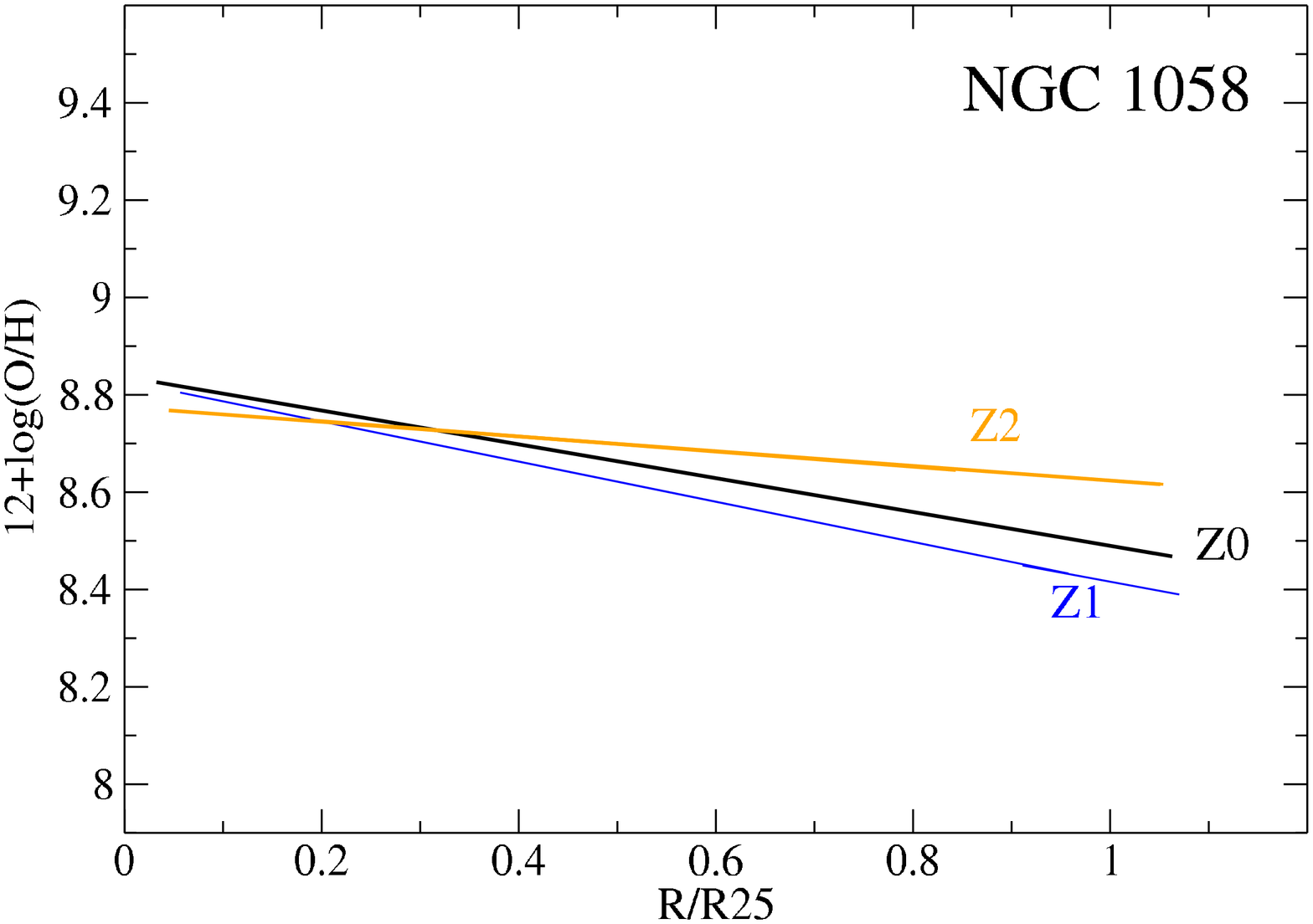}
   \includegraphics[width=0.34\linewidth,clip=true,bb=80 0 570 710,angle=-90]{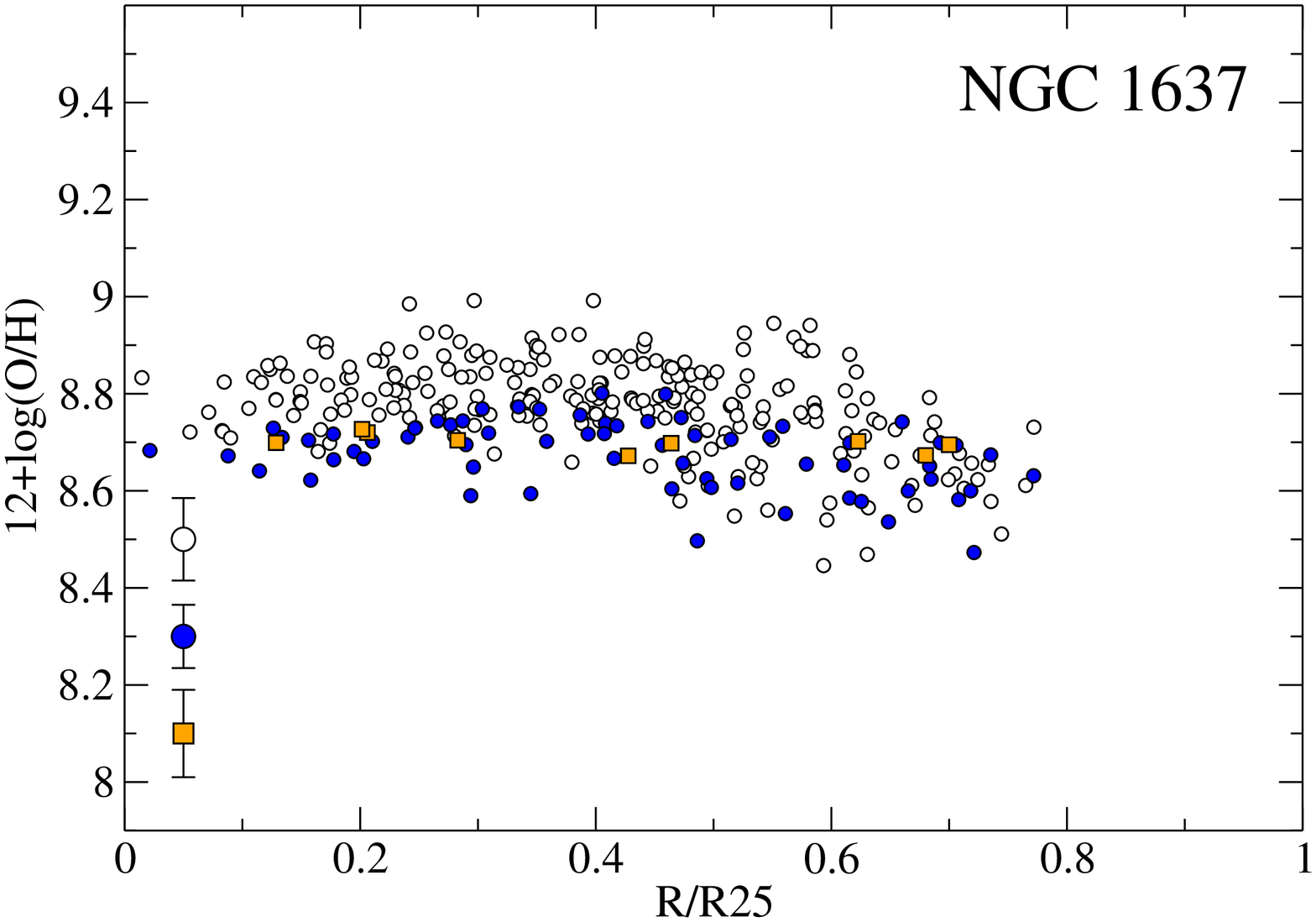}~
   \includegraphics[width=0.34\linewidth,clip=true,bb=80 0 570 710,angle=-90]{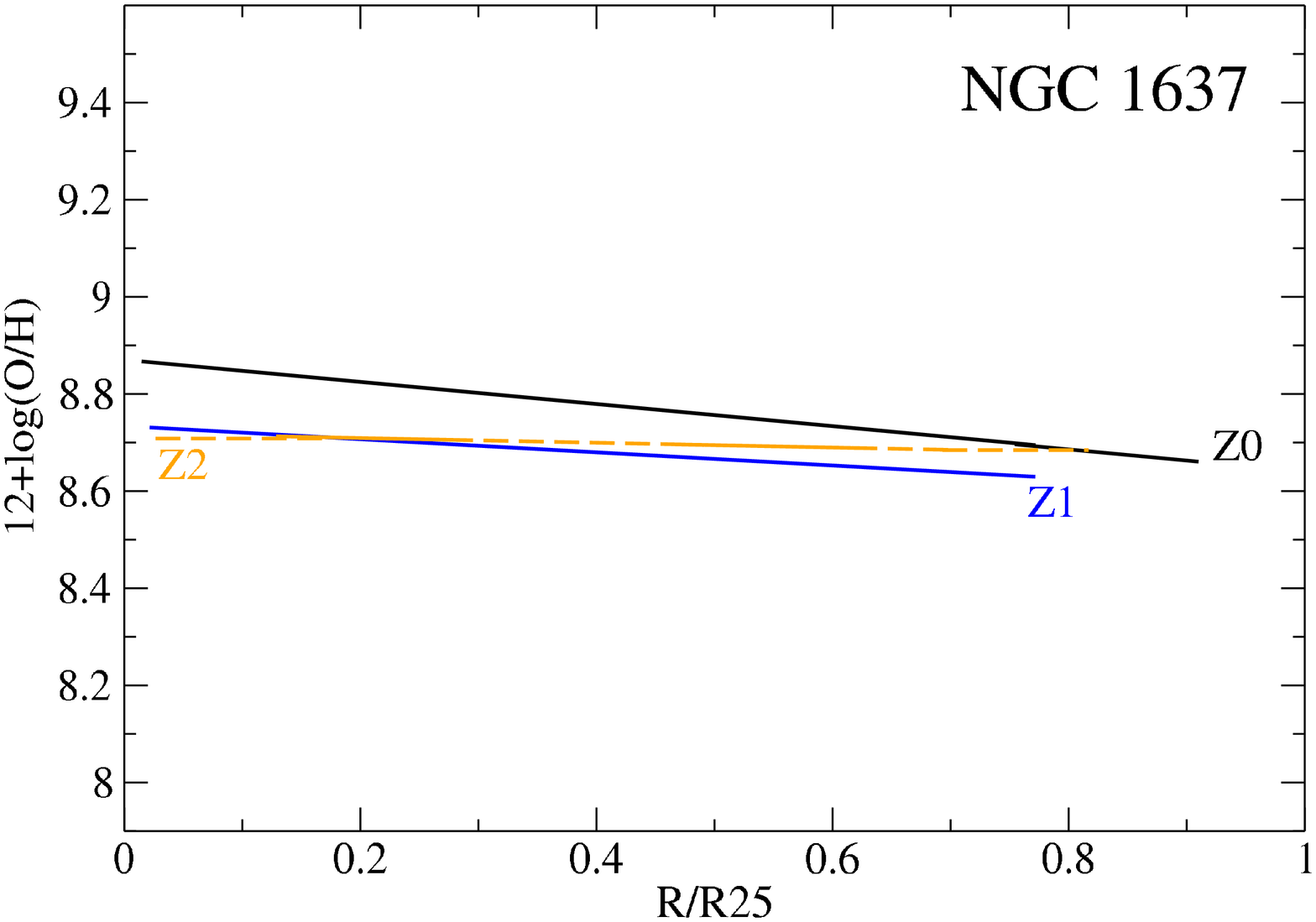}
   \caption{Radial abundance gradients derived using the relation O3N2 from
     \cite{2004MNRAS.348L..59P} for two of the galaxies. On each left graph the
     three redshift regimes are over-plotted, the empty circles correspond to
     $Z0$, the blue circles to $Z1$ and the orange squares to $Z2$. The
       three points on the down left corner show the average error bars. On
     the right we can see a linear regression fitting to each regime.}   
   \label{fig:abundances}
\end{figure*}

For the derivation of the oxygen abundance we adopted the O3N2 indicator using
the calibration provided by \cite{2004MNRAS.348L..59P}, defined by the
relation 12 + log(OH)$ = 8.73 - 0.32\,
log_{10}\left(\frac{\oiii\lambda5007/H\beta}{\nii\lambda6583/H\alpha}\right)$.
This calibration has the advantage of being relatively easy to measure
  (based on strong emission lines), is basically not affected by the effects of
  extinction (the involved line-ratios are very close in wavelength), and
  implies a simple linear conversion, reason for which this indicator is very
  popular for  high redshift studies\footnote{Note
    that this indicator has been recently re-calibrated for the
    high-metallicity regime by \citet{2013arXiv1307.5316M}. Although
  there is a known discrepancy in deriving metallicity with different
  calibrators \citep{2008ApJ...681.1183K}, \cite{Sanchez:2012b} showed that
  there is a correlation between the different determinations, being the
  discrepancies present at high redshift, to a large degree, a consequence of
  resolution effects. In any case, although the specific results obtained here
  are applicable to the O3N2 calibrator in particular, the use of one or other
  calibration does not modify qualitatively the results of the proposed
  analysis. Other methods will suffer from different systematics effects
  (e.g. absolute scale differences) which analysis are beyond the scope of
  this paper.} Fig. \ref{fig:ngc1058ratios} shows the radial distributions of
the ratios \nii/\ha\,and \oiii/\hb. We can see that the \oiii/\hb\, radial
gradient becomes flatter with redshift. This will have a clear effect on the
abundance determination. As an example, Fig. \ref{fig:abundances} shows the
radial abundances for regions in two of the galaxies (NGC 1058 and NGC 1637)
with fitted gradients for comparison purposes, corresponding the empty circles
to $Z0$, the blue circles to $Z1$ and the orange squares to $Z2$ (see
  Fig. \ref{fig:abundremaining} for the remaining objects).

Note that the oxygen abundance is not an additive but
a relative property, and it exhibits a ubiquitous radial gradient (e.g.,
\citealt{Sanchez:2012b}). Therefore,
neither the average nor a value at a fixed aperture are
characteristic of the full distribution (e.g., \citealt{2004ApJ...613..898T,2012ApJ...756L..31R}). Even more, it is still not
clear that the average of the values derived at different radii is
representative of a single value derived using an aperture that
encircles all the previous ones.

In all the galaxies we see a radial abundance gradient decreasing outwards,
with NGC 628 and NGC 1058 being the more evident cases. This effect is
expected as late-type galaxies are known to have a composition gradient
related to their star formation history (SFH). The capability of any survey of
being able to measure this gradient is essential if one of its goals is to
characterise the properties of galactic disks. 

\begin{figure}[t]
  \centering
  \includegraphics[width=0.75\columnwidth,clip=true,bb=40 0 590 710,angle=-90]{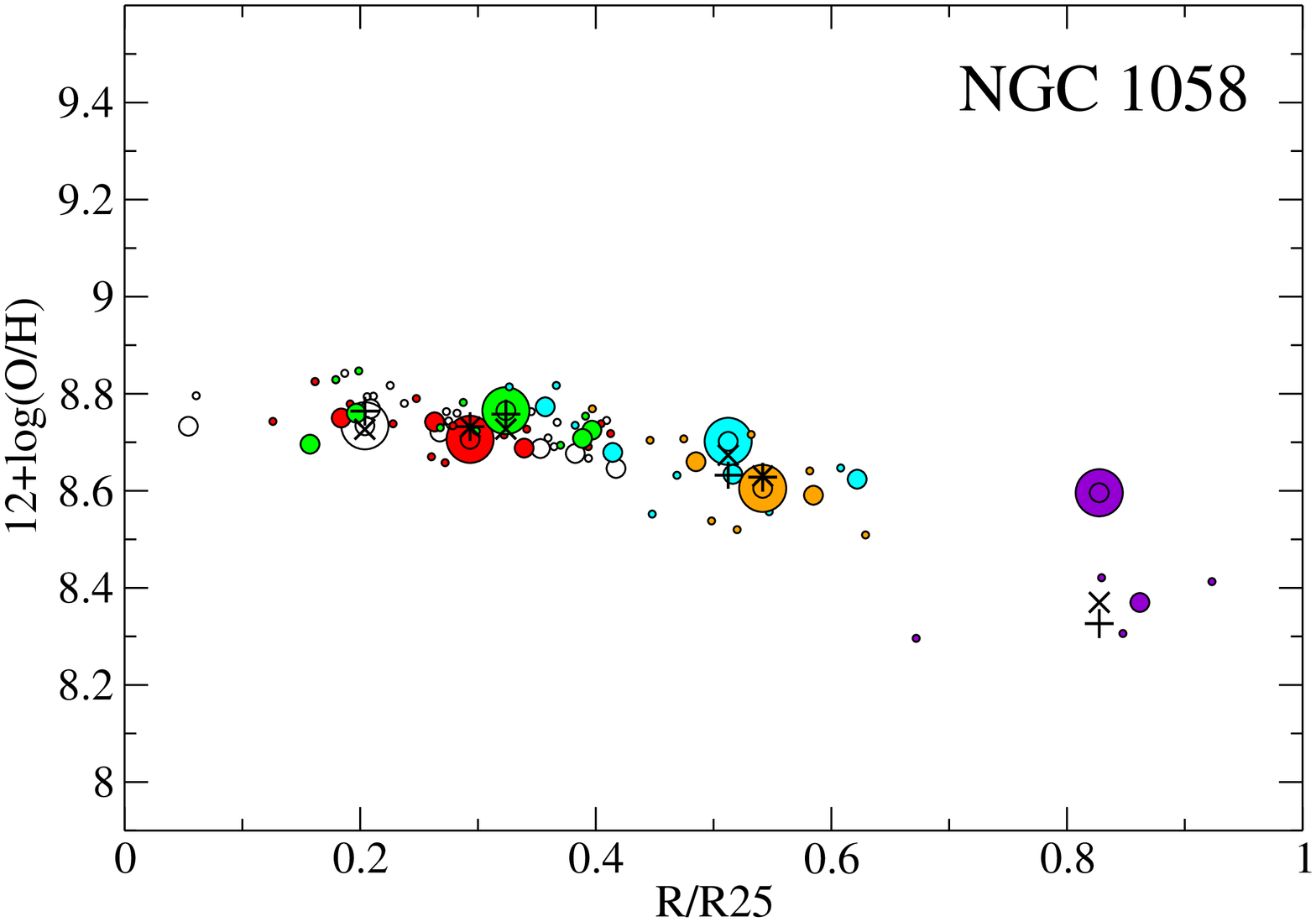}
  \caption{ Radial abundance gradient of NGC 1058 for the three redshift
      regimes. Color coded is each different $Z2$ region with its corresponding
      $Z0$ and $Z1$ components (same color). Small circles correspond to $Z0$,
      medium sized to $Z1$ and the bigger ones to $Z2$. {\em Plus} signs
      represent the abundance determined summing the flux of all the $Z0$
      regions inside the corresponding $Z2$ region, while crosses are the determination from the total flux of $Z1$
      regions (also the ones inside the $Z2$ region).}
\label{fig:fluxes}
\end{figure}

All galaxies show a nearly flat distribution in the innermost parts, until
$r\sim 0.2R_{25}$. From this point, the abundance measurements start to
exhibit higher values, reaching maxima at $r\sim 0.4R_{25}$ in 3 of the cases
($r\sim 0.2R_{25}$ for NGC 1058). After this, the slope of the radial
abundance distribution changes its sign, and decreases with increasing
galactocentric radius. In all  cases, $Z1$ reproduces notably the $Z0$ trend
with slightly lower abundances. The same radial gradients are visible, with
the only difference being the absence of the peak at  $r\sim 0.2R_{25}$ (in
the case of NGC1058) and $r\sim 0.4R_{25}$ (for NGC 628). For all the
galaxies, $Z2$ fails to reproduce the slope behaviour, as anticipated with
the \oiii/\hb\ line ratio, showing  a nearly flatter distribution (on average)
like in NGC 1058 and NGC 1637 (see Fig. \ref{fig:abundances}).

Fig. \ref{fig:abundances} shows that for the higher redshift simulation the
abundance flattens. To search for possible sources of this behavior on the
radial abundance distribution for $Z2$, we performed several tests. We used
the \hh regions catalogue of \cite{Sanchez:2012b} to simulate a spatial
binning adding together all the \hh regions inside different apertures. The
observed effect was a simple average of the \hh region properties as expected,
discarding the sum of \hh regions as source of the flattening. This can
  be also confirmed in Fig. \ref{fig:fluxes} that shows the radial abundance
  gradient of NGC 1058 for the three redshift regimes. Color coded is each
  different $Z2$ region with its corresponding $Z0$ and $Z1$ components (same
  color). Small circles correspond to $Z0$, medium sized to $Z1$ and the
  bigger ones to $Z2$. {\em Plus} signs represent the abundance determined
  summing the flux of all the $Z0$ regions inside the corresponding $Z2$ region, while crosses are the
  determination from the total flux of the $Z1$ regions inside that $Z2$ region.

It is well known that the derivation of Balmer emission line fluxes are affected by the
accuracy of the subtraction of the underlying stellar population. In
particular, \hb\, is strongly affected. Along the pilot studies for the CALIFA
survey \citep{2011A&A...534A...8M}, studied the capability of recovering this
emission line at different equivalent widths, underlying stellar populations,
and spectral resolutions. We noticed that at low equivalent widths \hb\, is
overestimated for old stellar populations and underestimated for young stellar
populations. This effect is stronger at lower spectral resolutions. The net
effect of the decrease of the spatial sampling produces these three combined
effects: (i) it reduces the original equivalent width by increasing more
regions dominated by low-intensity diffuse emission, increasing the underlying
continuum with respect to the gas emission; (ii) it increases the fraction of
old stellar populations in the inner regions, and the fraction of young
stellar populations in the outer ones; and (iii) it decreases the effective
spectral resolution by the beam effect, i.e., the co-adding of regions with
different kinematic properties. All together it produces an increase of
\oiii/\hb\, in the inner regions and its decrease in the outer ones. For
galaxies with a positive gradient in \nii/\ha\, it produces a flattening of
the radial abundance.

\begin{figure}[t]
  \centering
  \includegraphics[width=0.75\columnwidth,clip=true,bb=40 0 590 710,angle=-90]{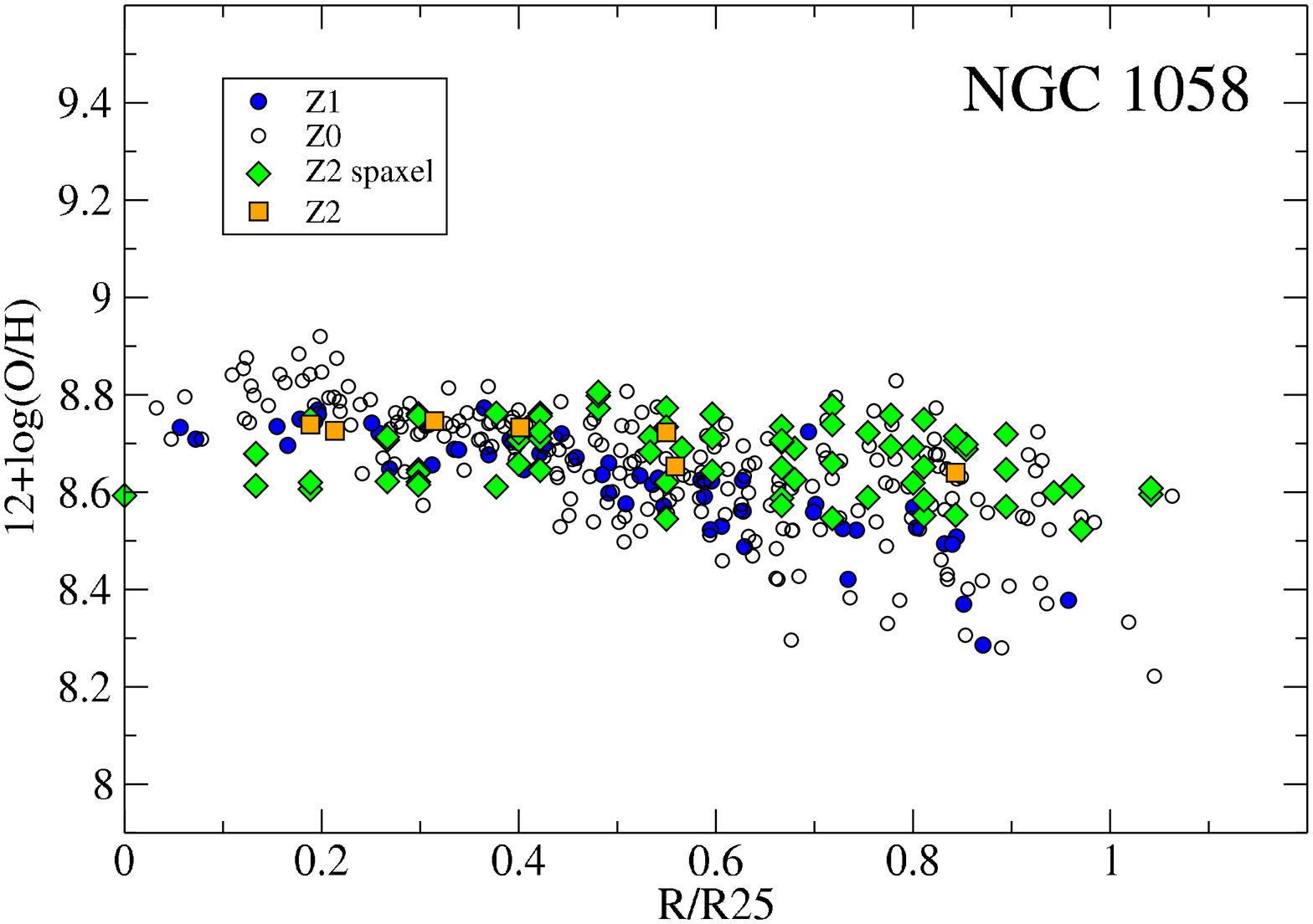}
  \caption{Comparison between the same analysis as Fig. \ref{fig:abundances}
    and an spaxel-spaxel determination above a S/N threshold (green) for NGC
    1058.}
  \label{fig:spaxel}
\end{figure}

This effect is also present in other abundance indicators, like R23, and less
affected by others like \nii/\ha. Evidently, the effect of the spatial
resolution degradation on the $Z2$ regime inhibits any  possibility of
measuring convincing radial abundance gradients at that redshift.

\cite{2013ApJ...767..106Y}  find similar results studying the \nii/\ha\,
ratio. They show that seeing-limited observations produce significantly
flatter gradients than higher angular resolution observations. They find a
critical angular resolution FWHM range ($< 0.02\arcsec$), which depends on the
intrinsic gradient of the galaxy, beyond which the measured metallicity is
significantly more flattened than the intrinsic metallicity.

Another effect that may be playing an important role on the radial abundance
gradients and the BPT diagrams, is that when we go to higher redshift, the
coarse resolution causes us to add more diffuse gaseous component to the \hh
aggregates, with potentially different dominating ionising sources than in the
\hh complexes. The difference between the {\em plus}, the cross and the
  bigger circles of Fig. \ref{fig:fluxes} would account for the different
  amount of diffuse medium considered on each region. The effect of this
contamination is very clear if we study cases where we can resolve the diffuse
medium at tens of parsec scale like in M33 or NGC 5253
(e.g. \citealt{2006ApJ...644L..29V,2010MNRAS.402.1635R,2010A&A...517A..27M,2011MNRAS.413.2242M}). In
all the mentioned cases, as we move away from the \hh region, with the maximum
local H$\alpha$ flux, as this flux decreases, the ratio \nii/\ha\, increases
while \oiii/\hb\, decreases. This will cause the abundance determination based
on the O3N2 method  to increase. If in a considered galaxy the \hh-region
spatial density is low, the contribution of the diffuse component to the line
ratios could become very important and even the dominant source of line
fluxes, i.e. as could be the case in the outer points of the radial abundance
distribution plotted in Fig. \ref{fig:abundances}. In any case, an estimation
of the contribution of the diffuse component is not trivial and depends on the
different galaxy components considered (arm, inter-arm, bulge, nucleus,
etc.).

\begin{figure*}[ht]
  \centering
  \includegraphics[width=0.33\linewidth,clip=true,angle=-90]{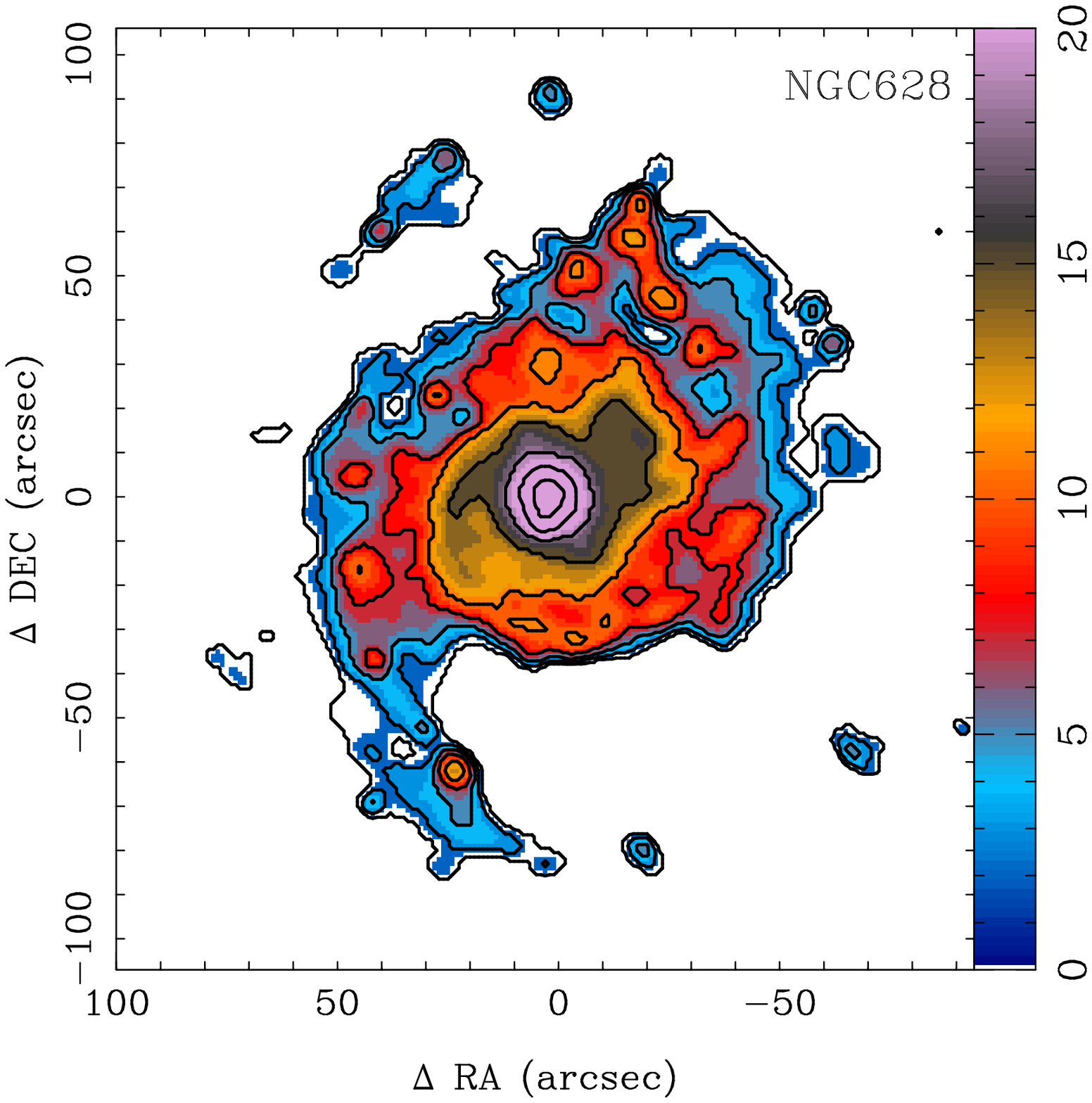}
  \includegraphics[width=0.33\linewidth,clip=true,angle=-90]{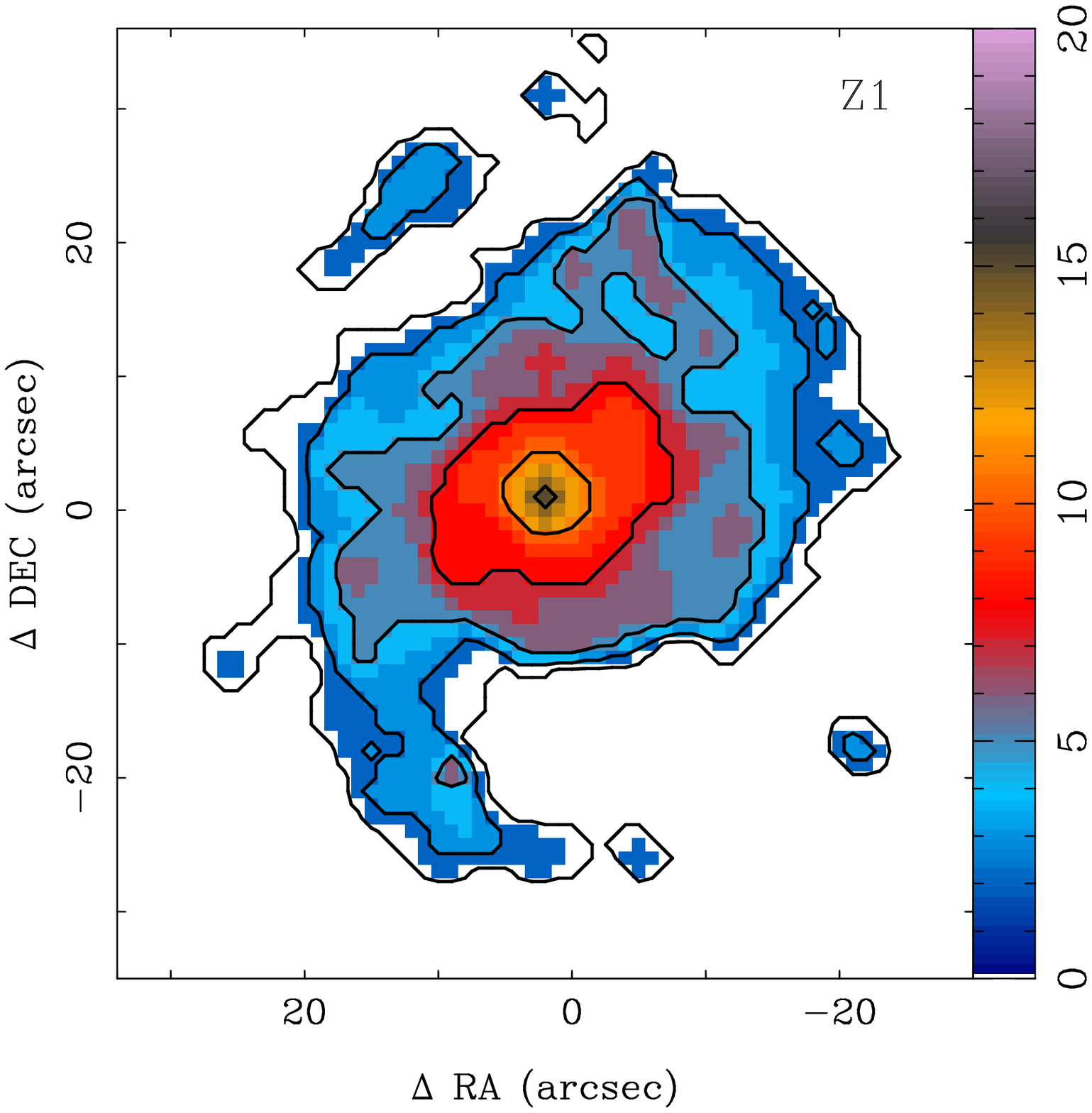}
  \includegraphics[width=0.33\linewidth,clip=true,angle=-90]{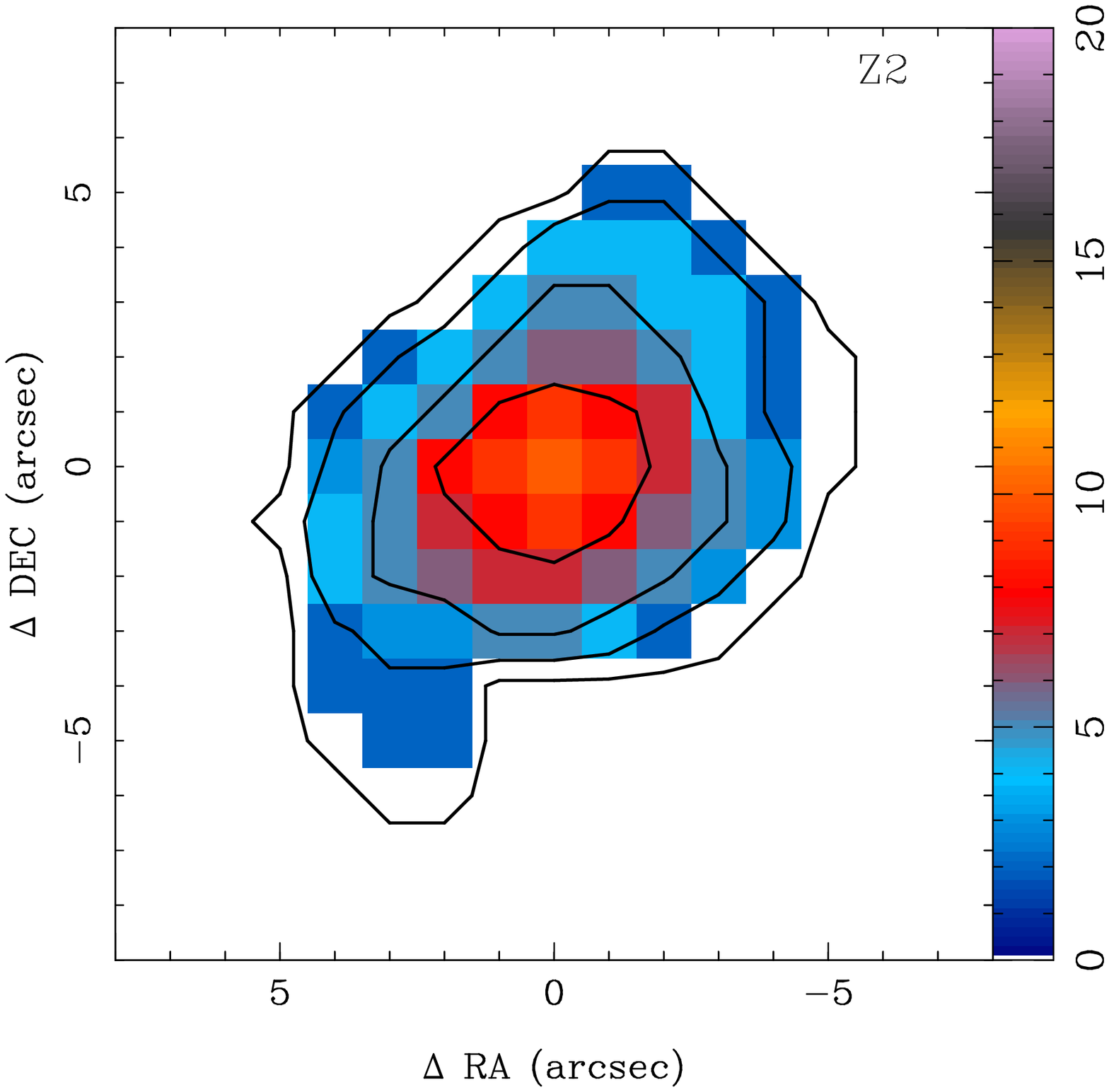}
  \caption{Isophotal fit performed to $V$-band images extracted from the NGC
    628 datacubes. Flux scale is in arbitrary units. For clarity, only
      some isophotes (black lines) are depicted on the plots. This fitting is
    later used as segmentation maps for the spectral extraction, i.e the
      extraction will be performed adding all the spectra between two
      isophotes separated a given flux step. North is up, East is to the
    left. \textit{Left:} $Z0$, \textit{center:} $Z1$, \textit{right:} $Z2$.}
  \label{fig:isophotal_fit}
\end{figure*}

Fig. \ref{fig:spaxel} shows the comparison between the same analysis as
Fig. \ref{fig:abundances} and an spaxel-spaxel determination above a S/N
threshold. The low abundance inner pixels may alter the gradient
determination. Although a deep analysis of the merging \hh complexes and the
contribution from diffuse ionized gas (DIG) is beyond the scope of this paper,
it is worth noting that they are key points in this kind of studies, and any
interpretation should consider them. Despite the fact that the DIG
contribution to the integrated flux in some lines has been quantified in
large values \citep[e.g., about 30\%--50\% of ionized gas, as traced by H$\alpha$, in
galaxies is found  outside of \hh
regions,][]{thilker:02,2007ApJ...661..801O,2009RvMP...81..969H,2010A&A...517A..28M,2010A&A...522A...7A}, its final
contribution to dust corrected luminosity is much lower \citep[e.g.,
5\%--30\%][]{2013ApJ...762...79C,Sanchez:2013a}. The line ratio of the DIG may differ from
that of the star-forming regions, being in general located in the LINER-like
region of the BTP diagram \citep{sing:13,papa:13},
reflecting ionizing conditions that depend on the morphology of the
galaxies \citep[e.g., post-AGBs for most of the earlier-type
galaxies][]{papa:13}. Therefore, the inclusion of larger fraction of DIG
emission as the resolution is degraded have a clear effect in the line ratios,
although it is the not the only effect to take into account.

When the spatial resolution is good enough to resolve the more luminous \hh
regions, it is still possible to use techniques of crowded field spectroscopy
\citep{2005A&A...437..217F,2009ApJ...704..842B,2013A&A...549A..71K}. 
These techniques are extensions to IFS of the well-known PSF-photometric techniques
(DAOPhot), and they rely on the basis that the considered regions are
basically unresolved at the spatial resolution of the data. Therefore, it is
possible to perform a multi-Gaussian 2D fitting, wavelength by wavelength, to
model each of the individual HII-regions, and a 2D polinomial function to take
into account the continuum. This procedure will effectively decouple the
diffuse from the resolved emission. However, for doing so it is required to
distinguish between both components, and, as we argue along this article, this
is not feasible with the spatial resolution of the Z2 data. We have still not
implemented this procedure in HIIexplorer, but we will have to add that
capability to perform more detailed analysis in the future.

It is clear that the effects in the line ratios and the corresponding
derivation of the abundance gradients are complex, and involve many different
components that are difficult to disentangle. However, the final result may
produce a change in this gradient induced by the degradation of the	
resolution.

With all the considerations made above, the parameters studied for the gaseous
phase show that, contrary to the naive picture, the effect of the resolution
degradation is not as intuitive as it would be if we were considering simple
additive quantities. In fact, some effects detected in abundance gradients or
BPT diagrams at high redshift could be due to spatial information loss.

\subsection{Stellar component}

\begin{figure*}
  \centering
   \includegraphics[width=0.32\linewidth,clip=true,bb=80 0 590 720,angle=-90]{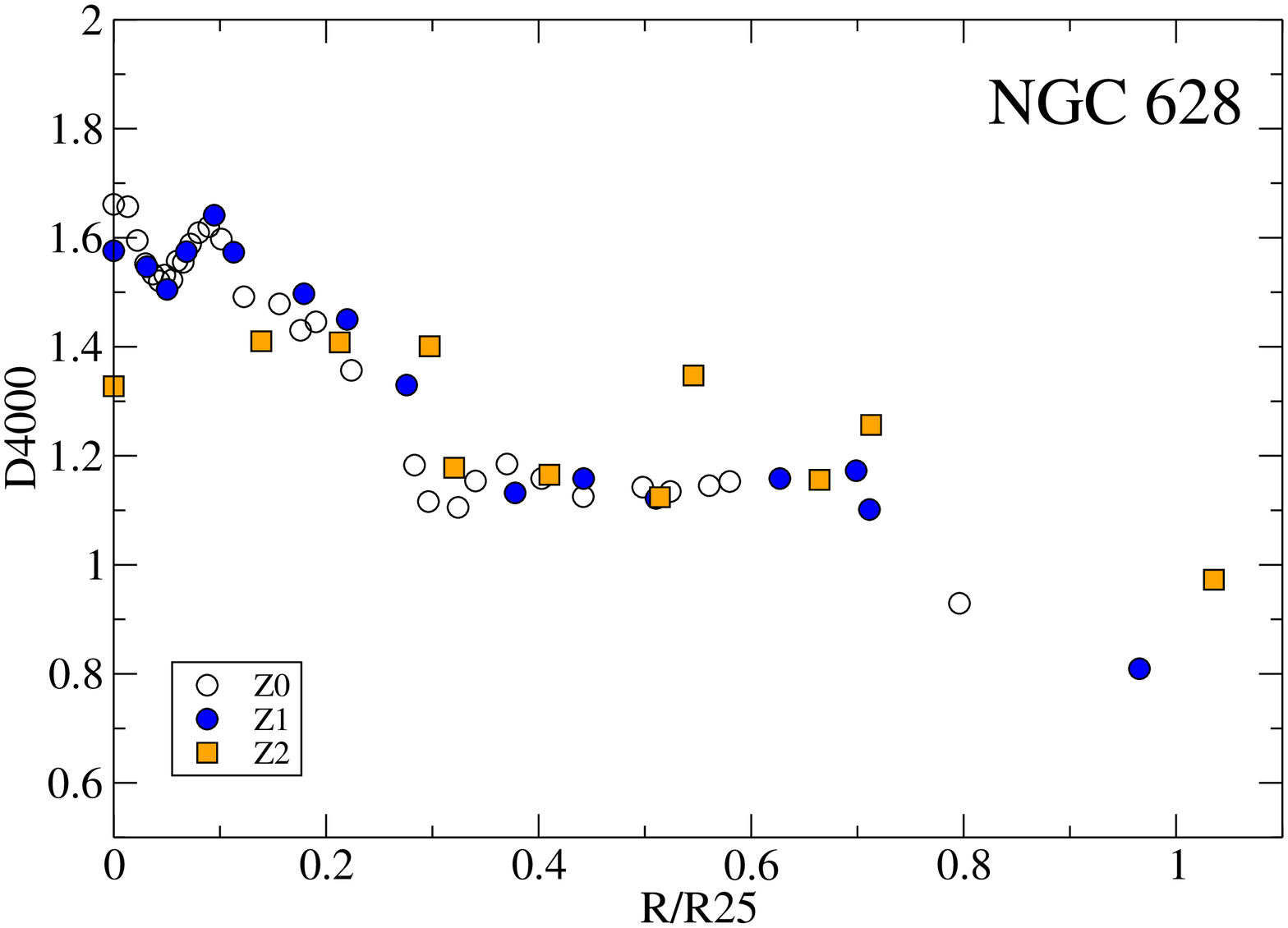} ~
   \includegraphics[width=0.32\linewidth,clip=true,bb=80 0 590 720,angle=-90]{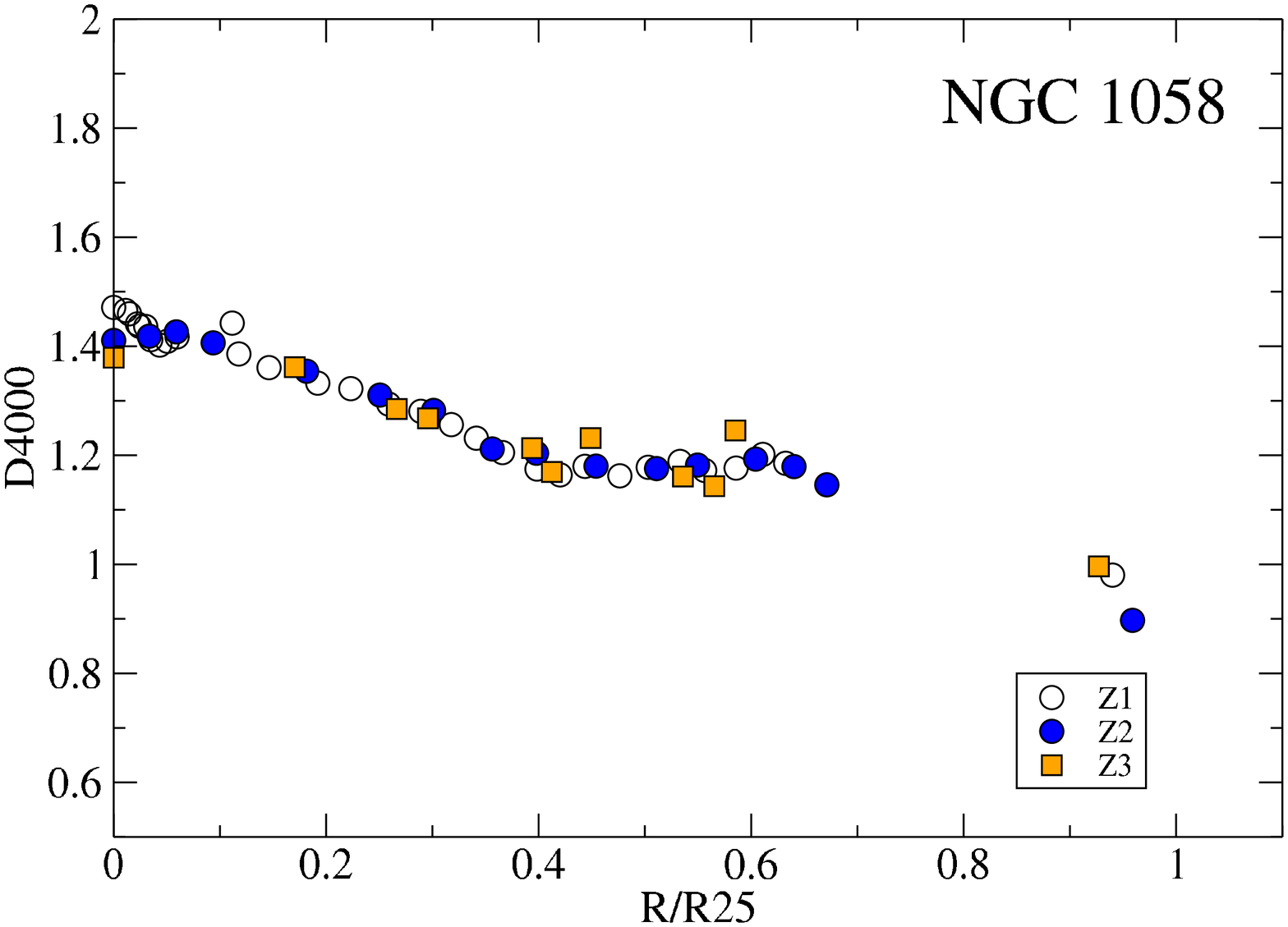} 
   \includegraphics[width=0.32\linewidth,clip=true,bb=80 0 590 720,angle=-90]{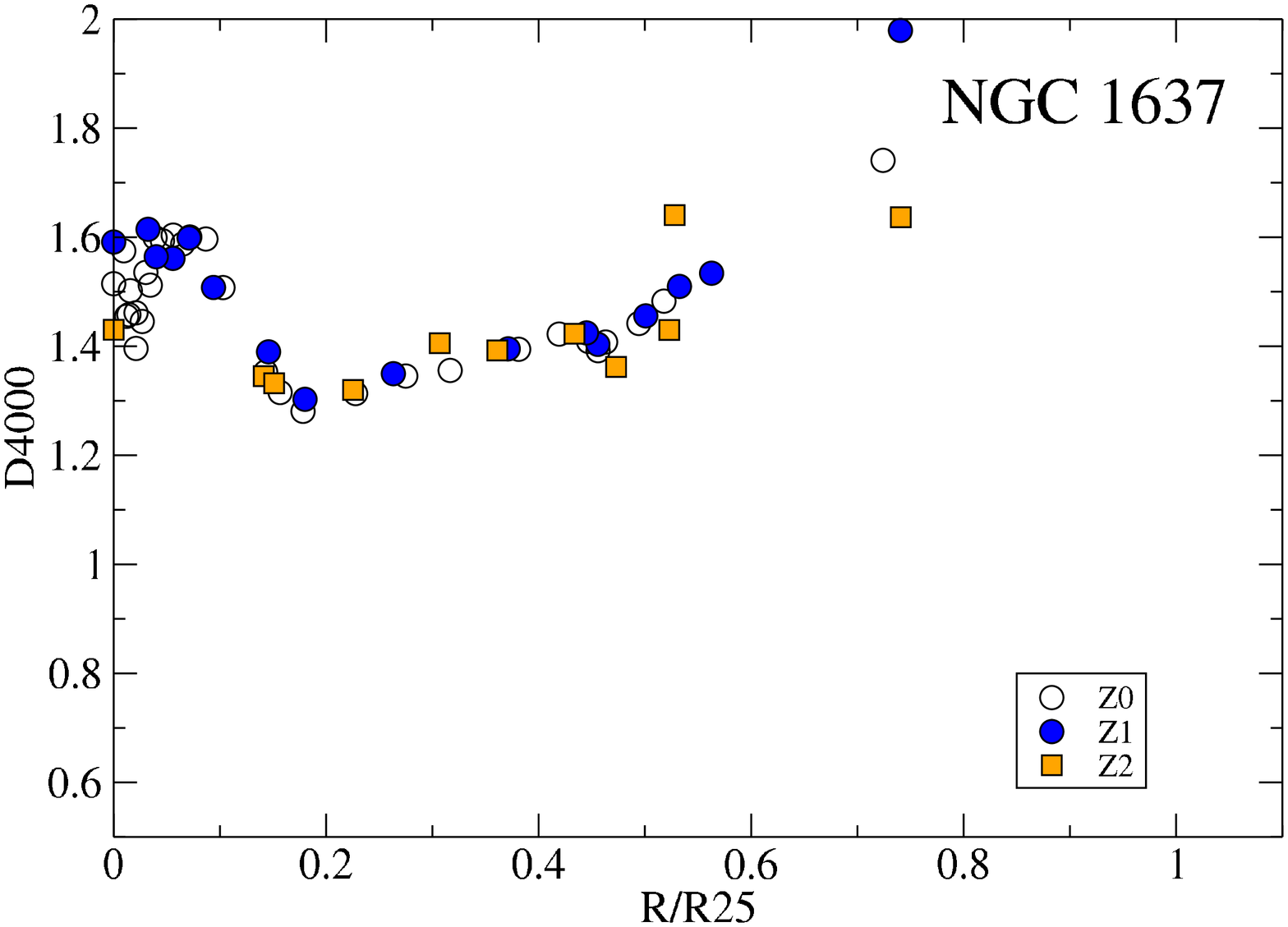} ~
   \includegraphics[width=0.32\linewidth,clip=true,bb=80 0 590 720,angle=-90]{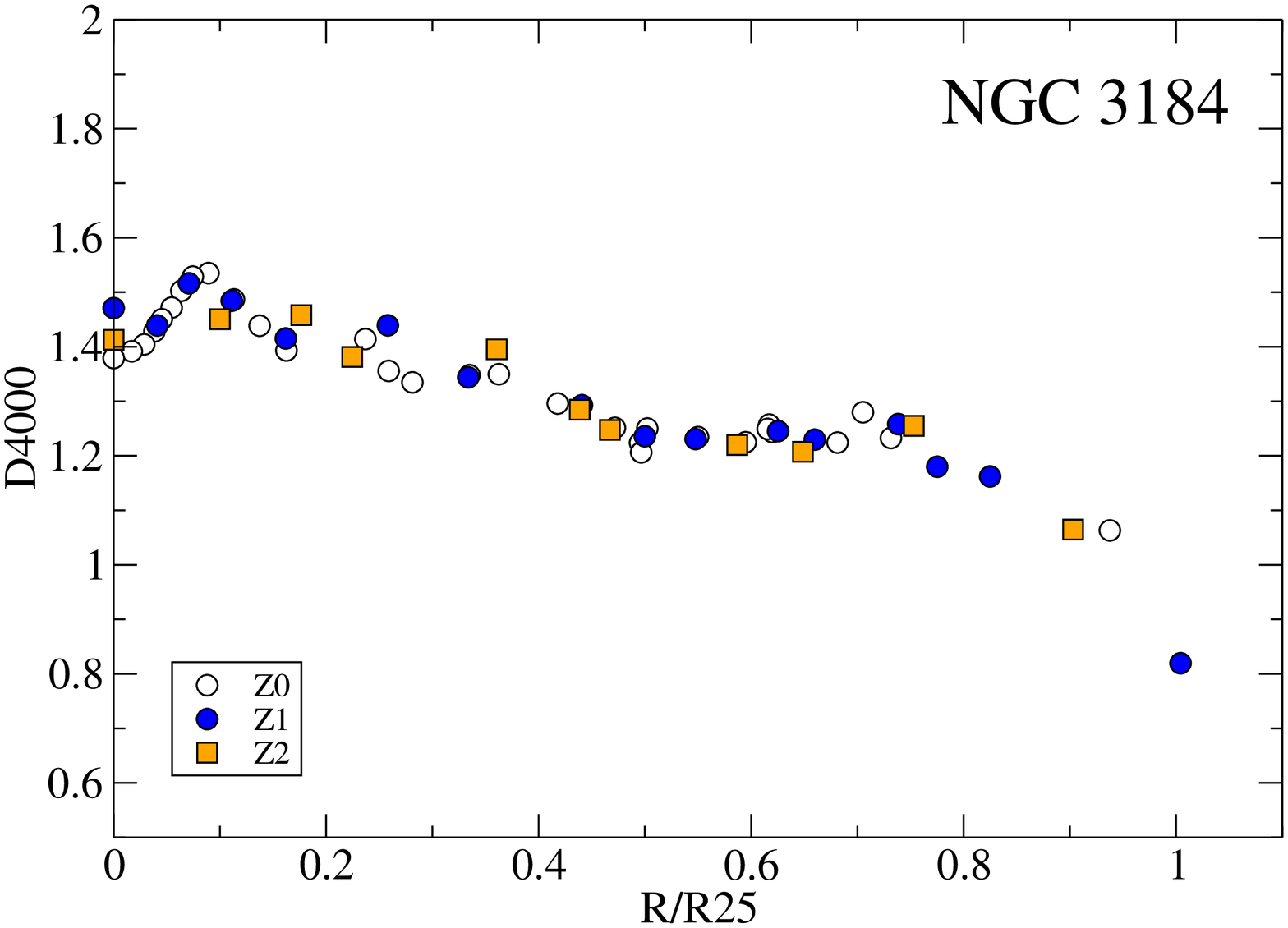} 
   \includegraphics[width=0.32\linewidth,clip=true,bb=80 0 590 720,angle=-90]{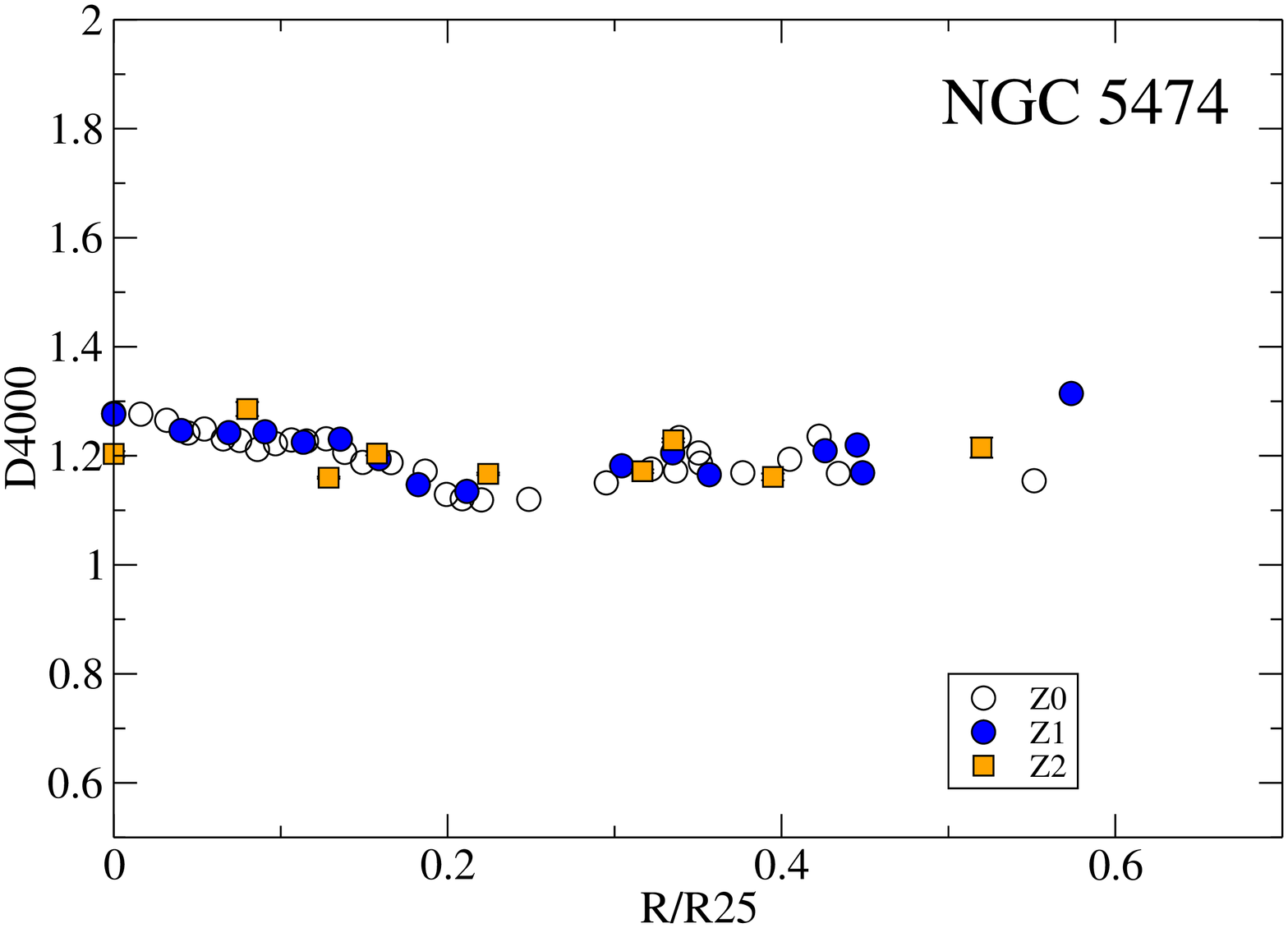} 
  \caption{Radial distribution of the D4000 index defined as the ratio of the
    flux in the red continuum to that in the blue continuum. Black empty
    circles correspond to $Z0$ regime, blue circles to $Z1$ and orange squares
    to $Z2$. The error bars are similar to the symbol sizes.}
  \label{fig:rad_D4000}
\end{figure*}

As we explained in Section \ref{analysis}, we used FIT3D for separating the
gaseous spectra from the underlying stellar population. In the previous
sections we showed the analysis for the gaseous phase, and in what follows we
are going to deal with the stellar component.

Spectral indices are widely used for the characterisation of stellar
population in galaxies \citep{2000AJ....120..165T,2005MNRAS.362...41G}. It is
beyond the scope of this paper to analyse which method is most suitable for
the detail study of stellar population. In the present study we chose the
D$_{4000}$ index to analyse how their derived properties vary in the different
redshift regimes. D$_{4000}$ is a good tracer of stellar age, besides being
model-independent. Errors were determined running a Montecarlo simulation
during the fitting procedure.

The chosen approach was to generate segmentation maps doing an isophotal fitting to the $V$-band images extracted from the datacubes. 
The basic idea of any binning for the analysis of data is to increase
the S/N as much as possible while minimising the degradation of the spatial
information. The most widely adopted binning procedure for the
analysis of IFS data is the one described by \cite{2003MNRAS.342..345C}. This
technique performs a binning of the data intended to 
increase the S/N of the low-surface brightness areas, considering
spatial vicinity. However, it does not take into account the
morphology (relative intensity) between adjacent pixels, which is
understandable since it was mostly adopted for the analysis of
early-type galaxies (SAURON/ATLAS3D data). For the analysis of
galaxies rich in structure (e.g., like the ones included in the
current study), we consider that a more representative binning should
take into account the relative intensity of adjacent pixels.

A similar effect is shown in the standard procedures to derive
isophotal distribution of properties (e.g., surface-brightness
profiles as the one described in Sect. \ref{sec:morpho}). They well describe the
azimuthal distribution of properties for early type galaxies, but they
do not work as well for late type galaxies.

Therefore, we propose a more simple method to derive isophotal
information similar to the one introduced by \cite{2002A&A...393..461P}.
Instead of assuming a certain shape for the isophote at
a certain radii, we just slice the intensity maps on bins of equal
intensity, within a certain percentage. Starting from the peak
emission of a galaxy, and selecting the range of adjacent pixels to be
aggregated within a selected percentage of flux with respect to the
peak emission, and iterating until reaching a certain surface
brightness, it is possible to bin the data in isophotal areas without
any assumption on their shape. Once segregated the galaxy in isophotal
areas, it is possible to extract the corresponding co-added spectra,
and analyse them. For a given isophotal bin, the distance considered is
  the distance from the galaxy center to the average distance to each
  isophote. This technique provides azimuthal distributions of
properties that describe better the actual shape of the galaxies.

If in addition to this simple criteria, a maximum distance between
adjacent pixels is considered, then we will end-up with a 2D binning
technique that preserves the actual shape of the galaxies. The details
of the technique and its comparison with other methods will be
described elsewhere.

On Fig. \ref{fig:isophotal_fit} we can see an example for this segmentation
method for NGC 628. As described in detail in \cite{2007MNRAS.376..125S},
FIT3D measures the stellar indices from the extracted stellar spectra,
normalised to the standard Lick/IDS resolution.

Fig. \ref{fig:rad_D4000} displays the radial distribution of the D4000 index for all of the sample. Being sensitive to stellar age, this index is
higher in the presence of an older stellar population.

For all the galaxies, the $Z0$ radial behaviour of the D4000 index (open
circles) is totally reproduced  by the $Z1$ regime (blue circles). Only the
innermost regions, when a sharp structure is present, it is slightly smeared
out in the $Z1$ case. The higher redshift regime (orange squares)  follows the
radial trend when the distribution is smooth, as is the case of NGC 1058. $Z2$
is unable to reproduce the innermost structures presented in NGC 628, NGC 3184
and NGC 1637.

All these indices are luminosity weighted. Summing two spaxels means performing
  a luminosity-weighted average of their D4000 break strength, so the observed
  behavior is more intuitive than the one for the more complex emission line
  ratios. Unlike the gaseous phase where we have the information in clumps,
the stellar component contains the information in a smooth, less clumpy
way. In the cases where the radial D4000 distribution is flatter, the
  spatial resolution loss has a small effect at higher redshifts. This
behavior allows the higher redshift regime to follow the radial trend without
much scatter, although loosing any fine detail present in the distribution.

We conducted similar test with other stellar indices as H$\delta$ index, i.e.,
the equivalent width of the Balmer line H$\delta$, and the [MgFe] index (
$\textnormal{[MgFe]} = \sqrt{ \textnormal{Mg}b\, (0.72\,
  \textnormal{Fe}_{5270} + 0.28\, \textnormal{Fe}_{5335})}$). Due to our low
spectral resolution we are not able to confirm if complex indices based on
absorption lines (as [MgFe]) are traceable to higher redshift, but we can
conclude that simple ones like H$\delta$ or D4000 are.

\section{Summary and conclusions}\label{conclusions}

In this paper we have studied how the information loss due to spatial
resolution degradation would affect IFS surveys at different redshifts.  For
this purpose we used a sample of five PINGS galaxies  ($Z0$) and simulated two
redshift regimes without taking into account surface brightness dimming
  or increase of noise. One associated with the ongoing CALIFA survey ($Z1$)
and the other with an hypothetical higher redshift survey ($Z2$). We then
performed the same analysis to the $Z0$ galaxies and their simulated
versions. We studied the behavior of the radial abundance (through the O3N2
method), BPT diagrams, and one spectral index D4000, in addition to the
H$\alpha$ emission and the morphology. Our main conclusions regarding each
measured quantity, can be summarized as follow:

\begin{itemize}

\item {\it Morphology:}  $Z1$ is able to reproduce all morphological
  signatures visible at lower redshift. Despite the detail loss (implying that
  several individual regions or hot spots are not detected as such), spiral
  arms, rings and bulge extensions can be traced perfectly. For the $Z2$
  regime the identification of spiral signatures is complicated, and only
  possible if the structure is a strong morphological feature at global scale
  in the galaxy. Several of the higher redshift examples shown in this
    paper present disturbed morphology that prevents from doing a reliable
    morphological classification.
 
\item {\it \hh regions detection:} the implemented method of \hh region detection with {\sc HIIexplorer}, showed that $Z1$ is detecting nearly $1/3$ of the number of original $Z0$ regions, and $Z2$ $\sim1/5$ of the number of regions detected at the $Z1$ redshift (hence, $\sim1/15$ of $Z0$). This illustrates what is the underestimation of the number of \hh regions due the loss of spatial resolution. 

\item {\it Diagnostic diagrams:} the BPT diagrams showed that the effect of spatial resolution degradation on this diagrams is to collapse the measured values into the denser regions of $Z0$ plots. $Z1$ is able to reproduce with acceptable accuracy the shape of the $Z0$ BPT. For the $Z2$ situation, only the most populated regions of the lower redshift regime are mapped. Despite the displacement observed, the main ionization mechanism of the observed galaxy at the putative higher redshift is mostly due to thermal ionization from hot massive young stars.

\item {\it Radial abundance:} The O3N2 method (commonly used at high redshift to derive gas metallicity) was applied to our sample. It showed that the flat inner part of the abundance distribution presented in the $Z0$ galaxies is observable in the $Z1$ case and, although with smoothed values, in $Z2$. $Z1$ and $Z2$ are able to reproduce the maximum value displayed around $(0.2-0.4)R_{25}$ in the $Z0$ galaxies. The gradient slopes in $Z0$ are slightly smoothed in $Z1$ but still present. In the higher redshift case global trends are traceable. Note that the degradation and contamination of \hh complexes at higher redshift may induce spurious radial trends.

\item {\it Spectral indices:}  For the analysed index D4000, global tendencies are correctly traced on both simulated cases, and $Z1$ is also capable, at some level, of reproducing fine structure of the $Z0$ distributions.

\end{itemize}

As a global conclusion, we showed that the information loss will depend on the level of detail contained in the analysed feature. In this sense, if the studied galaxy has a smooth profile, early-type, with smooth behaviour in their properties, that behaviour will still be visible at higher redshift. But any sharp structure will be lost. Our analysis allows us to conclude that CALIFA will be able to analyse to an acceptable scale, and with a good level of detail, all desired magnitudes, in its aim of characterising the Local Universe. For the hypothetical higher redshift survey, the perspectives are difficult but promising, since global trends, averaged values and, in some cases, local structures, are acceptably mapped and correctly interpreted in the considered framework.  

It is worth noting that the important figure of merit is the ratio between the spaxel size and the typical scale-length at a certain redshift. As examples of hypothetical $Z2$ surveys, we can consider SAMI  (1.6\arcsec/fibre) or MaNGA (3\arcsec/fibre). In this regard, the conclusions obtained for the $Z2$ simulation are also applicable to other redshift ranges depending on the instrument. E.g., if we are studying galaxies with $\sim0.25$ kpc/spaxel, we have shown that most of the common observables are perfectly traced, while this is not the case with a scale of $1\sim$kpc/spaxel. 

As real examples, we can consider the Gemini Multi-Object Spectrographs (GMOS) \citep{2002PASP..114..892A}. With 25$\times$35 spaxels and 0.2\arcsec/spaxel scale our results showed that at $z\sim 0.05$ is still possible to derive radial gradients, whenever the right depth and spatial coverage are granted. The same conclusion can be applied to VIMOS (VIsible MultiObject Spectrograph, \citealp{2003SPIE.4841.1670L}) in the Medium Resolution mode (a $13\arcsec\times13\arcsec$ FoV with 0.33\arcsec/spaxel). But, as we have already shown, at $z\sim0.4$, for example, none of these instruments is capable of measuring properly any of the observables studied. Another positive example could be SINFONI (Spectrograph for INtegral Field Observations in the Near Infrared, \citealp {2003SPIE.4841.1548E} ) in its $8\arcsec\times8\arcsec$ FoV with a $0.125\arcsec\times0.125\arcsec$ spaxel, or even with the $3\arcsec\times3\arcsec$ mode, provided that the Adaptive Optics Facility is used, $z\sim0.1$ can be reached with considerable success. In the usual redshift ranges that SINFONI has been used ($z\sim1$), we have shown that no convincing metallicity radial gradients can be obtained with the data.

\appendix
\onecolumn

\newpage

\section{Radial abundance determination for the remaining objects}
\label{sec:remaining}

\begin{figure*}[h]
  \centering
   \includegraphics[width=0.34\linewidth,clip=true,bb=50 0 590 720,angle=-90]{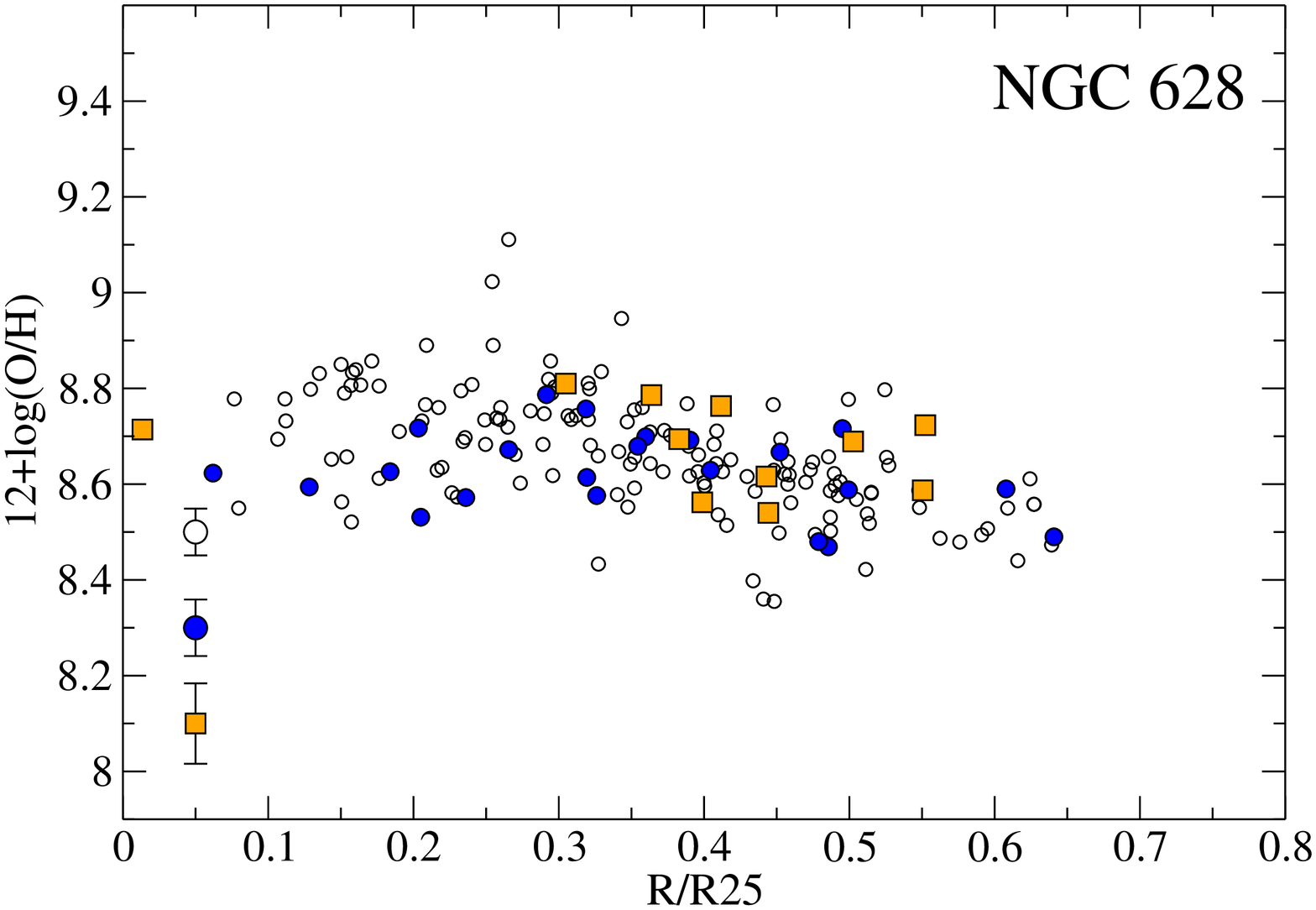}
  \includegraphics[width=0.34\linewidth,clip=true,bb=50 0 590 720,angle=-90]{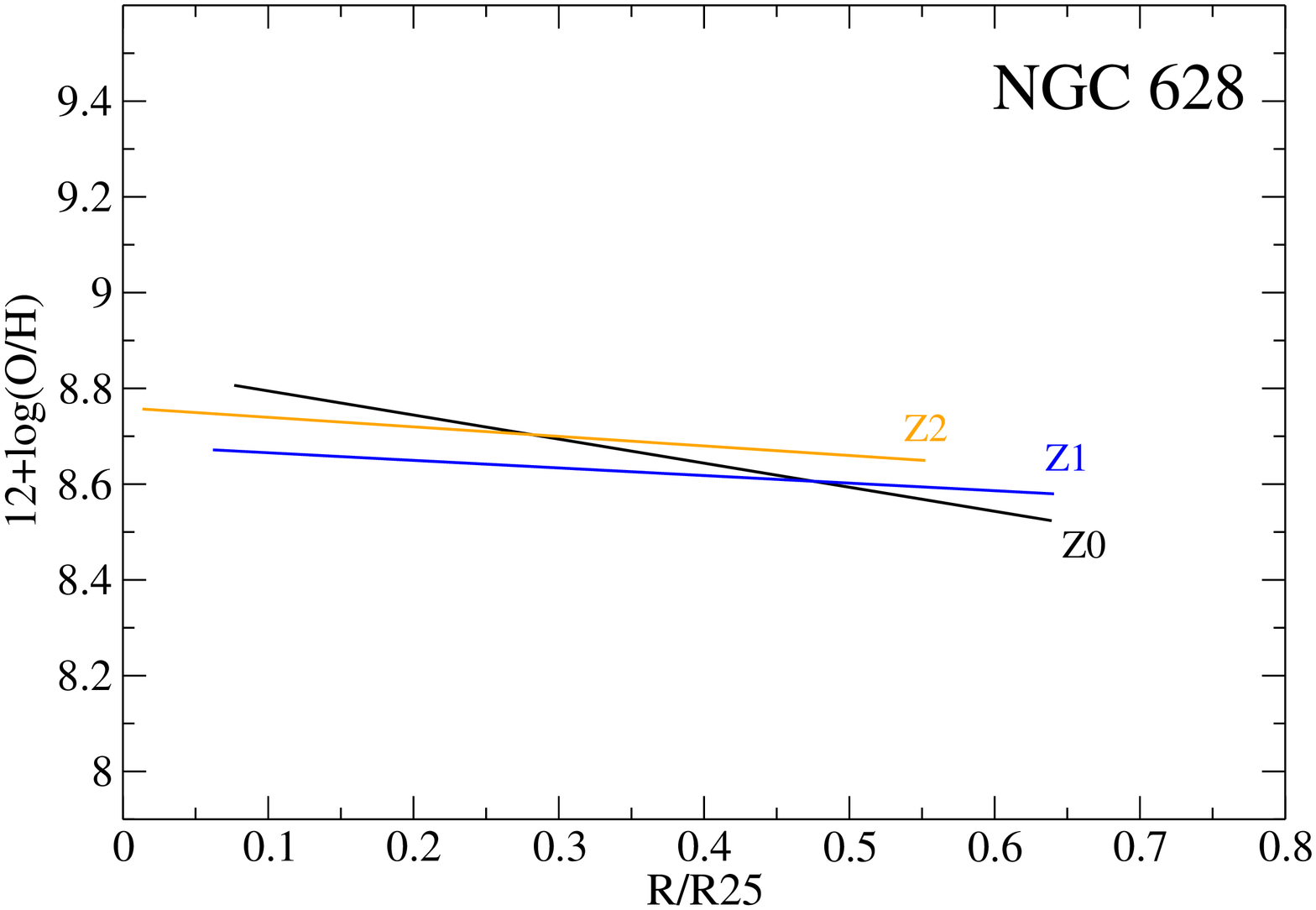}
  \includegraphics[width=0.34\linewidth,clip=true,bb=50 0 590 720,angle=-90]{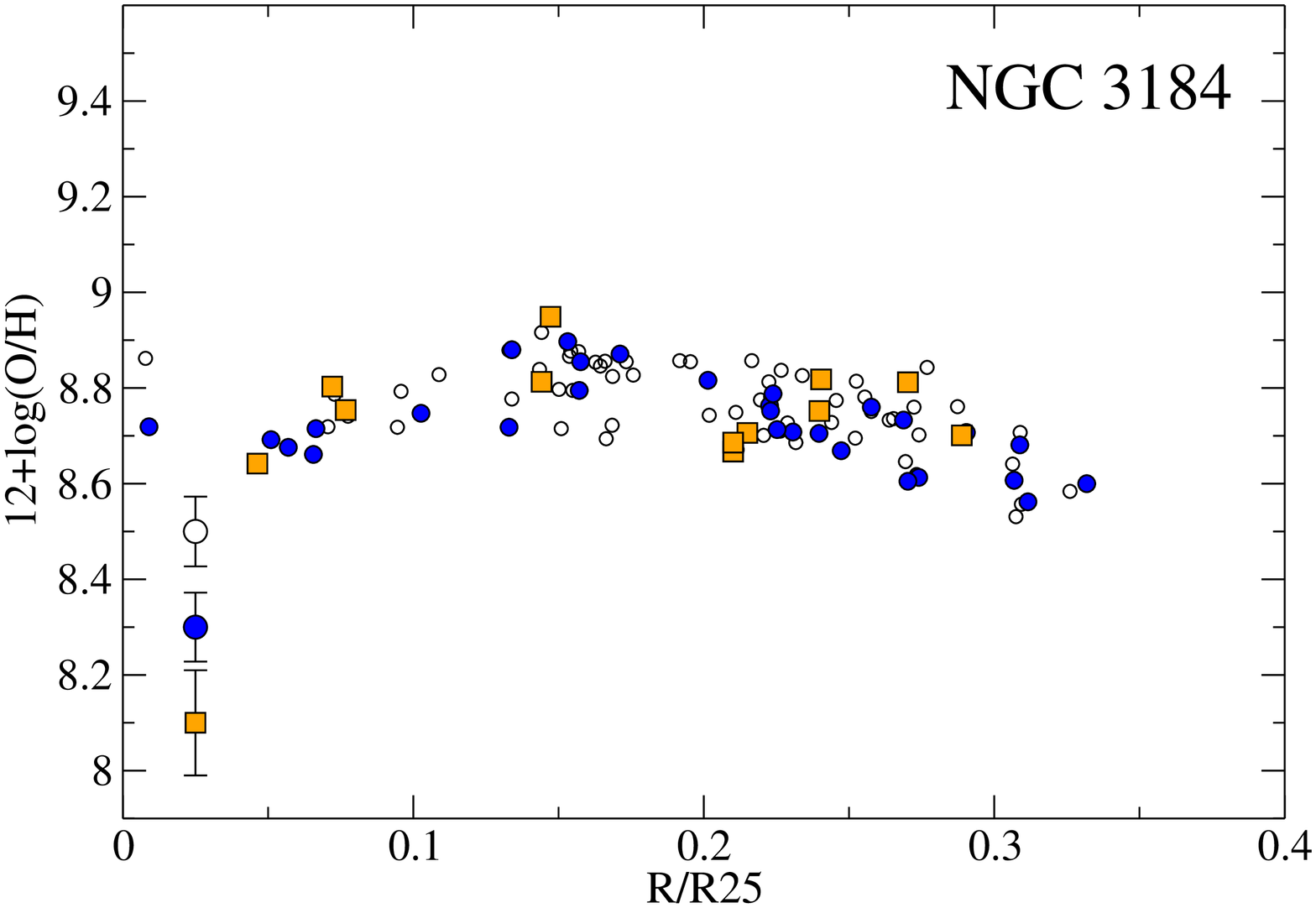}
  \includegraphics[width=0.34\linewidth,clip=true,bb=50 0 590 720,angle=-90]{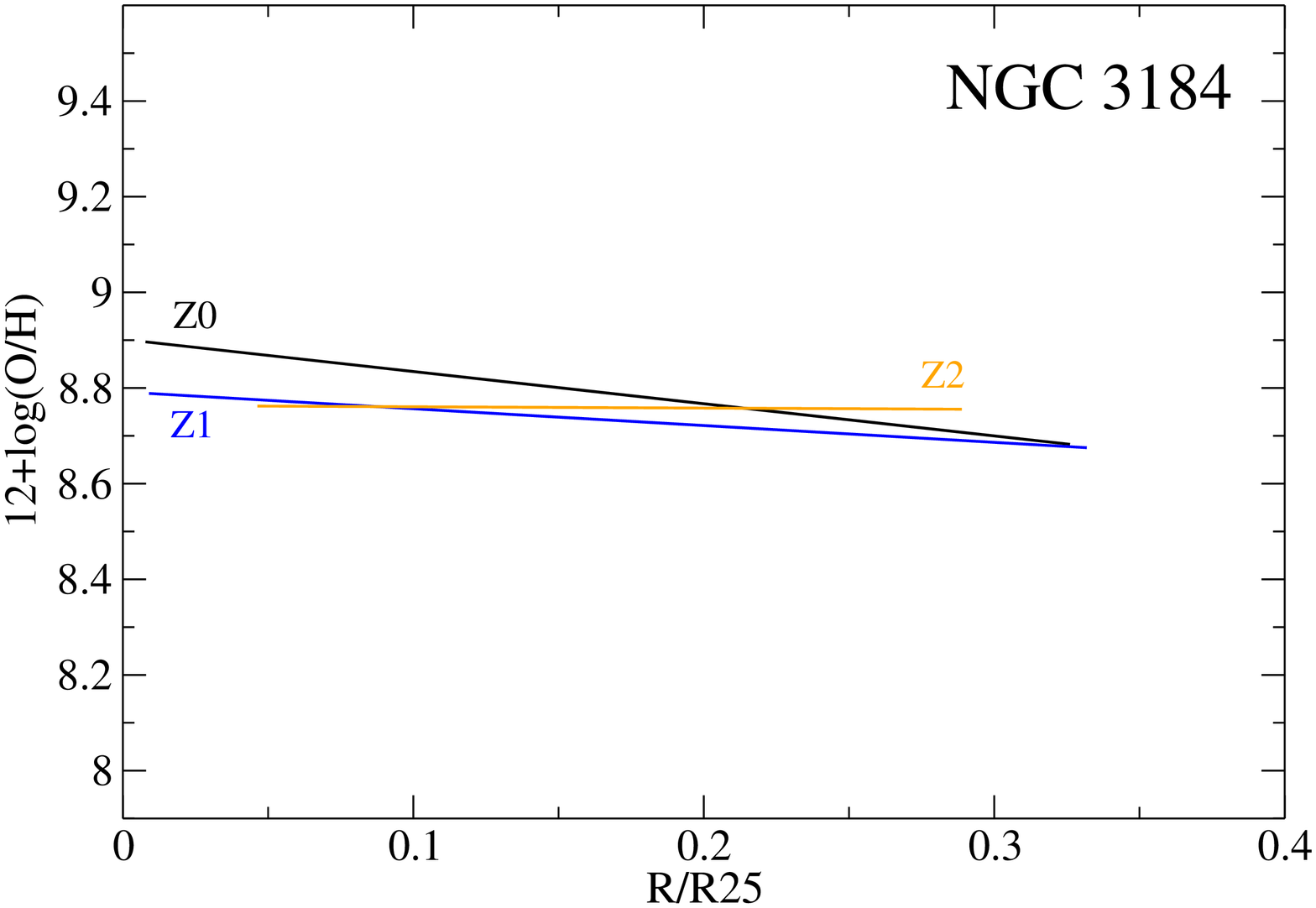}
  \includegraphics[width=0.34\linewidth,clip=true,bb=50 0 590 720,angle=-90]{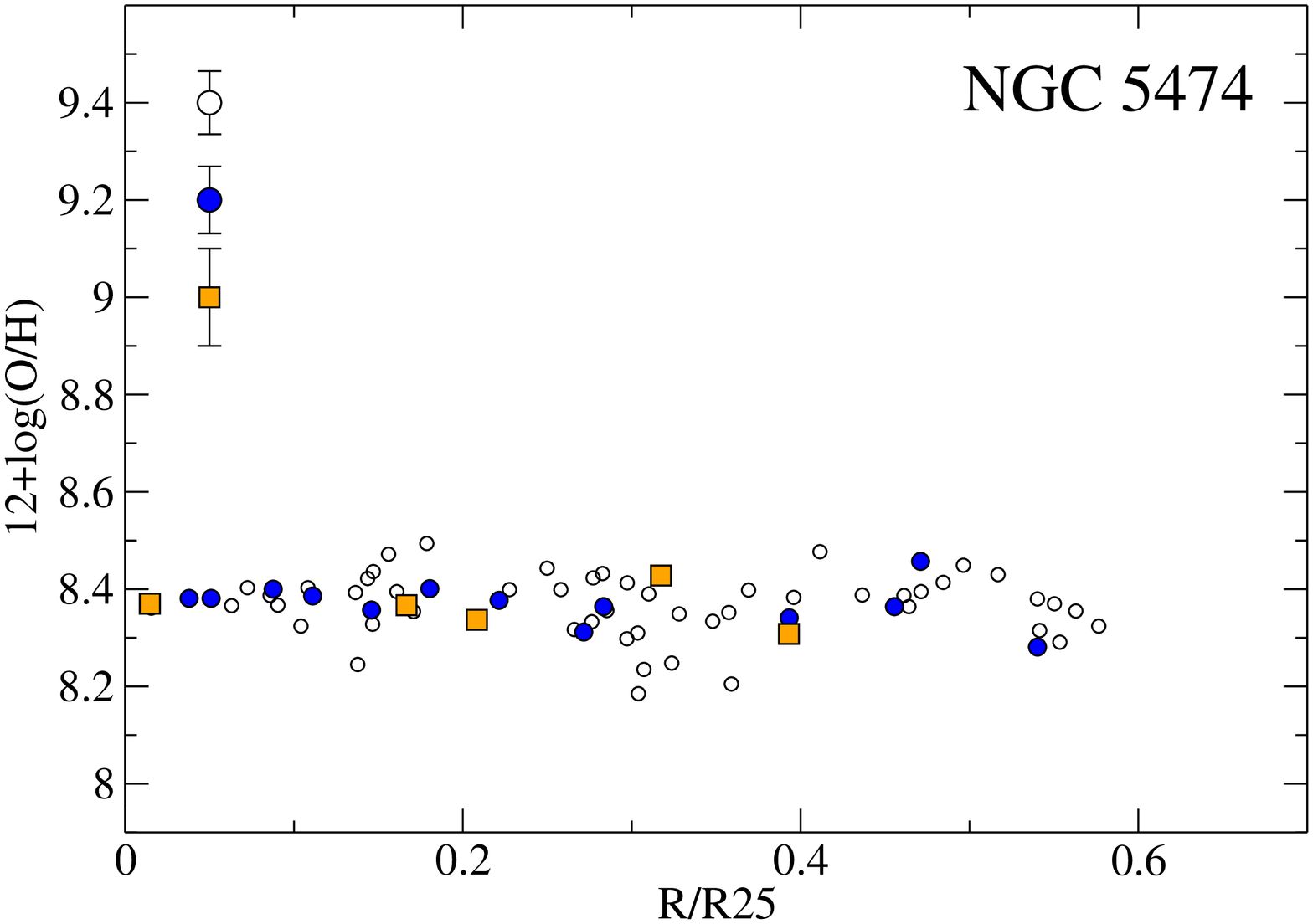}
  \includegraphics[width=0.34\linewidth,clip=true,bb=50 0 590 720,angle=-90]{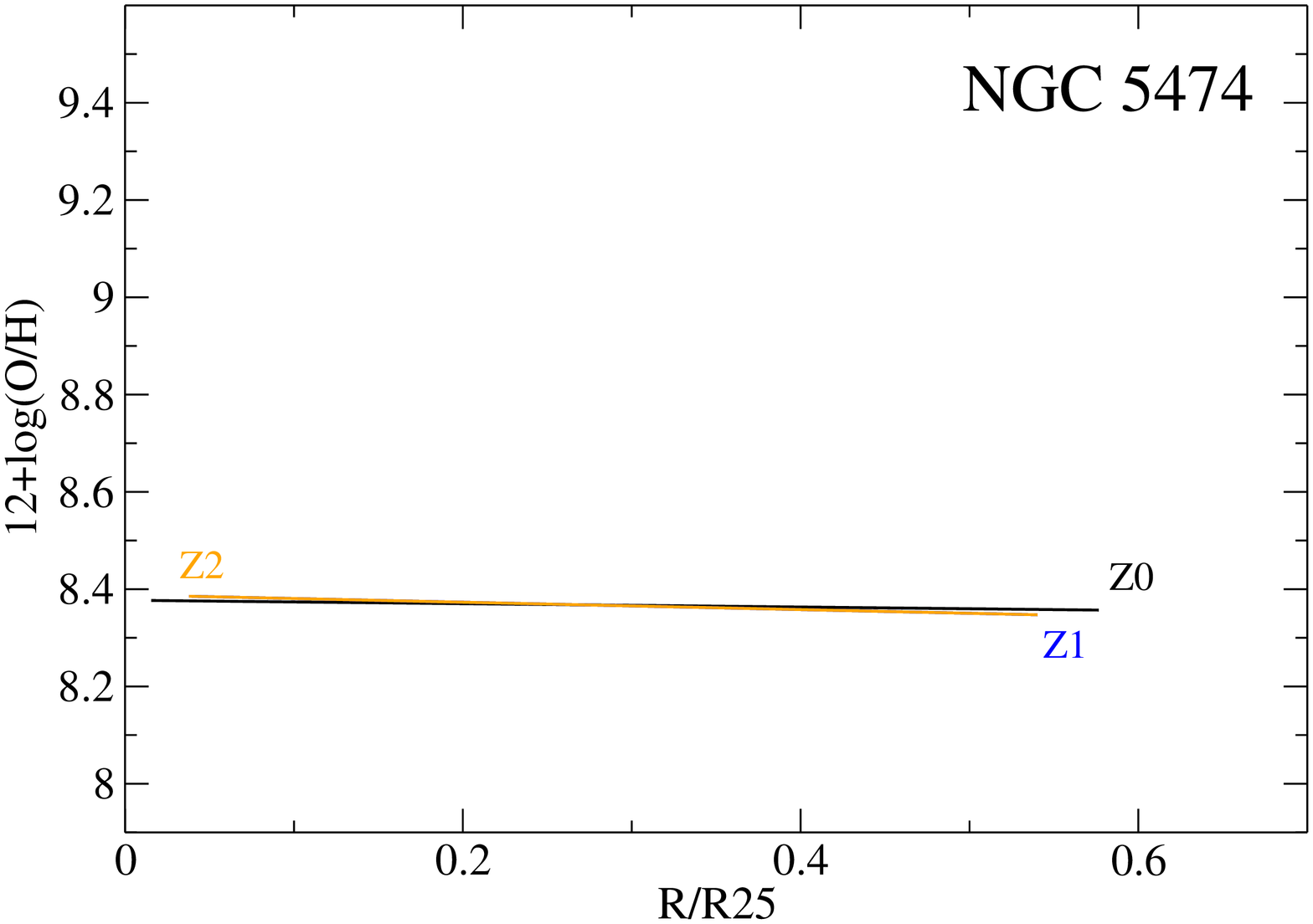}

  \caption{Radial abundance gradients derived using the relation O3N2 from
    \cite{2004MNRAS.348L..59P} for the three remaining galaxies. On each left
    graph the three redshift regimes are over-plotted, the empty circles
    correspond to $Z0$, the blue circles to $Z1$ and the orange squares to
    $Z2$. The three points on the down left corner (upper left for NGC 5474)
    show the average error bars. On the right we can see a linear regression fitting to each regime.}
  \label{fig:abundremaining}
\end{figure*}

In Sec. \ref{sec:abund}, for presentation purposes, we showed only 2
objects. In Fig. \ref{fig:abundremaining} the radial abundance gradients for
the remaining galaxies are shown.

\newpage

\section{Annular Binning}

\begin{figure*}[ht]
  \centering
  \includegraphics[width=0.305\linewidth,clip=true,bb=50 0 590 720,angle=-90]{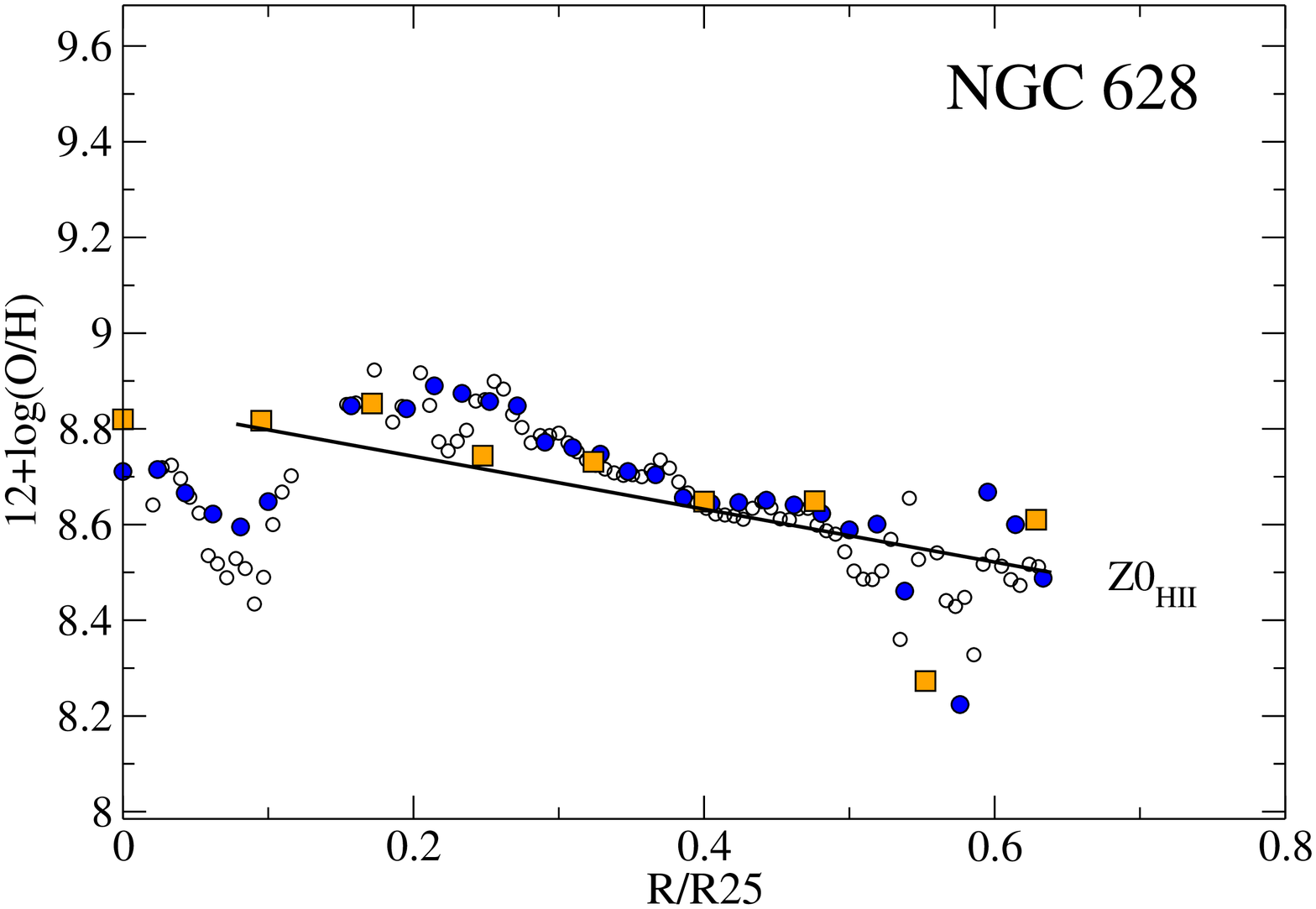}
  \includegraphics[width=0.305\linewidth,clip=true,bb=50 0 590 720,angle=-90]{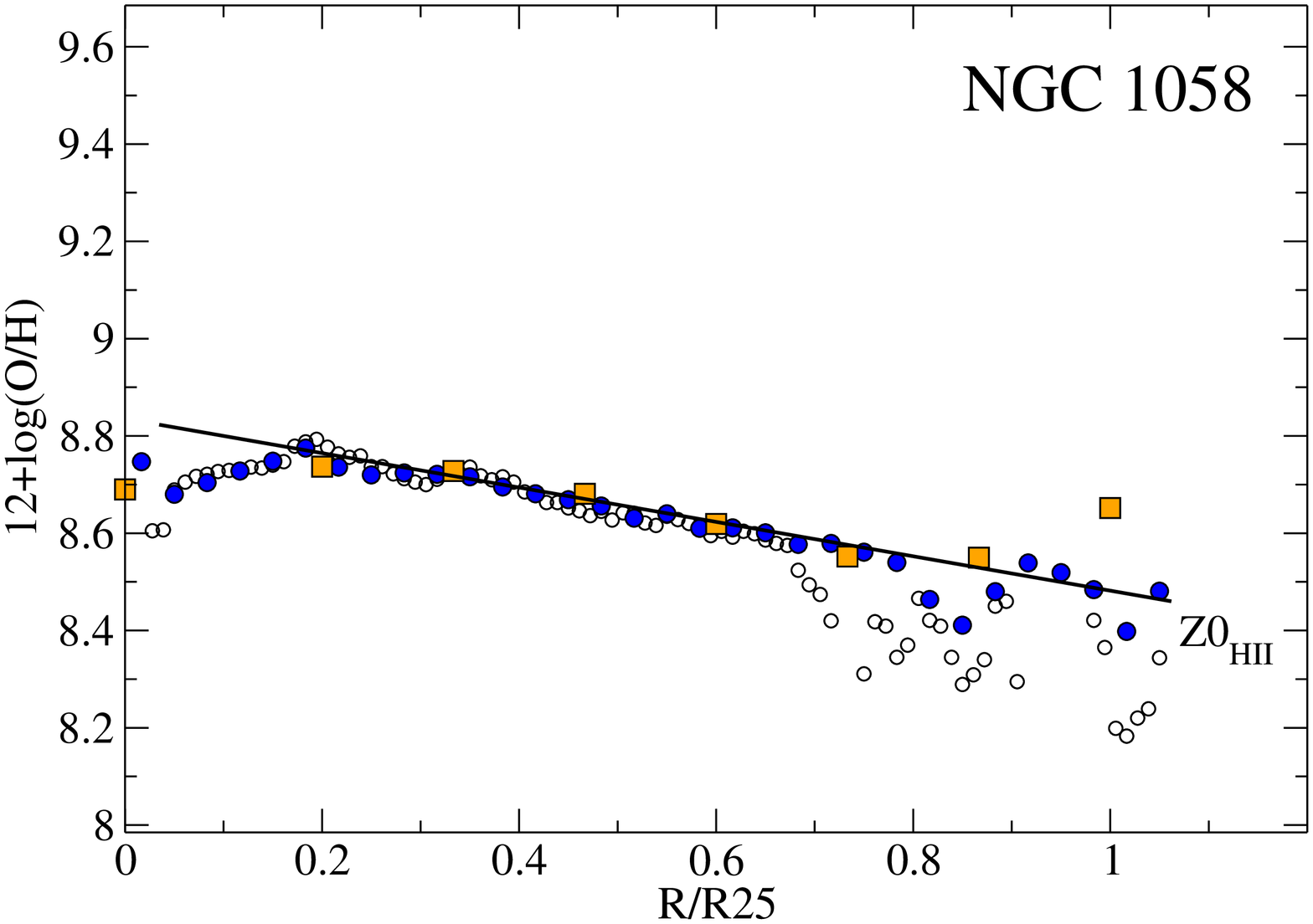}
  \includegraphics[width=0.305\linewidth,clip=true,bb=50 0 590 720,angle=-90]{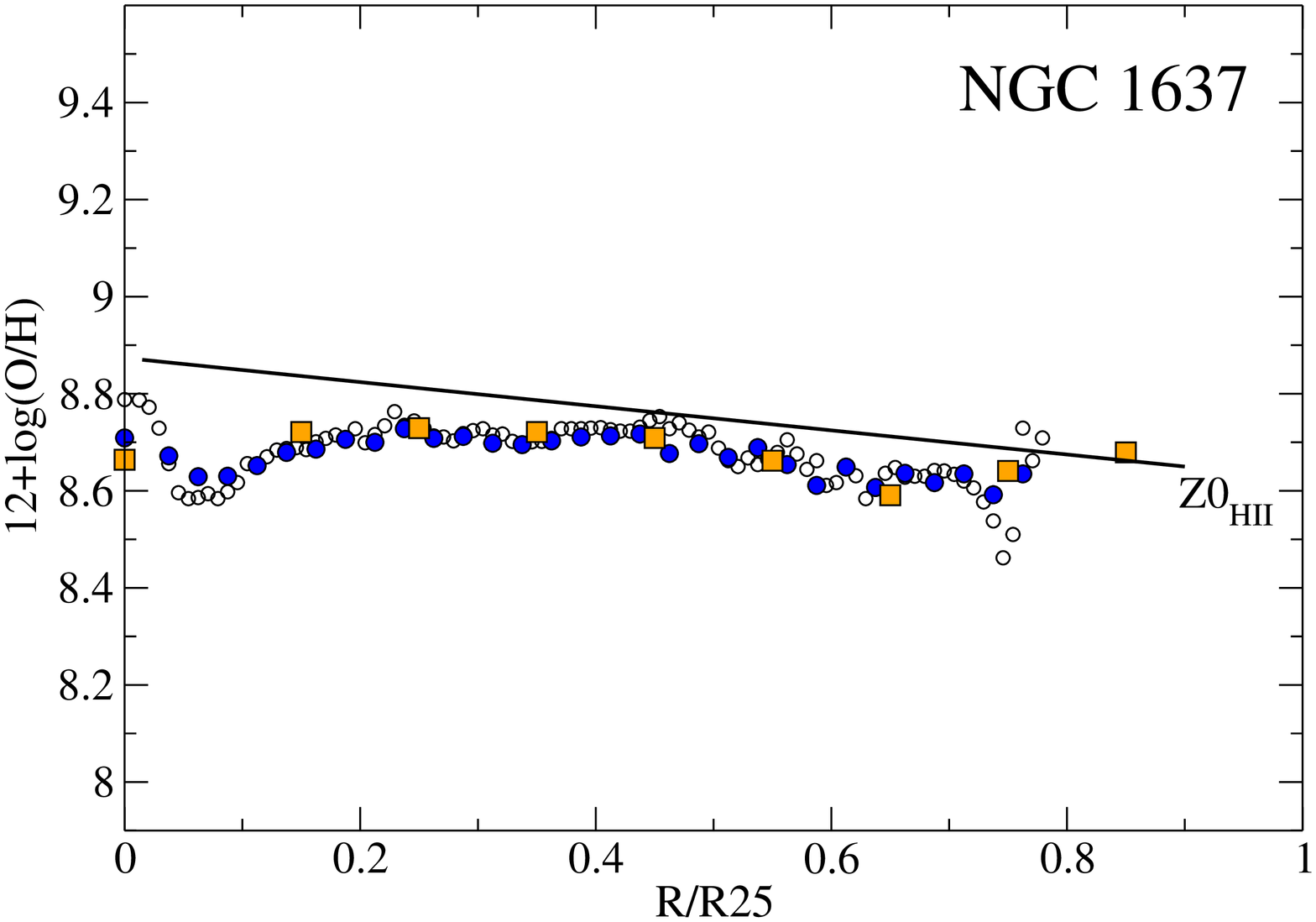}
  \includegraphics[width=0.305\linewidth,clip=true,bb=50 0 590 720,angle=-90]{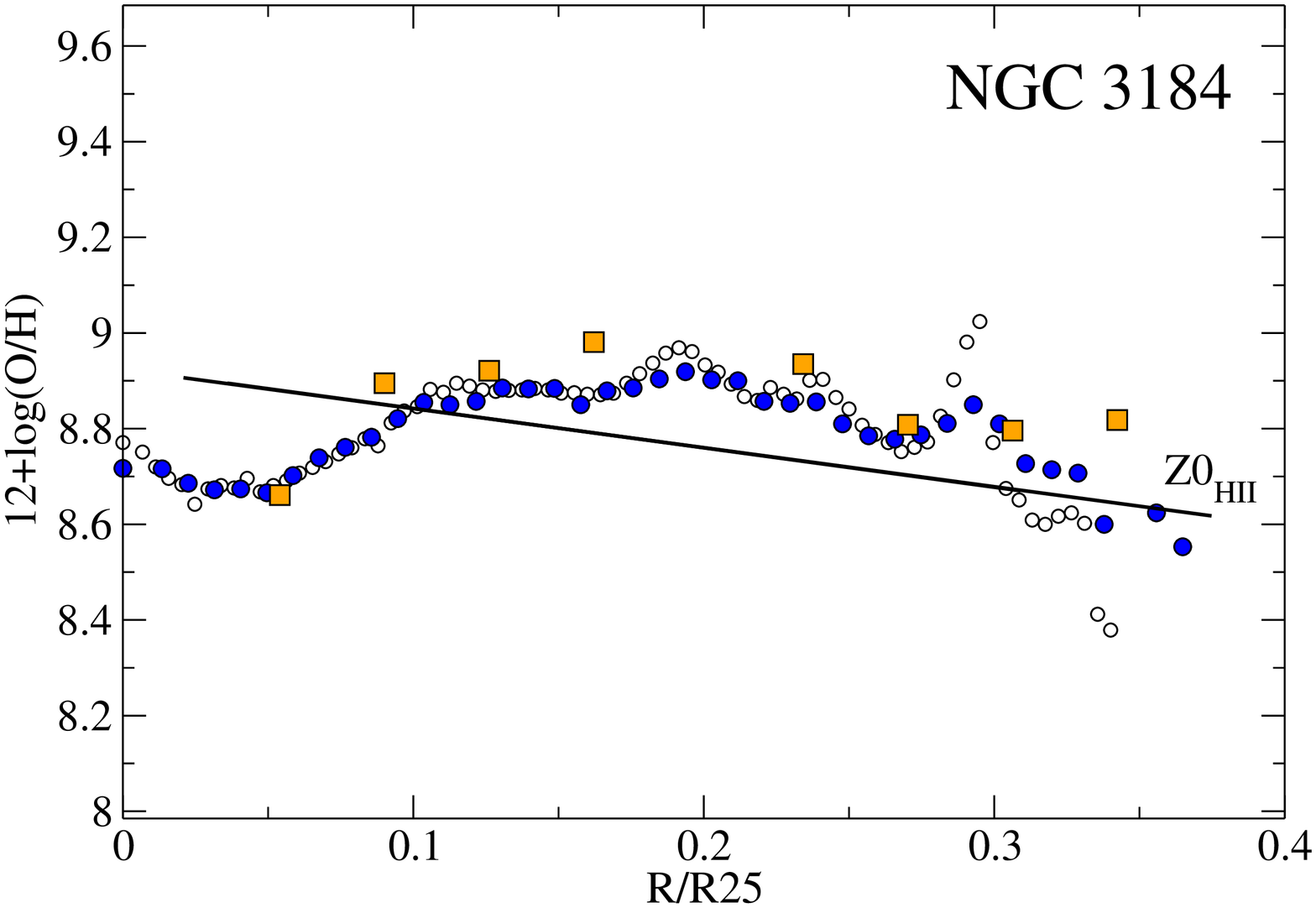}
  \includegraphics[width=0.305\linewidth,clip=true,bb=50 0 590 720,angle=-90]{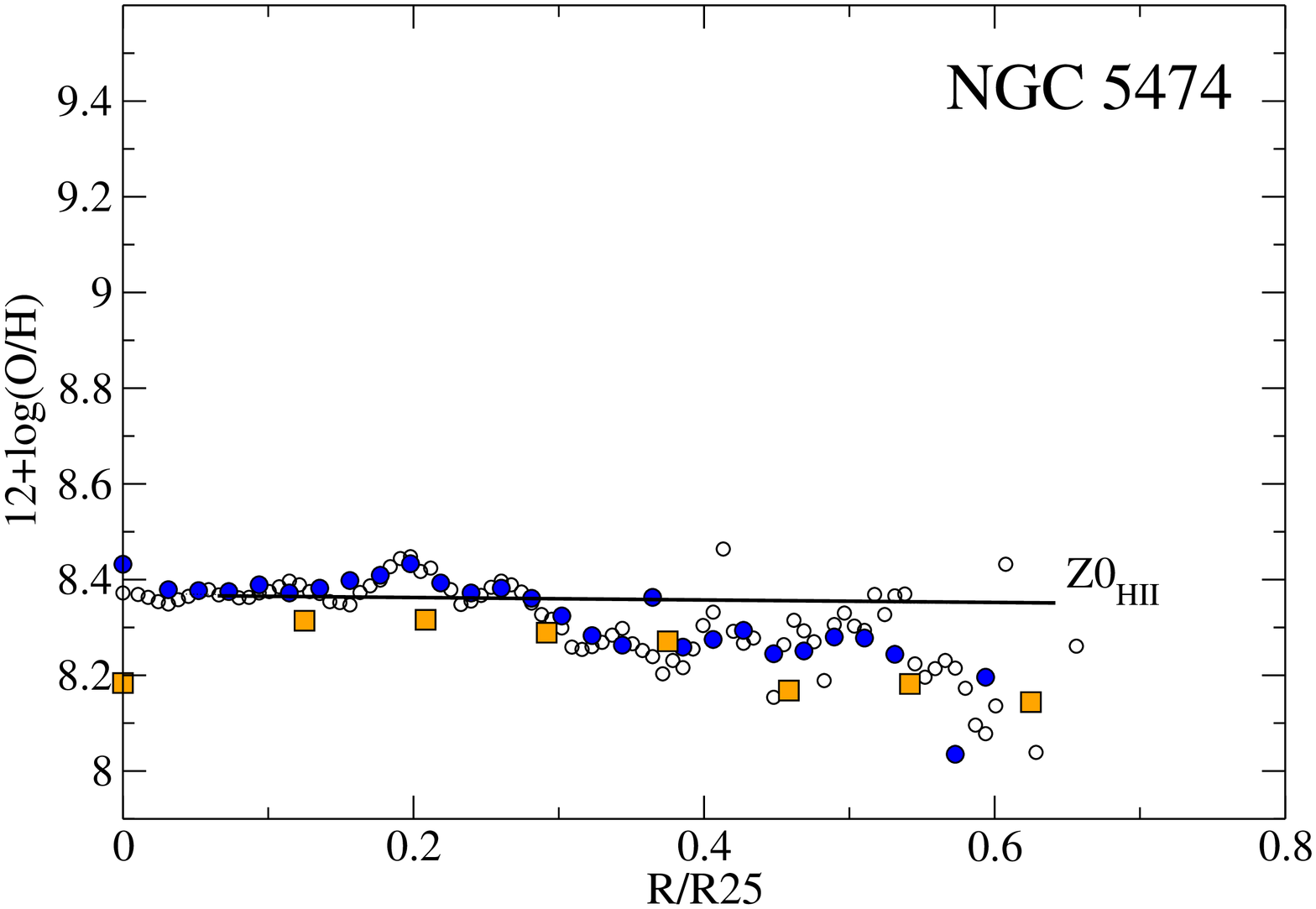}
  \caption{Radial abundance gradients for all the sample, using the
      annular binning scheme for the spectral extraction. On each redshift
      regime annuli are one pixel wide (on its corresponding scale). On each
      graph the three redshift regimes are over-plotted, the empty circles
      correspond to $Z0$, the blue circles to $Z1$ and the orange squares to
      $Z2$ as usual. The black line indicates the $Z0$ abundance gradient determined in Sec. \ref{sec:abund}.}
  \label{fig:annular}
\end{figure*}

One binning scheme widely used in different surveys and metallicity studies
is the annular binning extraction, method which is often used to derive
metallicity gradients, specially in low S/N high-z data 
\citep[e.g.][]{2010ApJ...725L.176J,2012A&A...539A..93Q,2012arXiv1209.1395S,2013ApJ...767..106Y}.
Fig. \ref{fig:annular} shows the radial abundance gradients for all the
sample, using this binning method. Annuli are one pixel wide for all the
redshift regimes. The black line overplotted to the graph represents the abundance determination of  Sec. \ref{sec:abund} for the $Z0$ regime using the  {\sc HIIexplorer}. As this method wipe out any azimuthal variation and add
together the \hh region and the diffuse emission (that in general do no have
the same ionizing conditions as the starforming-regions) for a given annulus
on all the three regimes, the result is not surprising. Radial distributions
have low dispersion compared with the ones in Sec. \ref{sec:abund} and small scale variations are smeared out as resolution gets
coarse. 

As can be seen from the plots, in the cases where the DIG is not dominated by LINER-like emission, the radial distribution for the three regimes are more similar (despite the small variations), indicating that for these cases the use of the annular binning extraction scheme would be more suitable at higher redshift. In any case it will depend on the morphology and distribution of the HII regions on each particular galaxy, information that in general it is not available for high-redshift objects.

\twocolumn
\begin{acknowledgements}

We would like to thank the anonymous referee for his/her useful comments that have significantly improved the first submitted version of this paper. Based on observations collected at the Centro Astron\'omico Hispano Alem\'an
(CAHA) at Calar Alto, operated jointly by the Max-Planck Institut f\"ur
Astronomie and the Instituto de Astrof\'isica de Andaluc\'ia(CSIC).

\noindent D.~M. and A.~M.-I. are supported by the Spanish Research Council within the
program JAE-Doc, Junta para la Ampliaci\'on de Estudios, co-funded by the
FSE. F.F.R.O. acknowledges the Mexican National Council for Science and
Technology (CONACYT) for financial support under the programme Estancias
Posdoctorales y Sab\'aticas al Extranjero para la Consolidaci\'on de Grupos de
Investigaci\'on, 2010-2012. I.M. acknowledges financial support from the
Spanish MINECO grant AYA 2010-15169, and from Junta de Andaluc\'ia TIC114 and
Proyecto de Excelencia P08-TIC-03531. J.~F-B. from the Ram\'on y Cajal
Program, grants AYA2010-21322-C03-02 and AIB-2010-DE-00227 from the Spanish
Ministry of Economy and Competitiveness (MINECO), as well as from the FP7
Marie Curie Actions of the European Commission, via the Initial Training
Network DAGAL under REA grant agreement n¡ 289313. CJW acknowledges support through the Marie Curie Career Integration Grant 303912. EMQ and J.~F-B. acknowledge
support from the Spanish Programa Nacional de Astronom\'ia y Astrof\'isica
under grant AYA2010-21322-C03-02. This work has been partially funded by the
Spanish PNAYA, project AYA2010-21887 of the Spanish MINECO. R.A. Marino was
also funded by the spanish programme of International Campus of Excellence
Moncloa (CEI).

\end{acknowledgements}

\end{document}